\documentclass[a4paper,11pt]{article}
\pdfoutput=1
\usepackage{jheppub}

\usepackage[utf8]{inputenc}
\usepackage{subfig}
\usepackage{color}
\usepackage{verbatim}
\usepackage{textcomp}
\usepackage{amsmath}
\usepackage{amssymb}
\usepackage{graphicx}
\usepackage{nicefrac}

\usepackage{array}
\usepackage{multirow}

\usepackage{slashed}
\usepackage{cancel}

\newcommand{\parentheses}[1]{\ensuremath{\left( #1 \right)}}
\newcommand{\average}[1]{\ensuremath{\left\langle #1 \right\rangle}}
\newcommand{\absolutevalue}[1]{\ensuremath{\left| #1 \right|}}
\newcommand{\commutator}[1]{\ensuremath{\left[ #1 \right]}}
\newcommand{\anticommutator}[1]{\ensuremath{\left\lbrace #1 \right\rbrace}}

\usepackage{color}
\definecolor{SR}{rgb}{0,0.2,0.8}
\definecolor{JHG}{rgb}{0.5,0,0.5}
\definecolor{TO}{rgb}{1,0,1}
\definecolor{JW}{rgb}{0,0.7,0.3}

\preprint{ADP-17-1/T1007}

\title{Full parameter scan of the Zee model: exploring Higgs lepton flavor violation}

\author[a,\, b]{\textbf{Juan Herrero-Garc\'ia,}}
\author[b]{\textbf{Tommy Ohlsson,}}
\author[b]{\textbf{Stella Riad}}
\author[b]{\textbf{and Jens Wir\'en}}

\affiliation[a]{ARC Center of Excellence for Particle Physics at the Terascale, University of Adelaide, Adelaide, SA 5005, Australia}

\affiliation[b]{Department of Physics, School of Engineering Sciences, KTH Royal Institute of Technology,\\ AlbaNova University Center, Roslagstullsbacken 21, 106 91 Stockholm, Sweden}

\emailAdd{juan.herrero-garcia@adelaide.edu.au}
\emailAdd{tohlsson@kth.se}
\emailAdd{sriad@kth.se}
\emailAdd{jenswir@kth.se}

\abstract{
We study the general Zee model, which includes an extra Higgs scalar doublet and a new singly-charged scalar singlet. Neutrino masses are generated at one-loop level, and in order to describe leptonic mixing, both the Standard Model and the extra Higgs scalar doublets need to couple to leptons (in a type-III two-Higgs doublet model), which necessarily generates large lepton flavor violating signals, also in Higgs decays. Imposing all relevant phenomenological constraints and performing a full numerical scan of the parameter space, we find that both normal and inverted neutrino mass orderings can be fitted, although the latter is disfavored with respect to the former. In fact, inverted ordering can only be accommodated if $\theta_{23}$ turns out to be in the first octant. A branching ratio for $h \to \tau \mu$ of up to $10^{-2}$ is allowed, but it could be as low as $10^{-6}$. In addition, if future expected sensitivities of $\tau\to \mu\gamma$ are achieved, normal ordering can be almost completely tested. Also, $\mu e$ conversion is expected to probe large parts of the parameter space, excluding completely inverted ordering if no signal is observed. Furthermore, non-standard neutrino interactions are found to be smaller than $10^{-6}$, which is well below future experimental sensitivity. Finally, the results of our scan indicate that the masses~of the additional scalars have to be below $2.5$~TeV, and typically they are lower than that~and therefore within the reach of the LHC and future colliders.}


\makeatother

\begin{document}

\maketitle

\section{Introduction}

In the Standard Model (SM), neutrinos are massless and lepton flavors are exactly conserved to all orders. However, from neutrino oscillation experiments, we know that neutrinos are not massless and that lepton flavor is not conserved in the neutrino sector. 

Whether lepton number is a good symmetry of Nature or not remains an open question. If the SM is considered an effective field theory (EFT), the only dimension-5 operator is the Weinberg operator~\cite{Weinberg:1979sa}, where lepton number is broken by two units, and which gives rise to Majorana masses for the neutrinos after electroweak symmetry breaking. There are different realizations of this operator, both at tree level and one-loop level. High-scale realizations of the Weinberg operator, for instance the type-I seesaw mechanism~\cite{Minkowski:1977sc,Yanagida:1980,Gell-Mann:1980vs,Glashow:1979vf,Mohapatra:1979ia}, are well motivated by grand unified theories (GUTs), such as $\rm SO(10)$. However, these models are difficult to test, and therefore, other avenues should be explored, in particular in light of new data from the LHC and low-energy experiments. 

Indeed, there is hope to test scenarios in which lepton number violation (LNV) occurs close to the electroweak scale. One such scenario is radiative neutrino mass models, where neutrinos are massless at tree level, but acquire mass at one or more loops. Thus, the new degrees of freedom involved in the generation of neutrino mass cannot be too heavy, and therefore, they can be searched for at the LHC. These new particles typically give rise to enhanced lepton flavor violating (LFV) signals in processes like $\mu \to e \gamma$, $\tau\to 3 \mu$, or $\mu e$ conversion, which we denote as charged lepton flavor violation (CLFV). Furthermore, with the discovery of the Higgs boson~\cite{Aad:2012tfa,Chatrchyan:2012xdj}, the ways to search for LFV have increased and one can look for Higgs lepton flavor violating (HLFV) decays, especially in the $\tau$-$\mu$ and $\tau$-$e$ sectors, which are subject to weaker constraints from low-energy probes than the $\mu$-$e$ sector.

The ATLAS and CMS experiments at the LHC have active programs to search for HLFV decays. Considering their 13 TeV data sets, no signal has been observed.\footnote{The CMS 8 TeV data showed a $2.4 \sigma$ excess in the channel $h\to \tau \mu$ \cite{Khachatryan:2015kon}, which is translated into a branching ratio ${\rm Br}(h\to \tau \mu)=(0.84^{+0.39}_{-0.37})~\%$. This corresponds to an upper limit $\mathrm{Br}(h \to \tau \mu)<1.51~\%$ at $95~\%$ C.L.~\cite{Khachatryan:2015kon}. Unfortunately, this small excess has disappeared with the CMS 13 TeV data. Of course, an excess at a lower level of $\mathcal{O}(10^{-3})$ could be observed at the LHC in the forthcoming years.} In table~\ref{tab:HLFVsignals}, we summarize the latest 13 TeV upper limits on HLFV decays from ATLAS and CMS. The LHC is sensitive to ${\rm Br}(h\to \tau \mu,\, \tau e)\gtrsim 0.001$, and therefore, these decays can be used to test the models of new physics with HLFV signals in such a range. 
\begin{table}[h]
\begin{center}
  \begin{tabular}{ | c | c | c|}
    \hline
   HLFV observable & ATLAS & CMS \\ \hline \hline
    $\mathrm{Br}(h \to \tau \mu)$ & $1.43~\%$ ~\cite{Aad:2016blu} & $1.20~\%$~\cite{CMS:2016qvi}\\ \hline
    $\mathrm{Br}(h \to \tau e)$ & $1.04~\% $~\cite{Aad:2016blu} & $0.69~\%$~\cite{Khachatryan:2016rke} \\ \hline
       \end{tabular}
  \caption{\label{tab:HLFVsignals} Experimental $95~\%$ C.L. upper bounds on HLFV decays from ATLAS and CMS in the tau sector using the 13 TeV data sets. In our numerical scan, we will use the strongest upper bounds from CMS shown in this table.}
\end{center}
\end{table}

In recent years, several studies have analyzed if a $\mathrm{Br}(h \to \tau \mu)\sim 1~\%$ is compatible with low-energy constraints, either using an EFT approach~\cite{Goudelis:2011un, DiazCruz:1999xe, Blankenburg:2012ex,Harnik:2012pb, Celis:2013xja, Banerjee:2016foh, Belusca-Maito:2016axk, Herrero-Garcia:2016uab} or in a type-III two-Higgs doublet model (2HDM)~\cite{Davidson:2010xv,Sierra:2014nqa,Dorsner:2015mja,Omura:2015nja, Botella:2015hoa, Guo:2016ixx} (see also ref.~\cite{Arganda:2015naa} for a study of a supersymmetric inverse seesaw scenario). Reference~\cite{Herrero-Garcia:2016uab} is particularly relevant to us, where it was shown that the only tree-level scenarios, which can accommodate the excess, are models with extra scalars. Furthermore, the connection between HLFV decays and neutrino masses was extensively discussed\footnote{See also refs.~\cite{Arganda:2004bz,Arana-Catania:2013xma,Arganda:2014dta, Arganda:2015naa,Arganda:2015uca,Arganda:2016qvs} for HLFV studies in supersymmetric and seesaw scenarios.} and it was found that the most general version of the Zee model~\cite{Zee:1980ai,Cheng:1980qt,Wolfenstein:1980sy} (see also refs.~\cite{Petcov:1982en,Zee:1985id,Bertolini:1987kz,Bertolini:1990vz,Smirnov:1994wj,Smirnov:1996hc,Frampton:1999yn,Jarlskog:1998uf,Ghosal:2001ep,Kanemura:2000bq,Balaji:2001ex,Koide:2002uk,Brahmachari:2001rn,Frampton:2001eu,Assamagan:2002kf,He:2003ih,Kanemura:2005hr,AristizabalSierra:2006ri,He:2011hs,Babu:2013pma} for different works and variations on the Zee model) was the most promising one. In this simple model, an extra Higgs scalar doublet and a new singly-charged scalar singlet are added to the SM and Majorana neutrino masses are generated at one-loop level. In order to describe leptonic mixing correctly, it is necessary that both scalar doublets couple to the charged leptons in a type-III 2HDM, see e.g.~refs.~\cite{Frampton:2001eu,Koide:2002uk,He:2003ih}. This is precisely the same requirement needed to have large HLFV~\cite{Herrero-Garcia:2016uab}, and therefore, a complete analysis, taking into account all phenomenological constraints and performing a full numerical scan of the parameter space, is of great interest. This is the aim of this work, including a reduction of the allowed parameter space of the model by taking into account recent data, like the discovery of the Higgs boson~\cite{Aad:2012tfa,Chatrchyan:2012xdj}, the determination of the leptonic mixing angle $\theta_{13}$~\cite{Abe:2011fz,An:2012eh,Ahn:2012nd}, the results from the latest global fits of neutrino parameters including the hint of leptonic CP violation and the uncertainty on the octant of the leptonic mixing angle $\theta_{23}$ (i.e.~if $\theta_{23}$ is smaller or larger than $\pi/4$)~\cite{Gonzalez-Garcia:2014bfa,Esteban:2016qun,NuFIT3.0}, the new limits on CLFV processes like the ones on $\mu\to e \gamma$~\cite{TheMEG:2016wtm}, and HLFV processes like those shown in table~\ref{tab:HLFVsignals}. Finally, the impact of future expected limits will also be studied, in particular those coming from $\tau \to \mu \gamma$, where Belle~II is expected to reach a sensitivity of $10^{-9}$~\cite{HerediadelaCruz:2016yqi}, and specially $\mu e$ conversion, which is expected to improve by several orders of magnitude in the near future, see e.g.~refs.~\cite{Carey:2008zz, Kutschke:2011ux, Donghia:2016lzt, ProjectX, Barlow:2011zza, Witte:2012zza, Cui:2009zz,Kuno:2013mha}.

The paper is structured as follows. In section~\ref{sec:Zeemodel}, we describe the Zee model and its relevant parameters. In section~\ref{sec:pheno}, we discuss the phenomenological constraints of the model. Then, in section~\ref{sec:numerics}, we perform a numerical scan and present our results for three different scenarios: (i) without neutrino masses (just a type-III 2HDM with an extra charged singlet) and with neutrino masses for both (ii) normal and (iii) inverted neutrino mass orderings. Finally, in section~\ref{sec:conc}, we summarize our results and give our conclusions. In addition, in appendices~\ref{sec:ewpt} and \ref{sec:PV}, we present the contributions of the model to the electroweak precision test parameters $S$, $T$, and $U$ and derive explicit analytical expressions for various loop functions that these parameters are constructed from and which can be used for any model.

\section{The general Zee model}
\label{sec:Zeemodel}

In addition to the SM content with a Higgs scalar doublet $\Phi_1$, the Zee model~\cite{Zee:1980ai,Cheng:1980qt,Wolfenstein:1980sy} contains an extra Higgs scalar doublet $\Phi_2$ and a singly-charged scalar singlet $h^+$. We start by discussing the most general scalar potential.

\subsection{The scalar potential}
\label{sec:potential}

The following analysis is similar to the ones performed for 2HDMs, see e.g.~refs.~\cite{Gunion:2002zf,Davidson:2005cw, Branco:2011iw, Djouadi:2005gj}. One can start in a generic basis, where both Higgs scalar doublets $\Phi_{1}$ and $\Phi_{2}$ take VEVs denoted by $v_1$ and $v_2$, respectively. Then, one can rotate to the Higgs basis, where only $H_1$ takes a VEV denoted as usual as $v=\sqrt{v^2_1+v^2_2} \simeq 246$~GeV. The rotation is given by the following transformation~\cite{Davidson:2005cw}:
\begin{equation}
\label{eq:rotation_HB}
\parentheses{\begin{array}{c} H_1 \\ H_2 \end{array}} = \parentheses{\begin{array}{cc} c_\beta & s_\beta \\ -s_\beta & c_\beta \end{array}} \parentheses{\begin{array}{c} \Phi_1 \\ \Phi_2 \end{array}} \,,
\end{equation}
where $\tan\beta\equiv v_2/v_1$ and the short-hand notations $s_x \equiv \sin x$ and $c_x \equiv \cos x$.\footnote{In type-III 2HDM, $\tan \beta$ is an unphysical parameter~\cite{Davidson:2005cw}. For the lepton sector, it can be defined as the ratio of the tau Yukawa coupling (times the vev) and its mass. In general, the definition of $\tan \beta$ will be different for up and down quarks. In our numerical scan (see section~\ref{sec:scan}), we will treat it as an arbitrary free parameter.} We will also use $t_x \equiv \tan x$. In the Higgs basis, the doublets take the form:
\begin{equation}
\label{eq:doublets_HB}
H_1 = \parentheses{\begin{array}{c} G^+ \\ \dfrac{1}{\sqrt{2}}\parentheses{v + \varphi_1^0 + i G^0} \end{array}} \,, \qquad H_2 = \parentheses{\begin{array}{c} H^+ \\ \dfrac{1}{\sqrt{2}}\parentheses{\varphi_2^0 + i A} \end{array}} \,,
\end{equation}
where $\varphi_1^0$ and  $\varphi_2^0$ are CP-even neutral Higgs fields, $A$ is a CP-odd neutral Higgs field, $H^+$ is a charged Higgs field, and $G^+$ and $G^0$ are the would-be Goldstone bosons, which are eaten by the $W^+$ and the $Z$.
The most general potential for the Zee model (see e.g.~ref.~\cite{Davidson:2005cw}) is given in the Higgs basis by
\begin{align}
\label{eq:scalar_potential}
V &= \mu_1^2 H_1^\dagger H_1 + \mu_2^2 H_2^\dagger H_2 -  \parentheses{\mu_3^2 H_2^\dagger H_1 + {\rm H.c.}} + \dfrac{1}{2}\lambda_1 \parentheses{H_1^\dagger H_1}^2 \nonumber\\
&+ \dfrac{1}{2}\lambda_2 \parentheses{H_2^\dagger H_2}^2 + \lambda_3 \parentheses{H_1^\dagger H_1} \parentheses{H_2^\dagger H_2} + \lambda_4 \parentheses{H_1^\dagger H_2} \parentheses{H_2^\dagger H_1} \nonumber\\
&+ \anticommutator{\dfrac{1}{2}\lambda_5 \parentheses{H_1^\dagger H_2}^2 + \commutator{\lambda_6 \parentheses{H_1^\dagger H_1} + \lambda_7 \parentheses{H_2^\dagger H_2}}H_1^\dagger H_2 + {\rm H.c.}} \nonumber\\
&+ \mu_h^2 \absolutevalue{h^+}^2 + \lambda_h \absolutevalue{h^+}^4 + \lambda_8 \absolutevalue{h^+}^2 H_1^\dagger H_1 + \lambda_9 \absolutevalue{h^+}^2 H_2^\dagger H_2 \nonumber\\
&+ \lambda_{10} \absolutevalue{h^+}^2 \parentheses{H_1^\dagger H_2 + {\rm H.c.}} + \parentheses{\mu \epsilon_{\alpha \beta} H_1^{\alpha} H_2^{\beta} h^- + {\rm H.c.} } \,,
\end{align}
where $\lambda_i$ ($i = 1,2,\ldots,10,h$) are the quartic couplings, $\mu^2_i$ ($i=1,2,3,h$) are bare mass-squared parameters, and $\mu$ is a trilinear coupling. In addition, $\epsilon_{\alpha \beta}$ is the rank two antisymmetric Levi-Civita tensor. In general, $\lambda_{5}$, $\lambda_{6}$, $\lambda_{7}$, $\lambda_{10}$, $\mu_3$, and $\mu$ can be complex. Note that one can choose $\lambda_5$ to be real by redefining $H_1$ and $H_2$~\cite{Davidson:2005cw}. Furthermore, without loss of generality, $\mu$ can be chosen to be real and positive by redefining the singlet $h^-$. In addition, we choose $\lambda_6$ to be real for simplicity. In the numerical scan (see section~\ref{sec:scan}), we will treat the three quantities $\mu$, $\mu_2$, and $\mu_h$ as free real parameters, except for the case when we will set $\mu = 0$. In section~\ref{sec:stability}, we will comment on the usage of the quartic couplings $\lambda_i$ in the numerical scan.

Since only $H_1$ takes a VEV, differentiating eq.~\eqref{eq:scalar_potential} with respect to $H_1$ and $H_2$, gives the following minimization conditions
\begin{equation}
\mu_1^2 = - \dfrac{1}{2}\lambda_1 v^2 \,, \qquad \mu_3^2 = \dfrac{1}{2}\lambda_6 v^2\,, \label{eq:mu1mu3}
\end{equation}
which can be used to eliminate $\mu_1^2$ and $\mu_3^2$ as independent variables. Equation~\eqref{eq:mu1mu3} applies to both the real and imaginary parts.
Inserting $\average{H_1} = (0, v/\sqrt{2})^T$ into eq.~\eqref{eq:scalar_potential}, we obtain the squared mass matrices of the charged and neutral CP-even Higgs states. For the charged ones, in the Higgs basis $\parentheses{H^+, h^+}$, we have
\begin{equation}
\mathcal{M}_c^2= \left( \begin{array}{cc}
M_{H^+}^2 & -\mu v /\sqrt{2} \\
-\mu v /\sqrt{2} & M_{33}^2
\end{array} \right),
\end{equation}
where
\begin{equation}
M_{H^+}^2 = \mu_2^2 +\dfrac{1}{2} v^2 \lambda_3\,, \qquad  M_{33}^2 = \mu_h^2 + v^2 \lambda_8\,.
\end{equation}
The mass eigenstates $h_1^+$ and $h_2^+$ are a mixing of $h^+$ and $H^+$ given by
\begin{equation} \label{eq:charged_rot}
\begin{pmatrix} h_1^+ \\ h_2^+ \end{pmatrix} = 
\begin{pmatrix} s_\varphi & \,c_\varphi \\
c_\varphi & -s_\varphi  \end{pmatrix} 
\begin{pmatrix} h^+ \\ H^+ \end{pmatrix}\,,
\end{equation}
where
\begin{equation}
s_{2\varphi} = \dfrac{\sqrt{2} v \mu}{m_{h^+_2}^2-m_{h^+_1}^2}
\label{eq:charged_mixing}
\end{equation}
and the masses are defined as
\begin{equation} \label{eq:charged}
m_{h^+_1,h^+_2}^2 \equiv \dfrac{1}{2} \left[M^2_{H^+} + M^2_{33} \mp \sqrt{\parentheses{M^2_{H^+} - M^2_{33}}^2 + 2v^2 \mu^2 }\right] \,.
\end{equation}
Similarly, the CP-even mass matrix is
\begin{equation}
\mathcal{M}_h^2= \left( \begin{array}{cc}
\lambda_1 v^2 & \lambda_6 v^2 \\
\lambda_6 v^2 &\, m_{A}^2+\lambda_5 v^2
\end{array} \right)\,,
\end{equation}
where the mass of the CP-odd Higgs state enters as
\begin{equation}
m^2_{A} = M^2_{H^+} - \dfrac{1}{2} v^2 \parentheses{\lambda_5 - \lambda_4}\,.
\end{equation}
Thus, in the Higgs basis, the mass eigenstates $h$ and $H$ are a mixture of the CP-even states $\varphi_{1}$ and $\varphi_{2}$
\begin{equation}
\label{eq:CP-even_mixing}
\parentheses{\begin{array}{c} h \\ H \end{array}} = \parentheses{
\begin{array}{cc}
s_{\beta -\alpha} & c_{\beta -\alpha} \\
 c_{\beta -\alpha} & -s_{\beta -\alpha} \\
\end{array}
} \parentheses{\begin{array}{c} \varphi_1^0 \\ \varphi_2^0\end{array}}
\end{equation}
with the masses defined as
\begin{equation}\label{eq:lambda6}
m^2_{H,h} \equiv \dfrac{1}{2} \left\{m^2_{A} + v^2 \parentheses{\lambda_1 + \lambda_5} \pm \sqrt{\commutator{m^2_{A}+ v^2 \parentheses{\lambda_5 - \lambda_1}}^2 + 4v^4 \lambda_6^2 }\right\}\,,
\end{equation}
where the CP-even mixing is given by
\begin{equation} \label{sinbminua}
s_{2(\beta -\alpha)}=  -\dfrac{2\lambda_6 v^2}{m^2_{H}-m^2_{h}}
\end{equation}
that needs to be sufficiently close to zero (i.e.~the alignment limit) to give rise to a SM-like Higgs boson \cite{Gunion:2002zf}.

\subsection{The lepton sector}
\label{sec:leptons}

As we will see, in order to describe leptonic mixing, both Higgs scalar doublets must couple to the charged leptons, and thus, we are considering a type-III 2HDM, see e.g.~ref.~\cite{He:2003ih}. The most general Yukawa Lagrangian in the generic basis, where both Higgs fields take VEVs, reads
\begin{equation}
-\mathcal{L}_{L}=
\overline{L}\, (Y^\dagger_1 \Phi_1 + Y^\dagger_2 \Phi_2)e_{\rm R}  +  \overline{\tilde{L}}f\,L h^{+}+\mathrm{H.c.}  \,, \label{eq:yuk2d}
\end{equation}
where $L=(\nu_{\rm L},\,e_{\rm L})^T$ and $e_{\rm R}$ are the SU(2) lepton doublets and singlets, respectively, and $\tilde{L} \equiv i \sigma_2 L^c = i \sigma_2 C \overline{L}^T$ with $\sigma_2$ being the second Pauli matrix. Due to Fermi statistics, $f$ is an antisymmetric Yukawa matrix in flavor space (i.e.~$f^{\alpha\beta} = - f^{\beta\alpha}$), while $Y_1$ and $Y_2$ are completely general complex Yukawa matrices. Furthermore, the charged-lepton masses are given by
\begin{equation} \label{mass_ch}
m_E=\frac{v}{\sqrt{2}}\parentheses{c_\beta Y^\dagger_1+s_\beta Y^\dagger_2}\,.
\end{equation}
Note that we will work in the basis where $m_E$ is diagonal with real and positive elements $m_e$, $m_\mu$, and $m_\tau$. Moreover, $Y_2$ will be a general complex matrix and $Y_1$ can be expressed completely in terms of $m_E$ and $Y_2$ using eq.~\eqref{mass_ch}.

In the Higgs basis, we can rewrite eq.~\eqref{eq:yuk2d} using eq.~\eqref{eq:rotation_HB} as
\begin{equation}
-\mathcal{L}_{L}=
\overline{L} \left[\dfrac{\sqrt{2} m_E}{v} H_1 + \parentheses{\dfrac{Y_2^\dagger}{c_\beta}-\dfrac{\sqrt{2} m_E t_\beta}{v}} H_2 \right] e_{\rm R}  +  \overline{\tilde{L}} f L h^{+} + \mathrm{H.c.} \label{eq:yuk2c}
\end{equation}
Without loss of generality, rotating the lepton doublets $L^\alpha$ and the lepton singlets $e^\alpha_{\rm R}$ by the same phase (so that $m_E$ remains diagonal and positive), three phases from $f$ can be removed. However, note that the phases from $Y_2$ cannot be removed by lepton field redefinitions.

In the mass basis (also for massive neutrinos), using eq.~\eqref{eq:yuk2c}, the most general leptonic Lagrangian reads
\begin{align}
\label{eq:zee_lagrangian}
-\mathcal{L}_{L} &=\, \overline{\nu_{\rm L}}\,U^\dagger \parentheses{\dfrac{-\sqrt{2} m_E t_\beta}{v} + \dfrac{Y_2^\dagger}{c_\beta}} e_{\rm R}\parentheses{c_\varphi h_1^+ - s_\varphi h_2^+} + 2\,\overline{\nu_{\rm L}^c}\,U^T f e_{\rm L} \parentheses{-s_\varphi h_1^+ - c_\varphi h_2^+} \nonumber\\
&+\overline{e_{\rm L}}  \parentheses{\dfrac{-m_E s_\alpha}{v c_\beta} + c_{\beta-\alpha}\dfrac{Y^\dagger_2}{\sqrt{2} c_\beta}}e_{\rm R} h + \overline{e_{\rm L}} \parentheses{\dfrac{m_E c_\alpha}{v c_\beta} - s_{\beta-\alpha}\dfrac{Y^\dagger_2}{\sqrt{2} c_\beta}}\,e_{\rm R} H \nonumber\\
 &+ i\overline{e_{\rm L}}  \parentheses{-\dfrac{m_E  t_\beta}{v} + \dfrac{Y^\dagger_2}{\sqrt{2} c_\beta}}e_{\rm R} A + {\rm H.c.}
\end{align}
We define the following effective couplings for the neutral Higgs fields $h^0=(h, \; H, \; A)$, which will turn out to be useful for CLFV processes:
\begin{eqnarray}
\label{eq:vertex_factors_neutral}
g_{h^0}^1& =& (g_h^1,\,g_H^1,\,g_A^1)= \left(-\dfrac{s_\alpha}{c_\beta},\, \dfrac{c_\alpha}{c_\beta},\,-i\,t_\beta \right)\,,\\
g_{h^0}^2&=& (g_h^2,\,g_H^2,\,g_A^2)= \left(\dfrac{c_{\beta-\alpha} }{\sqrt{2} c_\beta},\,- \dfrac{s_{\beta-\alpha}}{\sqrt{2} c_\beta},\, \dfrac{i}{\sqrt{2} c_\beta} \right)
\end{eqnarray}
and for the charged Higgs fields $h_c= (h_1^+,\,h_2^+)$:
\begin{eqnarray}
\label{eq:vertex_factors_charged}
g^1_{h_c}&=&(g_{h_1^+}^1,\,g_{h_2^+}^1) =  \left(-\sqrt{2} t_\beta c_\varphi,\, \sqrt{2} t_\beta s_\varphi\right)\,,\\
g^2_{h_c}&=&(g_{h_1^+}^2,\,g_{h_2^+}^2) =  \left(\dfrac{c_\varphi}{c_\beta} ,\, - \dfrac{s_\varphi}{c_\beta} \right)\,,\\
g^3_{h_c}&=&(g_{h_1^+}^3,\,g_{h_2^+}^3) =  \left(-2 s_\varphi,\, -2\, c_\varphi \right)\,.
\end{eqnarray}
One can observe that $g_{h^0}^1$ is flavor conserving and proportional to $m_E/v$.

\subsection{Neutrino parameters}
\label{sec:nus}

A general $3 \times 3$ Majorana neutrino mass matrix $\mathcal{M}_{\nu}$, which is defined as an effective mass term in the Lagrangian $\mathcal{L}_{\nu}\equiv-1/2\, \overline{\nu_{\rm L}^{c}}\mathcal{M}_{\nu}\nu_{\rm L}+\mathrm{H.c.}$, can be written as 
\begin{equation}
\mathcal{M}_{\nu}= U D_\nu U^T \,, 
\label{eq:numass}
\end{equation}
where $\nu_{\rm L}$ is the left-handed neutrino flavor eigenfield with three lepton flavors, $D_\nu$ is a $3 \times 3$ diagonal matrix with positive real eigenvalues and $U$ is the $3 \times 3$ unitary leptonic mixing matrix, which relates the neutrino mass eigenfields $\nu_i$
($i=1,2,3$) with definite masses $m_i$ and the neutrino flavor eigenfields $\nu_\alpha$ ($\alpha=e,\mu,\tau$):
\begin{equation}
\nu_\alpha = \sum_{i=1}^3 \,U_{\alpha i} \, \nu_i\,.
\end{equation} 
The standard parametrization for $U$ is \cite{Patrignani:2016xmw}
\begin{align}
U=\left(\begin{array}{ccc}
c_{13}c_{12} & c_{13}s_{12} & s_{13}e^{-i\delta}\\
-c_{23}s_{12}-s_{23}s_{13}c_{12}e^{i\delta} & c_{23}c_{12}-s_{23}s_{13}s_{12}e^{i\delta} & s_{23}c_{13}\\
s_{23}s_{12}-c_{23}s_{13}c_{12}e^{i\delta} & -s_{23}c_{12}-c_{23}s_{13}s_{12}e^{i\delta} & c_{23}c_{13}\end{array}\right)\left(\begin{array}{ccc}
1 & 0 & 0\\
0 & e^{i\phi_1/2} & 0\\
0 & 0 & e^{i\phi_2/2}\end{array}\right)\,,
\label{UPMNS}
\end{align}
where $c_{ij}\equiv\cos\theta_{ij}$ and $s_{ij}\equiv\sin\theta_{ij}$ ($\theta_{12}$, $\theta_{13}$, and $\theta_{23}$ being the three leptonic mixing angles and $\theta_{12}, \theta_{13}, \theta_{23} \in [0,\pi/2)$). Furthermore, in eq.~(\ref{UPMNS}), $\delta$ is the Dirac CP-violating phase ($\delta \in [0,2\pi)$) and $\phi_{1}$ and $\phi_{2}$ are two Majorana CP-violating phases ($\phi_1,\phi_2 \in [0,4\pi)$). Using the three definite neutrino masses $m_1$, $m_2$, and $m_3$, we also define the two linearly-independent neutrino mass squared-differences $\Delta m_{21}^2 \equiv m_2^2 - m_1^2$ and $\Delta m_{31}^2 \equiv m_3^2 - m_1^2$, known as the small and large mass squared-differences, respectively, where the sign of $\Delta m_{31}^2$ is still unknown. The case $\Delta m_{31}^2 > 0$ is generally referred to as `normal neutrino mass ordering' (NO), whereas the case $\Delta m_{31}^2 < 0$ is known as the `inverted neutrino mass ordering' (IO). Note that using neutrino oscillation experiments, it is not possible to determine $\phi_1$ and $\phi_2$ nor the absolute neutrino mass scale. The most up-to-date best-fit values from global analyses of the ordinary neutrino oscillation parameters (i.e.~the leptonic mixing parameters and the neutrino mass-squared differences) are $\theta_{12} \simeq 34^\circ$, $\theta_{13} \simeq 8.5^\circ$, $\theta_{23} \simeq 42^\circ$ for NO and $\theta_{23} \simeq 50^\circ$ for IO, $\delta \simeq 1.5 \pi$, $\Delta m_{21}^2 \simeq 7.5 \cdot 10^{-5} \, {\rm eV}^2$, and $\Delta m_{31}^2 \simeq 2.5 \cdot 10^{-3} \, {\rm eV}^2$ for NO and $\Delta m_{31}^2 \simeq -2.5 \cdot 10^{-3} \, {\rm eV}^2$ for IO \cite{Gonzalez-Garcia:2014bfa,Esteban:2016qun,NuFIT3.0}. Similar values can also be found in refs.~\cite{Capozzi:2016rtj,Forero:2014bxa}. Note that both the first and second octants of $\theta_{23}$ are allowed from the global analyses~\cite{Esteban:2016qun,NuFIT3.0} for both orderings with a mild preference of the first (second) octant for NO (IO).

In addition to the ordinary neutrino oscillation parameters, the following effective neutrino parameters appear naturally in different contexts~\cite{Patrignani:2016xmw}
\begin{align}
m_{ee} &\equiv \left| \sum_{i=1}^ 3 m_i U_{ei}^2 \right| = \left| \left( m_1 c_{12}^2 + m_2 s_{12}^2 e^{i \phi_1} \right) c_{13}^ 2 + m_3 s_{13}^2 e^{i (\phi_2 - 2 \delta)} \right| \,,\\
m_{\nu_e} &\equiv \sqrt{\sum_{i=1}^3 m_i^2 |U_{ei}|^2} = \sqrt{m_1^2 c_{13}^2 c_{12}^2 + m_2^2 c_{13}^2 s_{12}^2 + m_3^2 s_{13}^2} \,,\\
\sum m_i & \equiv \sum_{i=1}^3 m_i = m_1 +  m_2 + m_3 \,,
\end{align}
where $m_{ee}$ (or $m_{0\nu 2\beta}$) is the effective electron neutrino mass parameter that could be measured in neutrinoless double beta decay ($0\nu 2\beta$) experiments \cite{Wolfenstein:1981rk,Bilenky:1987ty} (see also ref.~\cite{DellOro:2016tmg} for a recent review), $m_{\nu_e}$ (or $m_\beta$) is the effective neutrino mass parameter measured in (single) beta decay experiments \cite{McKellar:1980cn}, and finally, $\sum m_i $ is the sum of the three neutrino masses, which, in the future, could be determined by cosmology, but at present it is only restricted by an upper bound, see e.g.~refs.~\cite{Hannestad:2010yi,GonzalezGarcia:2010un}.

\subsection{Neutrino masses in the Zee model}
\label{sec:numasses}

\begin{figure}
	\centering
        \includegraphics[width=0.5\textwidth]{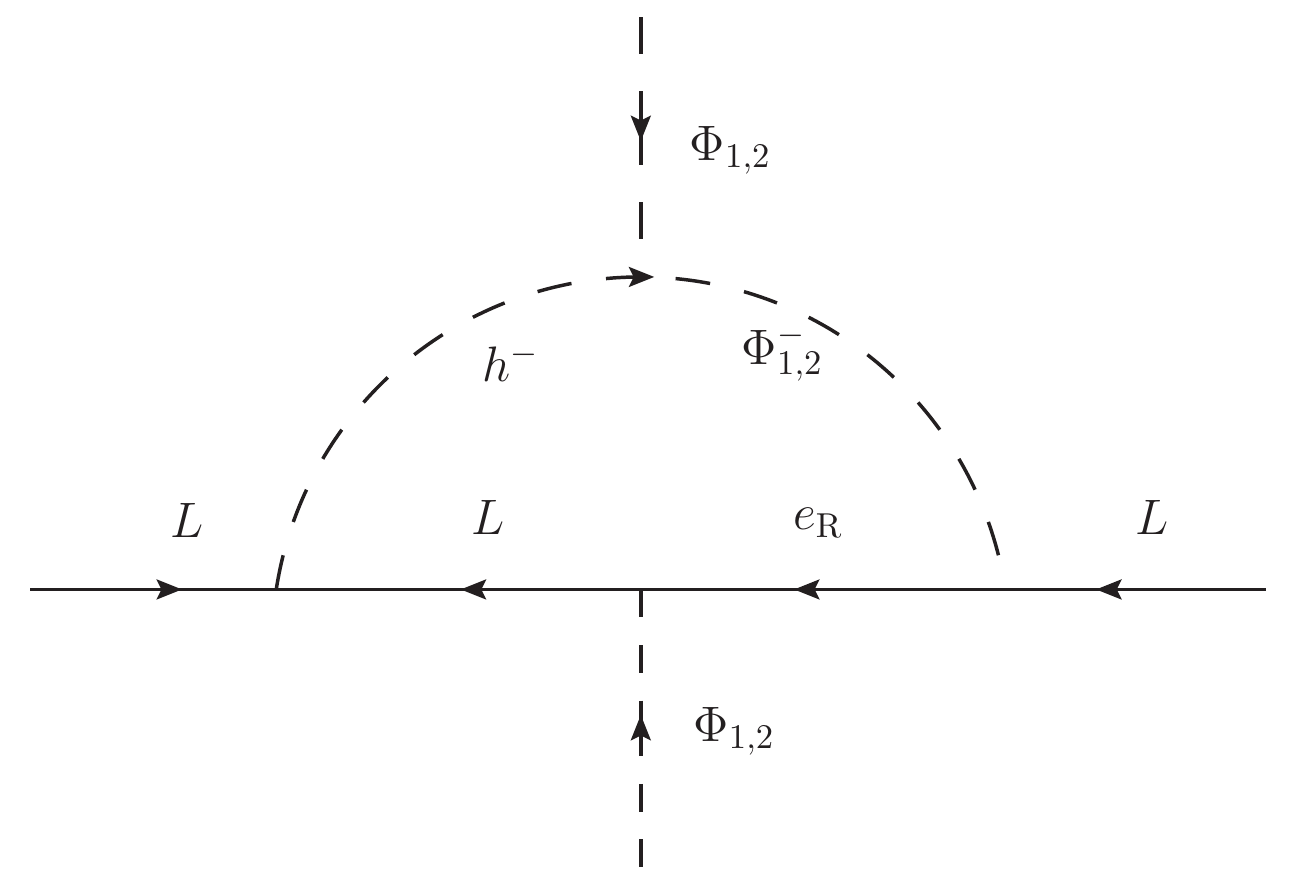}
	\caption{The Zee model diagram for neutrino masses.} \label{Zee}
\end{figure}

As can be seen from the potential and the Yukawa Lagrangian of the Zee model, eqs.~\eqref{eq:scalar_potential} and~\eqref{eq:zee_lagrangian}, respectively, in order to have LNV and therefore neutrino masses, we need the simultaneous presence of $Y_1$, $Y_2$, $f$, and $\mu$. In the Zee model, the one-loop diagram shown in figure~\ref{Zee}, where the charged scalars $h^+_1$ and $h^+_2$ run in the loop, generates neutrino masses. The complete neutrino mass matrix is then given by (see e.g.~ref.~\cite{He:2011hs})
\begin{equation}
{\cal M}_\nu= A\,\Big[f\,m_E^2+m_E^2f^T-\frac{v}{\sqrt{2}\,s_\beta}(f\,m_E\,Y_2+Y_2^T\,m_E\,f^T)\Big] \,,
\end{equation}
where we have defined
\begin{equation} \label{eq:Aeq}
A\equiv \frac{s_{2\varphi}\, t_\beta}{8 \sqrt{2}\pi^2\, v}\,\ln\frac{m^2_{h^+_2}}{m^2_{h^+_1}}
\end{equation}
with $\varphi$ being the mixing angle for the charged scalars given in eq.~\eqref{eq:charged_mixing}. Therefore, in the Zee model, due to loop and chiral suppression, the new physics scale can be light.

Assuming $f^{e\mu} = 0$, neglecting $m_e \ll m_\mu,\,m_\tau$, and keeping only the term proportional to $m_\mu$ in the 3-3 element,\footnote{Keeping the 3-3 element to order $m_\mu/m_\tau$ is phenomenologically relevant for the following two reasons. First, to have that all neutrinos are massive. Second, to obtain a constrain on $Y_2^{\mu\tau}$, which enters in $h\to \tau \mu$.  In our numerical analysis, we keep all terms proportional to $m_\mu$.} we obtain the following (symmetric) Majorana mass matrix
\begin{equation}
{\cal M}_\nu =A\, \frac{m_\tau v}{\sqrt{2}\,s_\beta} \left(\begin{array}{ccc}
-2 f^{e\tau} Y_2^{\tau e} & -f^{e\tau} Y_2^{\tau\mu} - f^{\mu\tau} Y_2^{\tau e} & \frac{\sqrt{2} s_\beta\,m_\tau}{v} f^{e\tau} - f^{e\tau} Y_2^{\tau\tau}\\
-f^{e\tau} Y_2^{\tau\mu} - f^{\mu\tau} Y_2^{\tau e} & -2 f^{\mu\tau} Y_2^{\tau\mu} & \frac{\sqrt{2} s_\beta m_\tau}{v} f^{\mu\tau} - f^{\mu\tau} Y_2^{\tau\tau}\\
\frac{\sqrt{2} s_\beta\,m_\tau}{v} f^{e\tau} - f^{e\tau} Y_2^{\tau\tau} & \frac{\sqrt{2} s_\beta m_\tau}{v} f^{\mu\tau} - f^{\mu\tau} Y_2^{\tau\tau} & 2 \frac{m_\mu}{m_\tau}f^{\mu\tau} Y_2^{\mu\tau}\end{array}\right)\,.
\label{Mnu}
\end{equation}
Note that in a simpler scenario, where one neglects terms proportional to $m_\mu$, one neutrino will be massless. On the other hand, taking terms proportional to $m_\mu$ into account, all neutrinos will obtain masses. 

In our analysis, we assume zero Yukawa couplings in the $e$-$\mu$ sector, i.e.~we assume $f^{e\mu} = 0$ and $Y_2^{\mu\mu} = Y_2^{\mu e} = Y_2^{e\mu} = Y_2^{ee} = 0$, which means that the non-zero Yukawa couplings are $f^{e\tau}$, $f^{\mu\tau}$, $Y_2^{\tau\tau}$, $Y_2^{\tau\mu}$, $Y_2^{\tau e}$, $Y_2^{\mu\tau}$, and $Y_2^{e\tau}$. We assume all these Yukawa couplings to be complex except for $f^{e\tau},\, f^{\mu\tau}$, and $Y_2^{e\tau}$ (which does not enter in neutrino masses), see the discussion in section~\ref{sec:leptons}. Thus, the Yukawa couplings will constitute eleven free real parameters in the numerical scan that will be described in section~\ref{sec:scan}.

In order to obtain correct mixing angles, we need both $Y_2^{\tau\mu}$ and $Y_2^{\tau e}$ different from zero, as they enter in the 1-2 submatrix of eq.~\eqref{Mnu}. Therefore, it is clear that reproducing the leptonic mixing angles correctly will imply restrictions on $\mathrm{Br}(h\to \tau \mu)$, $\mathrm{Br}(h\to \tau e)$, and other LFV processes. In fact, from this argumentation, a lower bound on the product $\mathrm{Br}(h\to \tau \mu)\cdot\mathrm{Br}(h\to \tau e)$ (in addition to an upper bound from other CLFV processes) is expected. 

\subsection{The (minimal) quark sector}
\label{sec:quarks}

Although the Zee model only deals with the lepton sector, the SM Higgs scalar doublet needs to couple to the SM quarks, like tops and bottoms, in order to be observed via its production and decay modes at the LHC~\cite{Khachatryan:2016vau}. In the generic basis, the most general Lagrangian in the quark sector is given by
\begin{equation}
-\mathcal{L}_{Q}=
\overline{Q}\, (Y^\dagger_{u1} \tilde \Phi_1 + Y^\dagger_{u2} \tilde \Phi_2)u_{\rm R}  +\overline{Q}\, (Y^\dagger_{d1} \Phi_1 + Y^\dagger_{d2} \Phi_2)\,d_{\rm R}+\mathrm{H.c.}  \,, \label{eq:yuk2dq}
\end{equation}
where $Q=(u_{\rm L},\,d_{\rm L})^T$ are the SU(2) quark doublets, $u_{\rm R}$ and $d_{\rm R}$ are the SU(2) quark singlets, and $\tilde{\Phi}_i \equiv i \sigma_2 \Phi_i^*$ ($i = 1,2$). However, flavor violation in the quark sector is severely constrained (see e.g.~ref.~\cite{Harnik:2012pb}), so we will assume the simplest scenario in which $Y_{d2} = Y_{u2}=0$. Then, we can use the basis, where the up-type quark mass matrix is diagonal. Furthermore, we assume the Yukawa couplings $Y_{d1}$ and $Y_{u1}$ to be Hermitian. The masses for the third generation quarks are given by $m_{b}= (Y^{33}_{d1})^* c_\beta v/\sqrt{2}$ and $m_{t}= (Y^{33}_{u1})^* c_\beta v/\sqrt{2}$.
Therefore, the interactions of the physical neutral Higgs fields with quarks are given by
\begin{align} \label{quarks}
g_{h\bar tt\,(h\bar bb)} &= -\dfrac{m_{t\,(b)} s_\alpha}{v c_\beta}\,, \qquad g_{H\bar tt\,(H\bar bb)} = \dfrac{m_{t\,(b)} c_\alpha}{v c_\beta}\,, \qquad g_{A\bar tt\,(A\bar bb)} = i\dfrac{m_{t\,(b)} t_\beta}{v}\,,
\end{align}
where the corresponding Feynman rule for the CP-odd scalar $A$ includes a $\gamma_5$. 

Note that if we had taken the couplings to quarks as general as possible, including the first generation, there would have been other phenomenological implications. In particular, related to neutrino masses, there would have been be new contributions to neutrinoless double beta decay and new universality and non-standard neutrino interactions with matter, stemming from interactions of the charged scalars $h^+_1$ and $h^+_2$. However, when naturality constraints are imposed on the Yukawa couplings to the leptons and the up and down quarks, see eq.~\eqref{nat}, these contributions are subdominant. We will therefore only discuss the universality constraints and the non-standard neutrino interactions generated through leptonic interactions, see section~\ref{sec:charged_lep}, and we will only consider the contributions to neutrinoless double beta decay mediated by $W$ bosons, i.e.~the contributions from the light neutrinos.

\section{Phenomenology}
\label{sec:pheno}

\subsection{Stability of the potential}
\label{sec:stability}

A Hamiltonian in quantum mechanics has to be bounded from below, which requires the quartic part of the scalar potential in eq.~\eqref{eq:scalar_potential} to be positive for all values of the fields and for all scales. 
Then, if two of the three fields $H_1,\,H_2$, and $h$ vanish, one immediately finds
\begin{equation}
\label{eq:stability}
\lambda_1 \geq 0 \,, \qquad \lambda_2 \geq 0 \,, \qquad \lambda_h \geq 0\,.
\end{equation}

For a general 2HDM potential with $\lambda_6=\lambda_7=0$, it has been shown in ref.~\cite{Ivanov:2008er} that the additional necessary conditions are
\begin{equation}
\label{eq:stability_2}
\lambda_3 > - \sqrt{\lambda_1\lambda_2}\,, \qquad \lambda_3+\lambda_4-|\lambda_5| > - \sqrt{\lambda_1\lambda_2} \,.
\end{equation}
However, when $\lambda_6, \lambda_7 \neq 0$, it has been shown that in addition to the previous conditions, the following condition~\cite{Ferreira:2009jb}
\begin{equation}
\label{eq:stability_extended}
2 \absolutevalue{\lambda_6 + \lambda_7} < \dfrac{\lambda_1 + \lambda_2}{2} + \lambda_3 + \lambda_4 + \lambda_5
\end{equation}
is both necessary and sufficient to ensure stability of the potential. Other stability conditions for similar potentials are discussed in refs.~\cite{Kanemura:2000bq,Kannike:2012pe,Herrero-Garcia:2014hfa}.

In this work, due to the large number of parameters, we will set $\lambda_4 = \lambda_7 = \lambda_8 = \lambda_9 = \lambda_{10} = \lambda_h = 0$, since they do not significantly impact phenomenology, even though their presence is expected to somewhat open the allowed parameter space. Thus, the four free Higgs couplings are $\lambda_1$, $\lambda_2$, $\lambda_3$, and $\lambda_5$, which we will treat as free real parameters, while $\lambda_6$ is a derived parameter that can be computed from eq.~\eqref{eq:lambda6}. In the numerical scan (see section~\ref{sec:scan}), we impose the conditions from eqs.~\eqref{eq:stability}--\eqref{eq:stability_extended}.

\subsection{Naturality and perturbativity}
\label{sec:naturality}

There are naturality and perturbativity constraints on the Yukawa couplings and on the quartic and trilinear couplings of the potential. In order not to have large fine-tuned cancellations between the different Yukawa couplings (see e.g.~refs.~\cite{Dorsner:2015mja,Herrero-Garcia:2016uab}), we demand that
\begin{equation} \label{nat}
\dfrac{v^ 2}{2} Y_2^{\tau\mu}Y_2^{\mu\tau} s_\beta^2 \leq m_\mu m_\tau \,,\qquad \dfrac{v^ 2}{2} Y_2^{\tau e}Y_2^{e\tau} s_\beta^2 \leq m_e m_\tau\,.
\end{equation}

One can also obtain an upper bound on $\mu$, which contributes to the scalar masses. In fact, using eq.~\eqref{eq:charged}, it is clear that naturality demands that $\mu \lesssim \sqrt{2} m^2_{h^+_2}/v$. Interestingly, we can also derive a natural upper bound using the $125$ GeV Higgs boson~\cite{Aad:2012tfa,Chatrchyan:2012xdj}, due to the fact that $\mu$ contributes at one-loop level to its mass. The relevant coupling of the light Higgs boson to the charged scalars induced by $\mu$ is
\begin{align}
-s_{\beta-\alpha}\, \frac{\mu}{\sqrt{2}v}\,\, \left[s_{2\varphi} (h_1^- h_1^+- h_2^- h_2^+)+
c_{2\varphi} (h_1^+ h_2^- + h_1^- h_2^+)\right]\,h\,.
\end{align}
We demand that the one-loop contribution to the Higgs mass fulfills $\delta m_h/m_h \lesssim \kappa$, where we choose $\kappa =1\, (10)$, which corresponds to \emph{no} (10~\%) fine-tuning.\footnote{The fine-tunings in the Higgs mass squared, which is the relevant parameter in the Lagrangian, would be 1~\% (100~\%) for $\kappa = 1\, (10)$.} Neglecting logarithms and factors of two in the Higgs self-energies, we obtain
\begin{equation}\label{eq:mu}
\mu\, \lesssim \kappa\,\frac{4 \pi\,m_h}{s_{\beta-\alpha}} \simeq1.5\,  \left(\frac{\kappa}{s_{\beta-\alpha}}\right)\, {\rm TeV}\,.
\end{equation} 
Taking $s_{\beta-\alpha} \sim 1$, we find an upper bound of $1.5\,(15)$~TeV for $\kappa=1\,(10)$. In addition, we impose that all the quartic couplings are perturbative:
\begin{equation}\label{eq:lambda_i}
|\lambda_i| \leq \sqrt{4 \pi}\,, \quad i = 1,2,3,5\,.
\end{equation} 

\subsection{Charged lepton flavor violation and electric and magnetic moments}
\label{sec:clfv}

\subsubsection{Trilepton decays}
\label{sec:trileptons}

The presence of the second Higgs doublet gives rise to tree-level trilepton decays $\ell_i \to \ell_j \overline{\ell_k} \ell_l$.\footnote{At one loop and two loops, there are dipole contributions which dominate the rate. However, these are strongly bounded by $\tau \to \mu \gamma$ and $\mu \to e \gamma$, see section~\ref{sec:muegamma}. Also, box diagrams are very suppressed, see ref.~\cite{Mitsuda:2001vh}.} The ratio of branching ratio reads
\begin{equation}
\label{eq:trileptondecay}
\frac{\mathrm{Br}(\ell_i \to \ell_j \overline{\ell_k} \ell_l)}{\mathrm{Br}(\ell_i \to \ell_j \overline{\nu_k} \nu_l)} = \dfrac{1}{32\,G_F^2} \parentheses{ \xi \absolutevalue{D_{\rm LL}}^2+ \xi \absolutevalue{D_{\rm RR}}^2+ \absolutevalue{D_{\rm LR}}^2 + \absolutevalue{D_{\rm RL}}^2}\,,
\end{equation}
where $G_F = g^2/(\sqrt{2} v^2) \simeq 1.166 \cdot 10^{-5} \, {\rm GeV}^{-2}$ is the Fermi coupling constant and the Wilson coefficients $D_{\rm PP'}$ (${\rm P,P' = L,R}$) are given by the coherent sum of the contributions from the neutral Higgs fields. In addition, $\xi=1/2\,(1)$ when there are two (no) indistinguishable particles in the final state. We are interested in tau decays.\footnote{Other processes, like $\mu\to e \overline{e} e$, are absent at tree level, since we assume $Y_2^{\mu e}=Y_2^{ e\mu}=0$. Also, tree level contributions to $\tau \to \mu e \bar e$ and $\tau \to e e \bar e$ are suppressed by $m_e$.} In this case, for $\tau \to \mu \mu \bar \mu$, the Wilson coefficient $D_{\rm LL}$ is given by
\begin{equation}
D_{\rm LL}= -\frac{1}{2\,m^2_{h^0}}\, (g^2_{h^0})^* Y_2^{\mu\tau} (g^1_{h^0})^* \frac{m_\mu}{v}\,,
\end{equation}
where $h^0=(h,\,H,\,A)$. Similarly, for $\tau \to e \mu \bar \mu$, one can simply substitute $Y_2^{\mu\tau} \to Y_2^{e\tau}$. Furthermore, $D_{\rm RL}$ is obtained from $D_{\rm LL}$ by changing $(g^1_{h^0})^* \to g^1_{h^0}$. Finally, $D_{\rm RR}$ ($D_{\rm LR}$) is obtained from $D_{\rm LL}$ ($D_{\rm RL}$) by making the replacement $Y_2^{\alpha\beta} \to (Y_2^{\beta\alpha})^*$ and conjugating the vertex factors.

As expected, these tree-level processes do not restrict the parameter space as much as $\tau \to \mu \gamma$, $\tau \to e \gamma$, or $\mu \to e \gamma$ does, since they always involve a muon mass suppression (squared) and are therefore irrelevant. In table~\ref{tab:upperbounds}, the upper bounds used in the numerical scan for the various observables are presented.

\subsubsection{$\ell_i \to \ell_j \gamma$ decays}
\label{sec:muegamma}

One of the most constrained CLFV process is the radiative process $\ell_i \to \ell_j \gamma$ with $\ell_i$ being the physical charged leptons $e$, $\mu$, and $\tau$. This process always arises at loop level and it can be viewed as stemming from an effective operator of the form \citep{Davidson:2010xv}
\begin{equation}
\mathcal{L}_{\rm eff} = \dfrac{C_{ij}'}{\Lambda^2} \dfrac{v}{\sqrt{2}}\,\overline{\ell}_i {\rm P_R} \sigma^{\mu \nu} \ell_j F_{\mu \nu} + {\rm H.c.} \,,
\end{equation}
where $\Lambda$ is the scale of new physics, $\sigma^{\mu \nu} = i [\gamma^\mu, \gamma^\nu]/2$, and $F_{\mu \nu}=\partial_\mu A_\nu - \partial_\nu A_\mu$ is the electromagnetic field strength tensor with $A_\mu$ being the photon field. It is useful to define
\begin{equation}
\label{eq:C}
\dfrac{C_{ij}'}{\Lambda^2} \dfrac{v}{\sqrt{2}} \equiv \dfrac{e m_{i} C_{\rm R}^{ij}}{2}\,,
\end{equation}
and similarly, $C_{\rm L}=C^\dagger_{\rm R}$. Then, it follows that
\begin{equation}
\dfrac{\mathrm{Br}(\ell_i\to \ell_j\gamma)}{\mathrm{Br}(\ell_i\to \ell_j \overline{\nu_j} \nu_i)} = \dfrac{48 \pi^3 \alpha}{G_F^2}\parentheses{\absolutevalue{C_{\rm L}}^2 + \absolutevalue{C_{\rm R}}^2}\,.
\end{equation}
We use the expressions for the Wilson coefficients given in refs.~\cite{Goudelis:2011un, Blankenburg:2012ex,Harnik:2012pb}, adapted to the Zee model. We also include the two-loop Barr--Zee contributions as given in ref.~\cite{Chang:1993kw}. At one-loop level, the dominant contribution for $\tau \to \mu\gamma$ reads
\begin{align}
C_{\rm L}^0 & \simeq \dfrac{1}{16 \pi^2} \sum_{h^{0}} \dfrac{1}{m_{h^0}^2} (g^{2}_{h^0})^* Y_2^{\mu\tau} \nonumber\\
&\times \left\{ \frac{1}{6} \left[ g^1_{h^0} \frac{m_\tau}{v} + g^2_{h^0} (Y_2^{\tau\tau})^* \right] + \left[(g^{1}_{h^0})^* \frac{m_\tau}{v} + (g^{2}_{h^0})^* Y_2^{\tau\tau}\right] \parentheses{\ln \dfrac{m_{h^0}^2}{m_\tau^2}-\dfrac{3}{2}} \right\}\,, \label{eq:tau_mu_gamma_neutral}
\end{align}
and similarly, we obtain $C_{\rm R}^{0}$ with the replacement $Y_2^{\alpha\beta}\to (Y_2^{\beta\alpha})^*$. For the charged scalars, we find (using $\sum_i |U_{\tau i}|^2\approx 1$)
\begin{align}
\label{eq:tau_mu_gamma_charged}
C_{\rm L}^{+} & \simeq -\dfrac{1}{16 \pi^2} \dfrac{1}{12}  \sum_{i=1,2}\dfrac{1}{m_{h_i^+}^2} \left[  g_{h_i^+}^1 \frac{m_\tau}{v} + g_{h_i^+}^2 (Y^{\tau \tau}_2)^* \right] g_{h_i^+}^1 Y^{\mu \tau}_2 \,, \\
C_{\rm R}^{+} & \simeq 0\,,
\end{align}
where the contribution to $C_{\rm R}^+$ is zero, since we assume $f^{e\mu}=0$. 

The total contribution to the Wilson coefficient is $C_{\rm L}=C_{\rm L}^0+C_{\rm L}^++C_{\rm L}^{\mbox{\scriptsize 2-loops}}$, and similarly for $C_{\rm R}$.
For $\mu \to e \gamma$, the dominant contributions, proportional to $m_\tau$, are given by the neutral Higgs fields and $\tau$ running in the loop
\begin{align}
C_{\rm L}^{0} & \simeq \dfrac{1}{16 \pi^2}\, \left(\frac{m_\tau}{m_\mu}\right)\,\sum_{h^{0}} \dfrac{1}{m_{h^0}^{2}}\commutator{(g^2_{h^0})^2 Y_2^{\tau\mu} Y_2^{e\tau}}\parentheses{\ln \dfrac{m_{h^0}^2}{m_\tau^2}-\dfrac{3}{2}}\,, \label{eq:muegamL}\\
C_{\rm R}^{0}  & \simeq \dfrac{1}{16 \pi^2}\, \left(\frac{m_\tau}{m_\mu}\right)\,\sum_{h^{0}} \dfrac{1}{m_{h^0}^2}\commutator{(g^2_{h^0})^2 (Y_2^{\mu\tau})^* (Y_2^{\tau e})^*}\parentheses{\ln \dfrac{m_{h^0}^2}{m_\tau^2}-\dfrac{3}{2}} \,. \label{eq:muegamR}
\end{align}
Note that we also add the contributions from $f$ to $\mu \to e \gamma$, which are proportional to $\,m_\mu$, with $\nu_\tau$ running in the loop:
\begin{align} \label{eq:muegamcharged}
C_{\rm L}^{+} & \simeq 0\,,\\
C_{\rm R}^{+} & \simeq -\dfrac{1}{16 \pi^2} \dfrac{1}{12}  \sum_{i=1,2}\dfrac{1}{m_{h_i^+}^2} (g_{h_i^+}^3)^2  (f^{e\tau})^*f^{\mu\tau} \,. \label{eq:CR+}
\end{align}
The contribution in eq.~\eqref{eq:CR+} strongly constraints the antisymmetric Yukawa coupling $f$ of the singly-charged scalar singlet.

\subsubsection{Electron and muon electric dipole and anomalous magnetic moments}
\label{sec:edms}

When flavor is conserved, anomalous magnetic moments (AMMs) are generated. The dominant contributions are given by loops with neutral Higgs fields and tau leptons, similar to those in eqs.~\eqref{eq:muegamL} and \eqref{eq:muegamR}. It can be defined as~\cite{Blankenburg:2012ex}
\begin{equation}
\dfrac{e a_\mu}{4 m_\mu} \equiv \dfrac{C'+C^{' \dagger}}{\Lambda^2}\dfrac{v}{\sqrt{2}}\,.
\end{equation}
Then, using eq.~\eqref{eq:C}, the muon AMM is given by
\begin{equation}
a_\mu = 2 m_\mu^2  \mathrm{Re} ( C_{\rm L} + C_{\rm R} )\,.
\end{equation}
Similarly, the electron AMM is obtained by replacing the indices $\mu \to e$ everywhere. 

For the electron AMM, there is no disagreement between theory and experiment, and in fact, it represents one of the most precisely measured quantities in all of physics with an experimental 90~\% C.L.~upper bound on possible new physics contributions of $2 \cdot 10^{-12}$ \cite{Patrignani:2016xmw}. On the other hand, for the muon AMM, there is an experimental deviation from the theoretical prediction of the SM, i.e.~$\absolutevalue{\delta a_{\mu}}=(2.88\pm 0.80)\cdot 10^{-9}$~\cite{Patrignani:2016xmw}. Unfortunately, neither a 2HDM nor the Zee model can accommodate this discrepancy~\cite{Dorsner:2015mja}. 

If there is CP violation, electric dipole moments (EDMs) are generated, which can be defined as
\begin{equation}
i \dfrac{d}{2} \equiv \dfrac{C'-C^{' \dagger}}{\Lambda^2}\dfrac{v}{\sqrt{2}}\,,
\end{equation}
which gives
\begin{equation}
\frac{d}{e} = m_\mu\, \mathrm{Im} \parentheses{ C_{\rm L} - C_{\rm R}}\,.
\end{equation}
The experimental 90~\% C.L.~upper bounds for the muon and electron EDMs are $1 \cdot 10^{-19} \, e$~cm \cite{Patrignani:2016xmw} and $8.7 \cdot 10^{-29} \, e$~cm \cite{Baron:2013eja}, respectively (see also table~\ref{tab:upperbounds}). Therefore, for the muon EDM, this bound is almost non-existing, whereas the electron one strongly constrains the imaginary parts of $Y_2^{\tau e}$ and $Y_2^{e\tau}$ (in our scenario, $Y_2^{e\tau}$ is assumed to be real for simplicity).

\subsubsection{$\mu e$ conversion in nuclei}
\label{sec:mueconv}

Finally, $\mu e$ conversion is a very interesting process because the sensitivity in the next generation experiments is expected to increase by around four orders of magnitude. As the dipole contribution is already heavily constrained by $\mu\to e\gamma$, we can use the current limits to restrict the monopole photonic contribution. In the limit, where the transferred momentum is zero, the conversion rate, relative to the muon capture rate, can be expressed in gold as~\cite{Kitano:2002mt}
\begin{equation}
\mathrm{Cr}(\mu \to e)_{\rm  Au} =\dfrac{(V^p_{\rm  Au})^2 \parentheses{\absolutevalue{g_{\rm LV}}^2 + \absolutevalue{g_{\rm RV}}^2}}{\Gamma_{\rm  Au}}\,,
\end{equation}
where $V^p_{\rm Au}=0.0974\,m_\mu^{5/2}$, $\Gamma_{\rm  Au}=8.6 \cdot 10^{-18}\,{\rm GeV}$~\cite{Kitano:2002mt}, and the vector current coefficient is given by~\cite{Dorsner:2015mja}
\begin{equation}
g_{\rm LV} = \sum_{h^0}  g_{h^0}^2 (g_{h^0}^2)^* \left(\frac{\alpha}{2\pi}\right) \left(\frac{-1}{9 m_{h^0}^2}\right)\left(4+3\,\ln\frac{m^2_\tau}{m^2_{h^0}}\right) Y_{2}^{\tau \mu} (Y_2^{\tau e})^*\,. \label{eq:gLV}
\end{equation}
Similarly, $g_{\rm RV}$ is obtained by replacing $Y_2^{\alpha\beta} \to (Y_2^{\beta\alpha})^*$ in eq.~\eqref{eq:gLV}. The currently best experimental limit is $\mathrm{Cr}(\mu \to e)_{\rm Au} < 7\cdot 10^{-13}$ at 90~\% C.L.~by the SINDRUM II experiment at PSI~\cite{Bertl:2006up}.

For titanium, the best experimental limit is $\mathrm{Cr}(\mu \to e)_{\rm Ti} < 4.3 \cdot 10^{-12}$ at 90~\% C.L.~\cite{Bertl:2006up}. In the future, PRISM/PRIME~\cite{Barlow:2011zza, Witte:2012zza} expects to achieve a sensitivity of $\mathcal{O}(10^{-18})$. For aluminium, the future experimental sensitivity is $\mathrm{Cr}(\mu \to e)_{\rm Al} < 6 \cdot 10^{-17}$ at 90~\% C.L.~in the Mu2e experiment at Fermilab~\cite{Carey:2008zz, Kutschke:2011ux, Donghia:2016lzt} and of $\mathcal{O}(10^{-17})$ in the COMET experiment at J-PARC~\cite{Cui:2009zz,Kuno:2013mha}. Mu2e may also run at Project X using either Al or Ti~\cite{ProjectX} with an expected sensitivity of $\mathcal{O}(10^{-19})$. In the numerical analysis, we will discuss how the Zee model will be constrained if the expected sensitivity of this process is achieved in the future.

\subsection{Leptonic interactions of the charged scalars} \label{sec:charged_lep}

Now, we study four-lepton interactions of the singly-charged scalar mass eigenstates $h_1^+$ and $h_2^+$ that involve two charged leptons and two neutrinos.\footnote{Since the new Higgs scalar doublet also couples to quarks (to the third generation in our scenario), there are four-fermion interactions between quarks and leptons, which we do not analyze any further, see the discussion at the end of section~\ref{sec:quarks}.} These give rise to muon and tau decays into lighter charged leptons and neutrinos as well as to non-standard neutrino interactions (NSIs), see e.g.~refs.~\cite{Bilenky:1993bt,Davidson:2003ha}.

At tree level, we can integrate out $h_1^+$ and $h_2^+$, see e.g.~ref.~\cite{Bilenky:1993bt}. Therefore, using eq.~\eqref{eq:zee_lagrangian} and expanding to first order in $p^2/m^2_{h_{1,2}^+}$, we obtain
\begin{align}\label{eq:zee_effective}
\mathcal{L}_{\rm eff} &=\, \dfrac{1}{\tilde{M}_1^2}[\overline{e_{\rm R}}\,(\mathcal{Y}_1^{\rm eff})^\dagger \, \nu_{\rm L}][\overline{\nu_{\rm L}}\,\mathcal{Y}_1^{\rm eff}\, e_{\rm R}] +\dfrac{1}{\tilde{M}_{2}^2} [\overline{e_{\rm L}}\,(\mathcal{Y}_2^{\rm eff})^\dagger\,  \nu^c_{\rm L}][\overline{\nu_{\rm L}^c}\,\mathcal{Y}_2^{\rm eff}\,  e_{\rm L}]\nonumber\\
&+ \dfrac{1}{\tilde{M}_{12}^2} \left\{[\overline{e_{\rm R}}\,(\mathcal{Y}_1^{\rm eff})^\dagger\,  \nu_{\rm L}][\overline{\nu_L^c}\,\mathcal{Y}_2^{\rm eff}\,  e_{\rm L}]+[\overline{e_{\rm L}}\,(\mathcal{Y}_2^{\rm eff})^\dagger\,  \nu^c_{\rm L}][\overline{\nu_{\rm L}}\,\mathcal{Y}_1^{\rm eff}\,  e_{\rm R}] \right\}\,,
\end{align}
where we have defined the effective Yukawa couplings
\begin{equation}
\mathcal{Y}_1^{\rm eff} \equiv -U^\dagger \parentheses{\dfrac{-\sqrt{2} m_E t_\beta}{v} + \dfrac{Y_2^\dagger}{c_\beta}}\,, \qquad
\mathcal{Y}_2^{\rm eff} \equiv -2 U^T f\label{eq:yukawas_charged_2}
\end{equation}
and the effective masses
\begin{equation}
\dfrac{1}{\tilde{M}_1^2}\equiv \dfrac{c^2_\varphi}{m^2_{h_1^+}} + \dfrac{s^2_\varphi}{m^2_{h_2^+}} \,,\qquad
\dfrac{1}{\tilde{M}_2^2} \equiv \dfrac{s^2_\varphi}{m^2_{h_1^+}} + \dfrac{c^2_\varphi}{m^2_{h_2^+}}\,,\qquad
\dfrac{1}{\tilde{M}^2_{12}} \equiv\dfrac{\sqrt{2} v \mu}{m_{h^+_1}^2 m_{h^+_2}^2} \label{eq:masses_charged_12}\,.
\end{equation}
Note that for the definition of $\tilde{M}_{12}^2$ we have used the charged-scalars mixing $s_{2\varphi}$ as defined in eq.~\eqref{eq:charged_mixing}.

\subsubsection{Universality}
\label{sec:univ}

The second operator in eq.~\eqref{eq:zee_effective}, which is second order in $f$, i.e.~$|\mathcal{Y}_2^{\rm eff}|^2\propto f^\dagger f$, couples to the left-handed leptons, like charged currents in the SM. This implies that it interferes constructively with the $W$ boson. In the SM, the Fermi constant extracted from muon decay $G_\mu^{\rm SM}$ and the one extracted from hadronic decays $G_\beta^{\rm SM}$ are tested to be equal with great precision. The presence of the charged scalars modifies the muon decay rate~\cite{Santamaria1987,Nebot:2007bc}, which implies that $G_{\beta}^{\rm SM}=G_{\mu}^{\rm SM} \neq G_{\mu}^{\rm Zee}$, where $G_\mu^{\rm Zee}$ is the Fermi constant from muon decay in the Zee model. Therefore, we have
\begin{equation}
\parentheses{\dfrac{G^{\rm Zee}_\mu}{G_\mu^{\rm SM}}}^2 = 1 + \dfrac{\sqrt{2}}{G_F \tilde{M}_2^2} \absolutevalue{f^{e\mu}}^2 + \mathcal{O}(m^{-4}_{h_{1,2}^+})\,,
\end{equation}
where $\tilde{M}_2$ is defined in eq.~\eqref{eq:masses_charged_12}.

In the SM, unitarity of the quark mixing matrix $V$ holds to great precision. In our scenario, as we assume $f^{e\mu}=0$, we also have that $V$ is unitary up to order $\mathcal{O}(1/m^4_{h_{1,2}^+})$:
\begin{equation}
\absolutevalue{V^{\rm exp}_{ud}}^2 + \absolutevalue{V^{\rm exp}_{us}}^2 + \absolutevalue{V^{\rm exp}_{ub}}^2 = \parentheses{\dfrac{G_\mu^{\rm SM}}{G^{\rm Zee}_{\mu}}}^2 = 1 + \mathcal{O}(m^{-4}_{h_{1,2}^+})\,.
\end{equation}
On the hand, other leptonic decays may not be universal (in the SM, they are mediated by gauge interactions and are therefore universal). The ratio of flavor violating decays can be tested among the different generations via the effective couplings given by
\begin{align}
\parentheses{\dfrac{g^{\rm exp}_{\tau}}{g^{\rm exp}_{\mu}}}^2 = & \parentheses{\dfrac{G^{\rm Zee}_{\tau \to e}}{G^{\rm Zee}_{\mu \to e}}}^2 \equiv g_{\tau \mu} \approx 1 + \dfrac{\sqrt{2}}{G_F \tilde{M}_2^2} \absolutevalue{f^{e\tau}}^2\,, \label{eq:gtm}\\
\parentheses{\dfrac{g^{\rm exp}_{\tau}}{g^{\rm exp}_{e}}}^2 = & \parentheses{\dfrac{G^{\rm Zee}_{\tau \to \mu}}{G^{\rm Zee}_{\mu \to e}}}^2 \equiv g_{\tau e} \approx 1 + \dfrac{\sqrt{2}}{G_F \tilde{M}_2^2} \absolutevalue{f^{\mu \tau}}^2\,,\label{eq:gte}\\
\parentheses{\dfrac{g^{\rm exp}_{\mu}}{g^{\rm exp}_{e}}}^2 = & \parentheses{\dfrac{G^{\rm Zee}_{\tau \to \mu}}{G^{\rm Zee}_{\tau \to e}}}^2 \equiv g_{\mu e}\approx 1 + \dfrac{\sqrt{2}}{G_F \tilde{M}_2^2} \parentheses{\absolutevalue{f^{\mu \tau}}^2 - \absolutevalue{f^{e \tau}}^2}\,. \label{eq:gme}
\end{align}
In our scenario the expressions~\eqref{eq:gtm}--\eqref{eq:gme} will generally, but not necessarily, be different from one, i.e.~they deviate from the SM prediction.

\subsubsection{Non-standard neutrino interactions}
\label{sec:NSI}

Apart from standard neutrino interactions (including neutrino oscillations), the new singly-charged scalar fields $h_{1}^+$ and $h_{2}^+$ introduced in the Zee model will induce NSIs at tree level. These NSIs are new LFV processes that are not allowed in the SM, but could be probed in future neutrino oscillation experiments, and are usually treated using an effective four-fermion operator. 

Interestingly, the operators in the second line of eq.~\eqref{eq:zee_effective} violate lepton number.\footnote{Their gauge invariant EFT operators are, of course, dimension 7~\cite{deGouvea:2014lva}, $LLL\overline{e_{\rm R}} \Phi_{1,2}$.} Indeed, they involve the same combination of four leptons that appears inside the neutrino mass diagram, see figure~\ref{Zee}, and thus, their coefficients are proportional to the same lepton-number combination appearing in the neutrino mass formula, see eq.~\eqref{Mnu}. Hence, they are subject to constraints from neutrino masses and therefore suppressed.

The operators in the first line of eq.~\eqref{eq:zee_effective} do not violate lepton number and, in principle, they give rise to NSIs that are not suppressed by neutrino masses. Applying Fierz identities one can express them in various ways. Using ref.~\cite{Nieves:2003in}, they can be written in the flavor basis with the usual NSI language as~\cite{Antusch:2008tz,Ohlsson:2009vk}
\begin{equation}
{\cal L}_{d=6}^{\rm NSI} = 2\sqrt{2} G_F \chi^{\rho \sigma}_{\alpha \beta} \parentheses{\overline{\nu_\alpha} \gamma^\mu {\rm P_L} \nu_\beta} \parentheses{\overline{e_\rho} \gamma_\mu {\rm P_R} e_\sigma}+ 2\sqrt{2} G_F \varepsilon^{\rho \sigma}_{\alpha \beta} \parentheses{\overline{\nu_\alpha} \gamma^\mu {\rm P_L} \nu_\beta} \parentheses{\overline{e_\rho} \gamma_\mu {\rm P_L} e_\sigma}\,, 
\end{equation}
where $\chi^{\rho \sigma}_{\alpha \beta}$ and $\varepsilon^{\rho \sigma}_{\alpha \beta}$ are the canonical NSI parameters given by
\begin{equation}
\chi^{\rho \sigma}_{\alpha \beta} = \dfrac{(Y_2^{\sigma \beta})^* Y_2^{\rho \alpha}}{4\sqrt{2} G_F\,c^2_\beta\, \tilde{M}_1^2}\,, \qquad \varepsilon^{\rho \sigma}_{\alpha \beta} = \dfrac{f^{\sigma \beta} (f^{\rho \alpha})^*}{\sqrt{2} G_F \tilde{M}_2^2} \,,
\label{eq:NSIdef}
\end{equation}
where $\tilde{M}_1$ and $\tilde{M}_2$ are defined in eq.~\eqref{eq:masses_charged_12}.

For neutrinos propagating in ordinary matter, these operators induce the following matter NSI parameters
\begin{equation}
\chi^{\rm m}_{\alpha \beta} \equiv \chi^{ee}_{\alpha\beta} = \dfrac{(Y_2^{e \beta})^* (Y_2)^{e \alpha}}{4\,\sqrt{2} G_F \,c^2_\beta\,\tilde{M}_1^2}\,, \qquad \varepsilon^{\rm m}_{\alpha \beta} \equiv \varepsilon^{ee}_{\alpha\beta} = \dfrac{f^{e \beta} (f^{e \alpha})^*}{\sqrt{2} G_F \tilde{M}_2^2} \,.\label{eq:NSI}
\end{equation}
In our scenario, the only relevant matter NSI parameters are $\chi^{\rm m}_{\tau\tau}$ and $\varepsilon^{\rm m}_{\tau\tau}$ (cf.~ref.~\cite{Ohlsson:2009vk}), since we assume $Y_2^{\mu e} = Y_2^{e\mu}=0$ and $f^{e\mu}=0$. Now, we can derive an upper bound on $\chi^{\rm m}_{\tau\tau}$ applying CLFV and HLFV limits. For illustration, let us impose the limits from ${\rm Br} (h\rightarrow \tau e)$ from table~\ref{tab:HLFVsignals}. Using a similar equation to eq.~\eqref{BRH}, but for the $\tau e$ channel, which depends on $Y_2^{e\tau}$, we obtain
\begin{equation}
\chi^{\rm m}_{\tau\tau} \leq \frac{4\pi\Gamma_h\,\mathrm{Br}(h\to \tau e) }{\sqrt{2}\,M_h\, G_F \,c^2_{\beta-\alpha}\,\tilde{M}_1^2}\lesssim \frac{2\cdot 10^{-4}}{c^2_{\beta-\alpha}}\,,
\end{equation}
which is below present and most probably future experimental sensitivity. Reproducing small neutrino masses and fulfilling other stronger CLFV constraints, e.g.~constraints on $\mu \to e \gamma$ and $\tau \to \mu \gamma$, imply that $\chi^{\rm m}_{\tau\tau}$ and $\varepsilon^{\rm m}_{\tau\tau}$ are much smaller than $10^{-8}$, which is beyond any future experimental sensitivity. This can be seen in our numerical scan. Other limits and future prospects on NSIs can be found in refs.~\cite{Biggio:2009nt,Blennow:2016etl}.

Finally, at a future neutrino factory, there could be source NSIs in the process $\mu\to e \overline{\nu_\beta} \nu_\alpha$. In our scenario, the only relevant source NSI parameters $\chi^{\mu e}_{\alpha\beta}$ and $\varepsilon^{\mu e}_{\alpha\beta}$ are those with tau neutrinos, i.e.~$\chi^{\mu e}_{\tau\tau}$ and $\varepsilon^{\mu e}_{\tau\tau}$. However, these are also very small, at least below $10^{-6}$.

\subsection{Higgs signals}
\label{sec:higgssignals}

In the Zee model, the couplings to SM particles are modified with respect to their SM values. For instance, the couplings to gauge bosons are
\begin{equation}
\label{eq:W_couplings}
g_{hWW}= \frac{2\,m^2_W}{v}\,s_{\beta-\alpha}\,,\qquad g_{HWW}= \frac{2\,m^2_W}{v}\,c_{\beta-\alpha}\,, \qquad g_{AWW} = 0\,,
\end{equation}
and similarly, for $g_{hZZ}$, $g_{HZZ}$, and $g_{AZZ}$, changing $m^2_W \to m^2_Z$. Clearly, close to the decoupling limit, $s_{\beta-\alpha}\to 1$, the light Higgs interactions are sufficiently SM-like~\cite{Gunion:2002zf}. As we will see in section~\ref{sec:hlfv}, in order to have HLFV, we cannot be exactly at the decoupling limit, but it needs to be close enough to fulfill the bounds on the Higgs decays measured at the LHC~\cite{Khachatryan:2016vau}. The other Higgs couplings in the Zee model are modified as in eq.~\eqref{eq:zee_lagrangian} (see also eq.~\eqref{BRH}) for leptons, eq.~\eqref{quarks} for quarks, and eq.~\eqref{eq:ggratio} for photons. 

The Higgs results at the LHC are usually given in terms of the global signal strength defined as
\begin{eqnarray}\label{eq:signalstrength}
\mu_{X Y} = \frac{\sigma_X(h)\cdot \text{Br}(h\to Y)}{\sigma_X(h)_{\rm SM}\cdot \text{Br}(h\to Y)_{\rm SM}} \,,
\end{eqnarray}
where $\sigma_X(h)$ is the cross section for the production mode $X$ and ${\rm Br} (h\to Y)$ is the Higgs branching ratio for the decay mode $Y$. By definition, in the SM, $\mu_{if}^{\rm SM}=1$ for all production modes $i$ and decay channels $f$. At the LHC, there are four production modes available for the Higgs boson, where the dominant one is gluon-gluon fusion (ggF), mainly through a top loop. The subdominant ones are vector boson fusion (VBF), associated production with a vector boson $Vh$ (where $V=W,Z$), and the associated production with a top-quark pair $t\bar{t}h$. The production modes are usually grouped into two effective modes according to ${\rm ggF}+t\bar{t}h$ and ${\rm VBF}+Vh$. We consider the five decay channels, where a signal has been detected, namely $\gamma\gamma$, $WW^*$, $ZZ^*$, $b\bar{b}$, and $\tau\bar\tau$. For instance, for $gg\to h \to b\overline{b}$, the signal strength is given by
\begin{equation}
\mu_{ggh+t\bar{t}h}^{b\overline{b}} = \left(\frac{s_\alpha}{c_\beta}\right)^4 \,. 
\end{equation}

For the contribution to the $\chi^2$ function (to be discussed in section~\ref{sec:numerics}) from the Higgs decay channels, we need to take into account correlations between different production modes. Thus, for each of the decay modes $f=\gamma\gamma, WW^*, ZZ^*, b\bar{b}, \tau\bar\tau$, the contribution to the $\chi^2$ function is defined as
\begin{equation}
\chi^2_f=\frac{1}{\hat{\sigma}_1^2(1-\rho^2)}(\mu_1^f-\hat{\mu_1}^f)^2+\frac{1}{\hat{\sigma}_2^2(1-\rho^2)}(\mu_2^f-\hat{\mu_2}^f)^2-\frac{2\rho}{\hat{\sigma}_1\hat{\sigma}_2(1-\rho^2)}(\mu_1^f-\hat{\mu_1}^f)(\mu_2^f-\hat{\mu_2}^f)\,,
\end{equation}
where $\mu^f_{1,2}$ are the results in the Zee model, $\hat{\mu}^f_{1(2)}$ are the measured Higgs signal strengths, $\hat{\sigma}^f_{1(2)}$ are the standard deviations, and $\rho$ is the correlation. The index 1 stands for the combination ${\rm ggF}+t\bar{t}h$ and the index 2 for the combination ${\rm VBF}+Vh$. The numerical values are given in refs.~\cite{Aad:2013wqa,ATLAS-CONF-2014-009,ATLAS-CONF-2013-079,Aad:2015vsa}.

\subsubsection{$h \to \gamma\gamma$ decays}
\label{sec:higgs_gamma}

In the Zee model, the decay of the Higgs boson to two photons is modified by two factors. First, the couplings to gauge bosons and top quarks are changed, since we have two Higgs doublets. Second, there are new extra charged scalars couplings to the Higgs boson. In ref.~\cite{Kanemura:2000bq}, a study of $h \to \gamma\gamma$ in the Zee model has been performed. However, $\lambda_7$ and $\lambda_{10}$ were set to zero. Thus, in the following, we will analyze this decay in our scenario. 

The value of the $h \to \gamma \gamma$ decay width in the Zee model with respect to the SM one is given by \cite{Ellis:1975ap,Shifman:1979eb,Carena:2012xa}
\begin{equation}
R_{\gamma \gamma} = \frac{\Gamma(h \to \gamma \gamma)_{\rm Zee}}{\Gamma(h \to \gamma \gamma)_{\rm SM}}= \left | \frac{s^2_{\beta-\alpha}A_1(\tau_W)+\frac{4}{3}\,s^2_{\alpha}/c^2_{\beta} A_{1/2}(\tau_t)+ \sum_{\cal S}\, \frac{\lambda_{{\cal S}H} \, v^2}{2  m_{\cal S}^2}\,A_0(\tau_{\cal S})}{A_1(\tau_W)+\frac{4}{3} A_{1/2}(\tau_t)} \right |^2\,, \label{eq:ggratio}
\end{equation}
where $\lambda_{{\cal S}H}$ is the coupling of a charged scalar ${\cal S}$ with mass $m_{\cal S}$ to the Higgs field, which is coming from a term in the potential of the form $(H^\dagger H) ({\cal S}^\dagger {\cal S})$. Note that we have used the modified couplings to tops and $WW$ given in eqs.~\eqref{quarks} and~\eqref{eq:W_couplings}, respectively. Here, $\tau_i\equiv 4 m_i^2/m_H^2$ and $A_i(x)$ ($i=0,1/2,1$) are loop functions:
\begin{align}
A_0(x) &= -x+x^2 \, f \left (\frac{1}{x}\right )\,, \\
A_{1/2}(x) &= 2x+ 2x (1- x) \, f \left (\frac{1}{x}\right )\,, \\
A_{1}(x) &= -2-3x-3x (2-x) \, f \left (\frac{1}{x}\right )\,.
\end{align}
We need to compute the couplings to charged scalars $\lambda_{{\cal S}H}$ in the Zee model. Since we are in the Higgs basis, the terms of the potential in eq.~\eqref{eq:scalar_potential}, which are coupling the Higgs boson to the charged scalars, are those involving $\lambda_{3}$, $\lambda_{7}$, $\lambda_{8}$, $\lambda_{10}$, and $\mu$. Using the rotations to the mass basis, i.e.~eqs.~\eqref{eq:charged_rot} and \eqref{eq:CP-even_mixing}, the relevant interactions $h h_1^+h_1^-$ and $h h_2^+h_2^-$ read
\begin{align}
\lambda_{{\cal S}_1 H} \equiv \lambda_{h h_1^+ h_1^-} &=  s_{\beta-\alpha}\, ( \lambda_3 c^2_\varphi +\lambda_8 s^2_\varphi)+c_{\beta-\alpha}\, (\lambda_{7} c^2_\varphi+\lambda_{10} s^2_\varphi)  -\frac{1}{\sqrt{2}}\,\frac{\mu}{v}\, s_{\beta-\alpha}\,\, s_{2\varphi}\,,\nonumber\\
\lambda_{{\cal S}_2 H} \equiv \lambda_{h h_2^+ h_2^-} &= s_{\beta-\alpha} \,(\lambda_3 s^2_\varphi+\lambda_8 c^2_\varphi)+c_{\beta-\alpha}\, (\lambda_{7} s^2_\varphi+\lambda_{10} c^2_\varphi)  +\frac{1}{\sqrt{2}}\,\frac{\mu}{v}\, s_{\beta-\alpha}\, s_{2\varphi}\,. \label{higgs_charged}
\end{align}
In the case when there is no mixing $\mu\, (\varphi)\to 0$, we obtain
\begin{align} 
 \lambda_{h H^+ H^-} &= s_{\beta-\alpha} \,\lambda_3+c_{\beta-\alpha} \, \lambda_7 \nonumber \,, \\
 \lambda_{h h^+ h^-} &= s_{\beta-\alpha}\, \lambda_8 +c_{\beta-\alpha}\, \lambda_{10}\,,
\end{align}
where the first equation agrees with eq.~(F1) in ref.~\cite{Gunion:2002zf}. Note that, in our scenario, only the terms proportional to $\lambda_3$ and $\mu$ in eq.~\eqref {higgs_charged} will contribute, as we set the other couplings to zero. 

\subsubsection{Higgs lepton flavor violation}
\label{sec:hlfv}

The Zee model predicts HLFV interactions that can be sizable. From the leptonic Lagrangian, i.e.~eq.~\eqref{eq:zee_lagrangian}, the branching ratio of $h\to \tau \mu$ is given by
\begin{equation}  \label{BRH}
\mathrm{Br}(h\to \tau \mu) = \frac{m_h}{8\pi\Gamma_h} \,\left(\frac{\,c_{\beta-\alpha}}{\sqrt{2}\,c_\beta}\right)^2\,(|Y_2^{\tau \mu}|^2+|Y_2^{\mu \tau}|^2)\,,
\end{equation}
and similarly, $\mathrm{Br}(h\to \tau e)\propto (|Y_2^{\tau e}|^2+|Y_2^{e \tau}|^2)$. We can expand around the decoupling limit, i.e.~$\beta-\alpha \approx \pi/2$, by using eq.~\eqref{sinbminua}, to obtain \cite{Bizot:2015qqo}
\begin{equation}  \label{BRH2}
\mathrm{Br}(h\to \tau \mu) \approx\frac{m_h}{16\pi\Gamma_h} \frac{\lambda^2_6 v^4}{c^2_\beta m^4_{H}} (|Y_2^{\tau \mu}|^2+|Y_2^{\mu \tau}|^2)\,.
\end{equation}
Thus, in order to have large HLFV, i.e.~$\mathrm{Br}(h\to \tau \mu)\sim 1\,\%$, we need
\begin{equation}  \label{BRH3}
\frac{\lambda_6}{c_\beta} \frac{v^2}{m^2_{H}} \sqrt{|Y_2^{\tau \mu}|^2+|Y_2^{\mu \tau}|^2} \simeq 0.004\,.
\end{equation}
In principle, this can be achieved quite easily. For instance, choosing $c_\beta\sim 0.5$, $\lambda_6\sim 0.05$, and $m_{H}\sim 2 v$, we can obtain the desired branching ratio for $\sqrt{|Y_2^{\tau \mu}|^2+|Y_2^{\mu \tau}|^2}\sim 0.002$. In order to have a sizable $\mathrm{Br}(h\to \tau \mu)$, the correct neutrino mass scale can be obtained with very small singly-charged Yukawa couplings $f^{e\tau}$ and $f^{\mu\tau}$.

Note that for a type-III 2HDM there is an upper bound on $\mathrm{Br}(h\to \tau \mu)\cdot \mathrm{Br}(h\to \tau e)$ from combining the rates of $\mu\to e\gamma$ and $\mu e$ conversion ~\cite{Dorsner:2015mja} (which currently saturates the bound), as all combinations of couplings relevant to these HLFV processes enter in CLFV with tau leptons running in the loop, see secs.~\ref{sec:muegamma} and \ref{sec:mueconv}:
\begin{equation}  \label{BR}
\mathrm{Br}(h\to \tau \mu)\cdot \mathrm{Br}(h\to \tau e) \lesssim 10^{-6}\,.
\end{equation}
In the Zee model, as we will see, reproducing the leptonic mixing angles correctly implies that there are also lower bounds on the HLFV processes.

\section{Numerical analysis}
\label{sec:numerics}

\subsection{Scan of the parameter space}
\label{sec:scan}

In order to study the large parameter space of our scenario of the Zee model and to be able to investigate how large CLFV and HLFV processes can be, we perform a full numerical scan using the software {\sc MultiNest} \cite{Feroz:2007kg,Feroz:2008xx,Feroz:2013hea}. {\sc MultiNest} is a Bayesian inference tool that uses so-called nested sampling and especially suitable when there are possibly several maxima in the parameter space. It is designed to determine the Bayesian evidence, but as a byproduct, it also yields the posterior distribution that is relevant for a Bayesian analysis. Nevertheless, it also maximizes the likelihood, which is relevant for a frequentist analysis. We are interested in the maximization of the likelihood and we perform a fully frequentist analysis. We scan over all free parameters in the model, in total 19 real parameters, which are given in table~\ref{tab:scan} together with their chosen allowed parameter ranges. The plots, including best-fit points and $1\sigma$ and $2\sigma$ confidence regions, are produced using the graphical interface {\sc Superplot}~\cite{Fowlie:2016hew}. 
\begin{table} [h]
\begin{center}
  \begin{tabular}{ | c | c | }
    \hline
     Parameter & Range  \\ \hline \hline
   Complex: $Y_2^{\tau\tau}$, $Y_2^{\tau\mu}$, $Y_2^{\tau e}$, $Y_2^{\mu\tau}$ & $[10^{-12}, 10^{-1}]$  \\ \hline
    Real: $f^{\mu\tau}$, $f^{e\tau}$, $Y_2^{e\tau}$& $[10^{-12}, 10^{-1}]$   \\ \hline
    $\tan\beta$ & $[0.3,50]$  \\  \hline
    $\lambda_1$, $\lambda_2$, $|\lambda_3|$, $|\lambda_5|$ & $[10^{-5},\sqrt{4\pi}]$  \\  \hline
    $\mu_h$, $\mu_2$~[GeV] & $[1,10^{7}]$  \\   \hline
    $\mu$~[GeV] & $[1,10^{7}]$ \\ \hline
  \end{tabular}
  \caption{\label{tab:scan} Priors on the 19 free real parameters used in the scan. For $Y_2^{\tau\tau}$, $Y_2^{\tau\mu}$, $Y_2^{\tau e}$, and $Y_2^{\mu\tau}$, we scan real and imaginary parts independently, while $f^{\mu\tau}$, $f^{e\tau}$, and $Y_2^{e\tau}$ can be assumed to be real without loss of generality. We use logarithmic priors for all parameters except for $\tan\beta$, where uniform priors are used.}
\end{center}
\end{table}
We impose the stability conditions on the scalar couplings given in eqs.~\eqref{eq:stability_2}, \eqref{eq:stability_extended}, and \eqref{eq:lambda_i}. Direct searches on singly-charged scalars from LEP II imply $m_{h^+_1}, m_{h^+_2}> 80$~GeV \cite{Abbiendi:2013hk}. In the scan, we assume that all Higgs bosons, except the light one (with the mass fixed to $m_h = 125.5$~GeV), are heavier than $100$~GeV, i.e.~not only the charged ones. This means that $m_A, m_H, m_{h^+_1}, m_{h^+_2} > 100$~GeV. The scan over the free parameters is performed for three cases: (i) $\mu=0$, i.e.~no neutrino masses,\footnote{In this case, we have only 18 free real parameters, since $\mu$ is set to zero.} (ii) $\mu\neq 0$ with neutrino masses in NO, and (iii) $\mu\neq 0$ with neutrino masses in IO.

The quantity that is maximized is the likelihood $L$, which is equivalent to minimizing the $\chi^2$ function: $\chi^2=-2\ln L$. We assume Gaussian likelihoods, and thus, the contribution to the $\chi^2$ function from the $N$ signals is given by
\begin{equation}
\chi_{\rm signals}^2=\sum_{i=1}^{N} \chi_i^2\,(\mathcal O^i_{\rm th})=\sum_{i=1}^{N} \left(\frac{\mathcal O^i_{\rm th}-\mathcal O^i_{\rm obs}}{\sigma_i}\right)^2\,,
\label{eq:chi2signals}
\end{equation}
where $\mathcal O^i_{\rm th}$ ($\mathcal O^i_{\rm obs}$) is the theoretical prediction (experimental measurement) of the observable $i$ with the respective standard deviation $\sigma_i$. For $s_{23}^2$ and $\delta$, we use their combined two-dimensional distribution function (therefore including their correlation) given in refs.~\cite{Esteban:2016qun,NuFIT3.0}. For the $M$ upper bounds, we use
\begin{equation}
\chi_{\rm bounds}^2=
\left\{\begin{array}{l}0\,, \ \mathcal O^j_{\rm th}<B^j,\\ \sum_{j=1}^{M} \left(\frac{\mathcal O^j_{\rm th}}{B^j}\right)^2\,, \ \mathcal O^j_{\rm th}\geq B^j
\end{array}\right.\,,
\label{eq:chi2bounds}
\end{equation}
where $B^j$ is the experimental upper limit at $1 \sigma$ significance level of the observable $j$. According to the definitions for the standard Gaussian distributions assumed for the observables, the 90~\%~C.L.~and 95~\%~C.L.~upper limits are normalized (i.e.~divided) by the factors 1.645 and 1.949, respectively. The limits for the signals are rescaled in the same way. The choice of eq.~\eqref{eq:chi2bounds} for the $\chi^2$ function is made so that we do not penalize deviation from zero when this is not supported by data.

We take into account all the relevant bounds and observables described in section~\ref{sec:pheno}, except for the muon AMM discrepancy with respect to the SM, which cannot be accommodated in our present scenario of the Zee model. 
Thus, adding eqs.~\eqref{eq:chi2signals} and \eqref{eq:chi2bounds}, the total $\chi^2$ function reads 
\begin{equation}
\chi^2=\chi_{\rm signals}^2 + \chi_{\rm bounds}^2\,.
\end{equation}
In tabs.~\ref{tab:HLFVsignals} and \ref{tab:signals}, we present different observables that yield positive signals (see also appendix~\ref{sec:ewpt}). In table~\ref{tab:neutrinoNO}, we show the values for the neutrino oscillation parameters from a global fit~\cite{Esteban:2016qun,NuFIT3.0} (see also section~\ref{sec:nus}). Furthermore, in table~\ref{tab:upperbounds}, observables for which there are only upper bounds are presented, and in table~\ref{tab:neutrinoIOobs}, we show the upper bounds for different neutrino mass parameters (see also section~\ref{sec:nus}). Finally, we also treat the naturality constraints of eqs.~\eqref{nat} and~\eqref{eq:mu} as bounds and include them in $\chi_{\rm bounds}^2$.
\begin{table} [h]
\begin{center}
  \begin{tabular}{ | c | c | }
    \hline
   Observable & Central value~$\pm$~$1\sigma$ error \\ \hline \hline
    $S$ & $0.05\pm 0.11$~\cite{Baak:2014ora} \\ \hline
    $T$ & $0.09\pm 0.13$~\cite{Baak:2014ora} \\ \hline
    $U$ & $0.01\pm 0.11$~\cite{Baak:2014ora} \\ \hline
    $\left|g_{\tau}^{\rm exp}/g_{\mu}^{\rm exp}\right|$ & $1.0011 \pm 0.0015$~\cite{Pich:2013lsa} \\ \hline
    $\left|g_{\tau}^{\rm exp}/g_{e}^{\rm exp}\right|$ & $1.0030 \pm 0.0015$~\cite{Pich:2013lsa} \\ \hline
    $\left|g_{\mu}^{\rm exp}/g_{e}^{\rm exp}\right|$ & $1.0018 \pm 0.0014$~\cite{Pich:2013lsa} \\ \hline
  \end{tabular}
  \caption{\label{tab:signals} The current experimental values and $1\sigma$ errors for electroweak precision tests (see appendix~\ref{sec:ewpt}) and universality (see section~\ref{sec:univ}). The limits on the parameters $S$, $T$, and $U$ are derived from a fit to different electroweak precision data, see ref.~\cite{Baak:2014ora} for more details. The correlation coefficients among the obervables $S$, $T$, and $U$ are $\rho_{\rm ST}=0.90$, $\rho_{\rm SU}=-0.59$, and $\rho_{\rm TU}=-0.83$~\cite{Baak:2014ora}.}
\end{center}
\end{table}
\begin{table} [h]
\begin{center}
  \begin{tabular}{ | c | c | c | }
    \hline
    Observable & NO & IO\\ \hline \hline
    $\sin^2\theta_{12}$ & $0.306 \pm 0.012$ & $0.306 \pm 0.012$\\ \hline
    $\sin^2\theta_{13}$ & $0.02166 \pm 0.00075$ & $0.02179 \pm 0.00076$\\ \hline
    $\sin^2\theta_{23}$ & $0.441 \pm 0.027$ & $0.587 \pm 0.024$\\ \hline
    $\Delta m^2_{21}$~$[\mathrm{eV^2}]$ & $(7.50 \pm 0.19) \cdot 10^{-5}$ & $(7.50\pm 0.19) \cdot 10^{-5}$\\ \hline
    $\Delta m^2_{3\ell}$~$[\mathrm{eV^2}]$ & $(2.524 \pm 0.040) \cdot 10^{-3}$ & $-(2.514 \pm 0.041) \cdot 10^{-3}$\\ \hline
    $\delta$~$[{}^\circ]$ &$261 \pm 59$ & $277 \pm 46$\\
  \hline
  \end{tabular}
  \caption{\label{tab:neutrinoNO} The best-fit values and $1\sigma$ errors for the leptonic mixing parameters and mass-squared differences from the NuFIT group (NuFIT 3.0, November 2016)~\cite{Gonzalez-Garcia:2014bfa,Esteban:2016qun,NuFIT3.0}. See also the discussion in section~\ref{sec:nus}. Note that we use symmetric lower and upper errors (choosing the largest of the two when different asymmetric errors are present). Here $\Delta m^2_{3\ell}=\Delta m^2_{31}>0$ for NO and $\Delta m^2_{3\ell}=\Delta m^2_{32}<0$ for IO. Note that for $\sin^2\theta_{23}$ there are two minima in their distribution~\cite{Esteban:2016qun,NuFIT3.0} for both orderings, corresponding to the first and second octants. For $\delta$ the distribution is also not $\chi^2$ distributed. Therefore, for $\sin^2\theta_{23}$ and $\delta$, the two-dimensional complete distribution is used.}
\end{center}
\end{table}
\begin{table} [h]
\begin{center}
  \begin{tabular}{ | c | c | }
   \hline
   Observable & Upper bound \\ \hline \hline
   $\mathrm{Br}(\tau^-\to \mu^- \gamma)$& $4.4\cdot 10^{-8}$~\cite{Patrignani:2016xmw}\\ \hline
   $\mathrm{Br}(\tau^-\to e^- \gamma)$ & $3.3\cdot 10^{-8}$~\cite{Patrignani:2016xmw} \\ \hline
   $\mathrm{Br}(\mu^+\to e^+ \gamma)$& $4.2\cdot 10^{-13}$~\cite{TheMEG:2016wtm}\\ \hline
   $\mathrm{Br}(\tau^-\to \mu^- \mu^+ \mu^-)$& $2.1 \cdot 10^{-8}$~\cite{Patrignani:2016xmw}\\ \hline
   $\mathrm{Br}(\tau^-\to \mu^- \mu^+ e^-)$& $2.7 \cdot 10^{-8}$~\cite{Patrignani:2016xmw} \\ \hline
    ${\rm Cr}(\mu \to e)_{\rm Au} $ & $7\cdot 10^{-13}$~\cite{Bertl:2006up}\\ \hline
   $\absolutevalue{\delta a_{e}}$ & $2 \cdot 10^{-12}$~\cite{Patrignani:2016xmw} \\ \hline
   $\absolutevalue{d_{\mu}/e}$~[cm] & $1 \cdot 10^{-19}$~\cite{Patrignani:2016xmw} \\ \hline
   $\absolutevalue{d_{e}/e}$~[cm] & $8.7\cdot 10^{-29}$~\cite{Baron:2013eja}
    \\ \hline
  \end{tabular}
\caption{\label{tab:upperbounds} The current experimental 90~\% C.L.~upper bounds for relevant charged lepton flavor violating processes and electric and magnetic dipole moments.}
\end{center}
\end{table}
\begin{table} [h]
\begin{center}
  \begin{tabular}{ | c | c | }
    \hline
    Observable & Upper bound\\ \hline \hline
    $m_{ee}$~$[\mathrm{meV}]$ & $ [190,450] $~\cite{Albert:2014awa}\\  \hline
    $m_{\nu_e}$~$[\mathrm{eV}]$ & $ [2.05,2.3]$~\cite{Lobashev:2003kt,Kraus:2004zw,Aseev:2011dq} \\ \hline
    $\sum m_i$~$[\mathrm{eV}]$ & $ 0.23$~\cite{Ade:2015xua}\\
    \hline
  \end{tabular}
  \caption{\label{tab:neutrinoIOobs} The current experimental 95~\% C.L.~upper bounds on parameters related to the neutrino masses, see also section~\ref{sec:nus}. The two upper bounds on $m_{ee}$ are due to the sensitivity on the nuclear matrix elements, whereas for $m_{\nu_e}$, we show results coming from two different experiments. We use the most stringent values of these two upper bounds in the scan.}
\end{center}
\end{table}

\subsection{Results of scan}

In this section, we discuss the main results of our numerical scan, performed for $\kappa=1$ with the naturality upper limit of $\mu \lesssim 1.5$~TeV, see eq.~\eqref{eq:mu}, unless otherwise stated. At the end of the section, we discuss how the results would change for $\kappa=10$ ($\mu\lesssim15$~TeV). The numerical scan is performed for three different cases of our scenario of the Zee model, i.e.~for $\mu=0$, NO, and IO. For these cases, the values for the minima of the $\chi^2$ function, $\chi_{\rm min}^2$, at the respective best-fit points are 
\begin{enumerate}
\item[(i)] $\mu=0$, i.e.~massless neutrinos: $\chi_{\rm min}^2 \simeq 5.1$, 
\item[(ii)] $\mu \neq 0$, massive neutrinos in NO: $\chi_{\rm min}^2 \simeq 10.7\,(11.0)$ for $\kappa = 1\,(10)$, 
\item[(iii)] $\mu \neq 0$, massive neutrinos in IO: $\chi_{\rm min}^2 \simeq 21.7 \,(21.5)$ for $\kappa = 1\,(10)$.
\end{enumerate}
Thus, in our scenario of the Zee model, neutrino masses and leptonic mixing can be accommodated in both NO and IO. However, IO is disfavored compared to NO. We also note that if $\theta_{23}$ will turn out to be in the second octant, IO cannot be accommodated.

\begin{figure}
\centering
\includegraphics[scale=0.5]{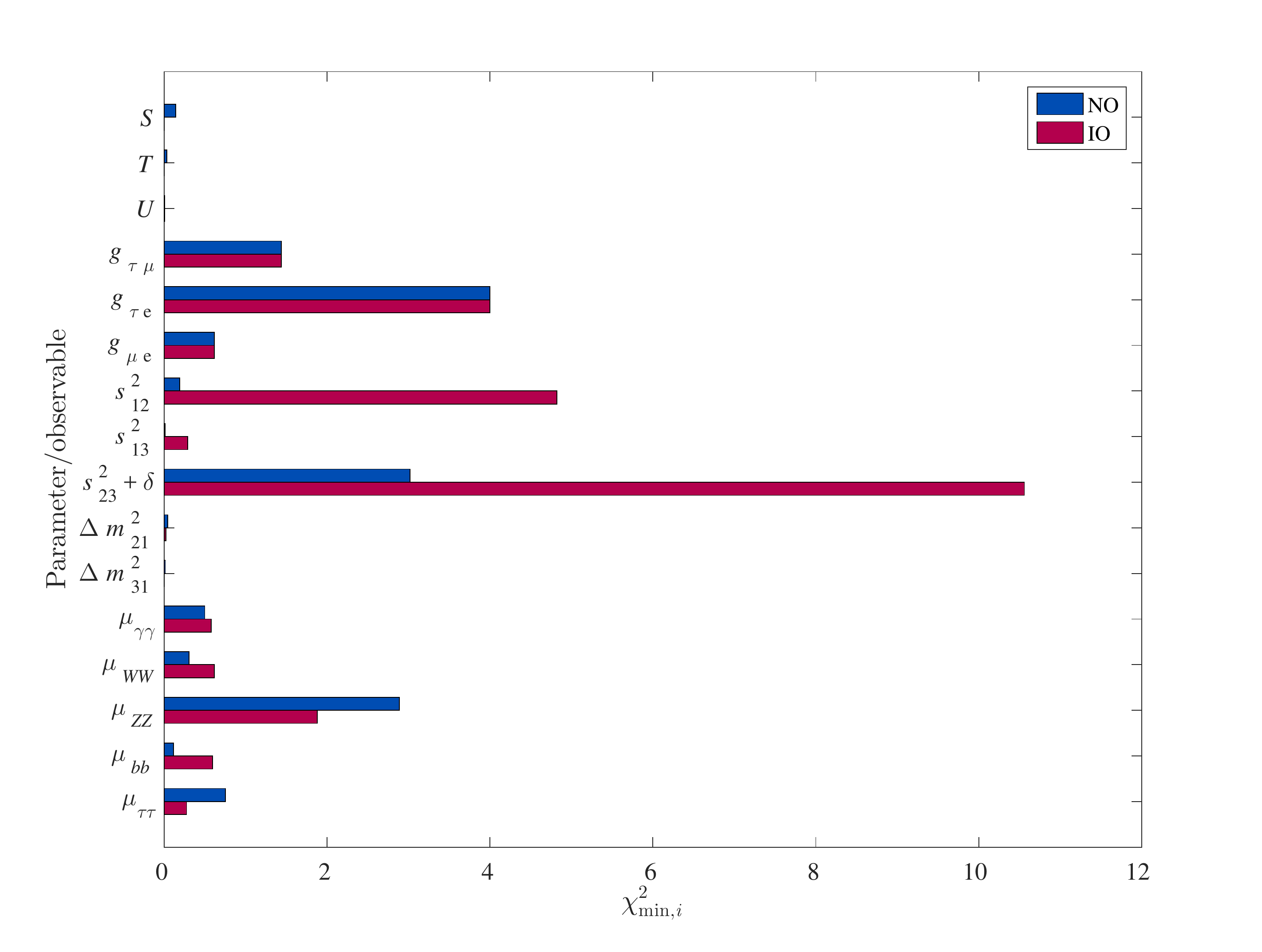}
\caption{Individual contributions to $\chi_{\rm min}^2$ from the different parameters and observables in NO and IO. The number of standard deviations that each observable $i$ is away from the observed value is given by the respective pull $\sqrt{\chi_{{\rm min},i}^2}$.}\label{fig:chi2}
\end{figure}

In figure~\ref{fig:chi2}, we present the contributions to $\chi_{\rm min}^2$ from the observables in both NO and IO. Note that we do not show the contributions from the upper bounds, since these are always satisfied, and thus, the corresponding contribution to $\chi_{\rm min}^2$ is exactly zero. Note that we present the combined contribution from $s^2_{23}$ and $\delta$ to $\chi^2_{\rm min}$. In IO, the largest contributions stem from $s^2_{12}$ and $s^2_{23}+\delta$, although they are all within $3\sigma$ of their experimental values. In NO, the corresponding contributions are small. These observables account for the fact that the fit in IO is much worse than in NO. All other observables in IO are of the same size as those in NO and within $2\sigma$. In addition, we perform a run assuming $Y_2^{\mu\tau}=0$, which renders one neutrino massless, see eq.~\eqref{Mnu}. In this case, we find that IO is significantly better than NO, which is in agreement with the results of ref.~\cite{He:2011hs}, and the value of $\chi_{\rm min}^2$ in IO is of the same size as for $Y_2^{\mu\tau}\neq 0$ (i.e.~quite large), whereas in NO, the difference is more than one order of magnitude with $\chi_{\rm min}^2\sim \mathcal{O}(100)$. Thus, in this scenario, both NO and IO are basically excluded. Moreover, there are non-negligible contributions to $\chi_{\rm min}^2$ from the Higgs signals, especially $h\to ZZ$. Regarding $h\to \gamma\gamma$, we find that it is around the SM value. Note that in this case we choose some of the couplings of the scalar potential to be zero, see eqs.~\eqref{higgs_charged}, so its value can be modified by turning them on. The values for the universality parameters $g_{ij}$, see eqs.~\eqref{eq:gtm}--\eqref{eq:gme}, are always very close to one, even closer than the experimental values, and compatible with observations at $2\sigma$. However, these also give non-negligible contributions to $\chi_{\rm min}^2$.

\begin{figure}[ht!]
\centering
\subfloat[NO]{\includegraphics[width=0.45\textwidth]{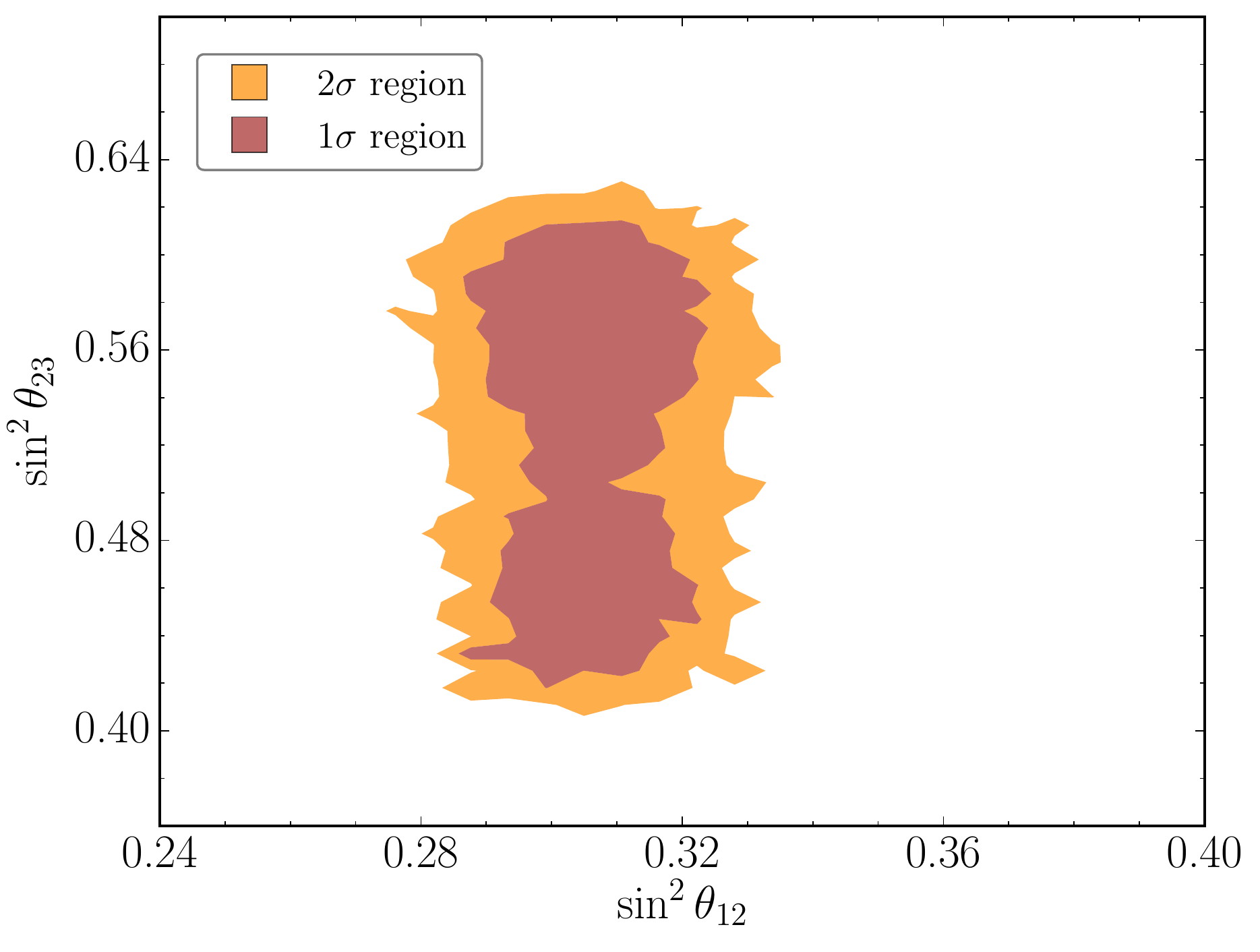}}
\subfloat[IO]{\includegraphics[width=0.45\textwidth]{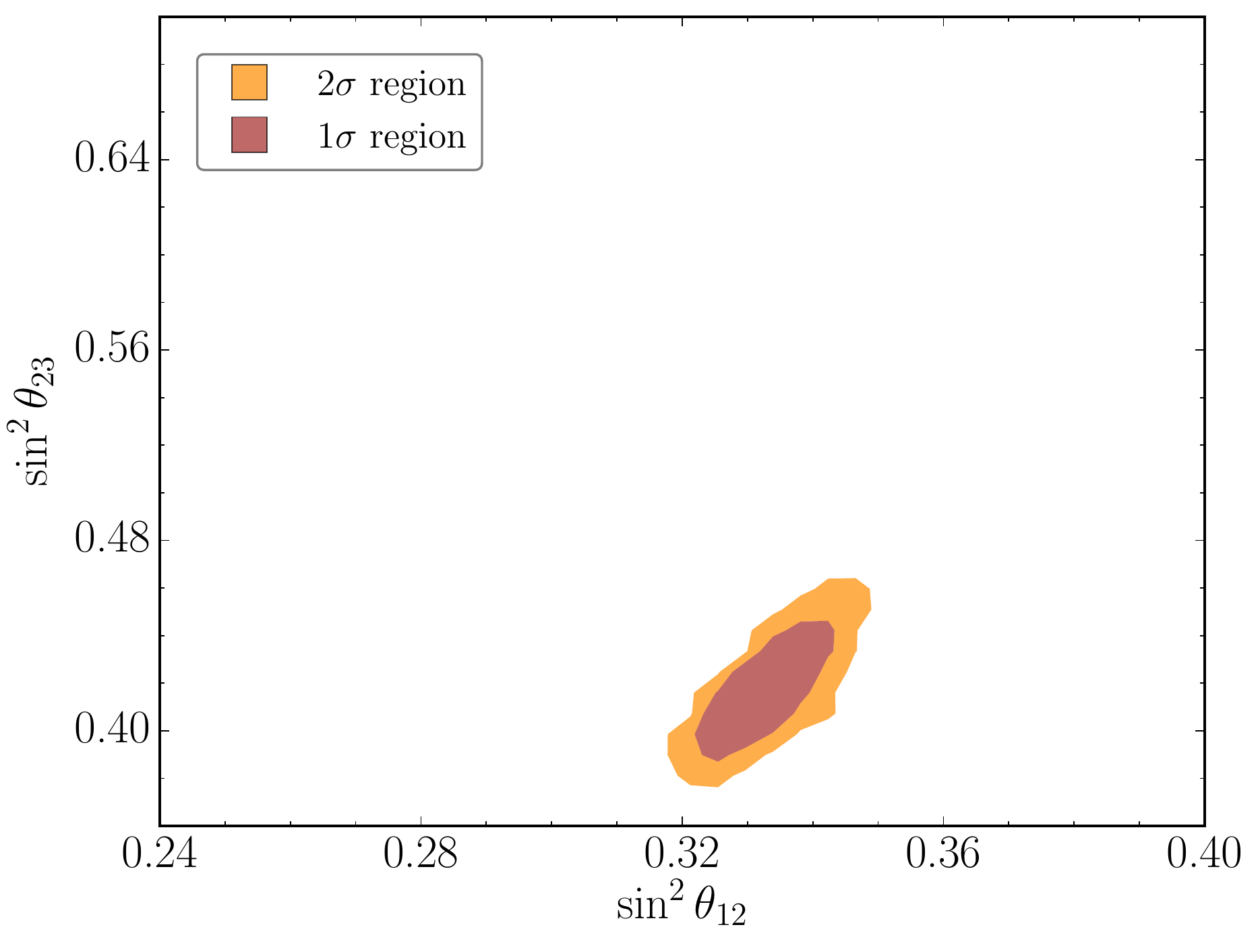}}
\caption{Allowed regions of the leptonic mixing parameters $\sin^2 \theta_{12}$ and $\sin^2 \theta_{23}$ for (a) NO and (b) IO. The $3\sigma$ C.L.~ranges from global fits to neutrino oscillation data~\cite{Esteban:2016qun,NuFIT3.0} are $0.271<\sin^2 \theta_{12}<0.345$ for both orderings and $0.385\,(0.393)<\sin^2 \theta_{23}<0.635\,(0.640)$ for NO (IO), see also table~\ref{tab:neutrinoNO}.}\label{fig:mixing_angles}
\end{figure}

In the following, we present the allowed regions for the most interesting parameters and observables. In figure~\ref{fig:mixing_angles}, we plot the leptonic mixing parameters $s^2_{12}$ and $s^2_{23}$ for NO (left panel) and IO (right panel), where one can clearly see that the fit is very good for NO, while for IO it crucially depends on the octant of $\theta_{23}$. In fact, the fitted value of $s^2_{12}$ would be $5\sigma$ away from its experimental best-fit value if $\theta_{23}$ lies in the first octant. For NO, it is clear that both octants are viable. 

\begin{figure}[ht!]
\centering
\subfloat[$\mu=0$]{\includegraphics[width=0.33\textwidth]{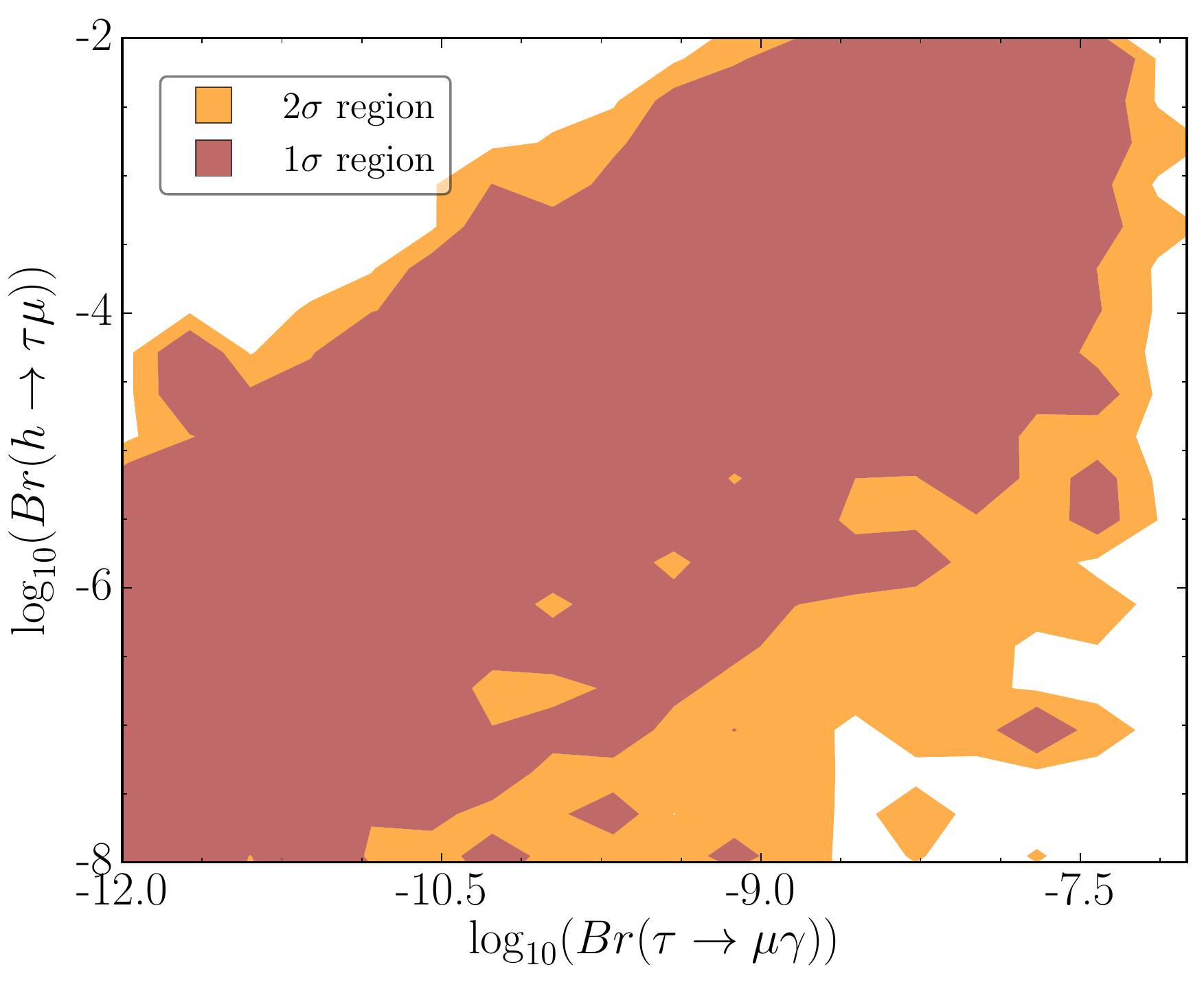}}
\subfloat[NO]{\includegraphics[width=0.33\textwidth]{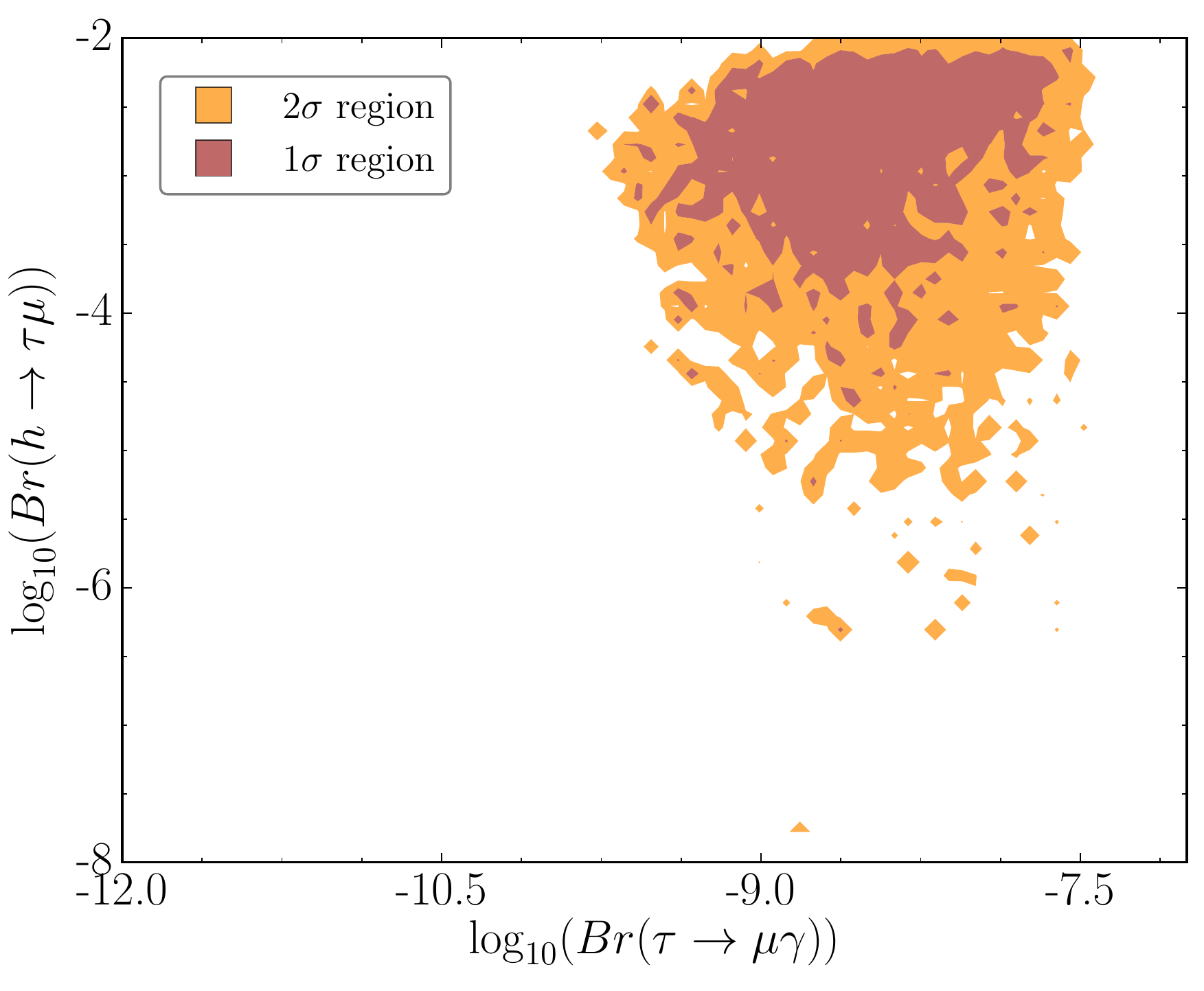}}
\subfloat[IO]{\includegraphics[width=0.33\textwidth]{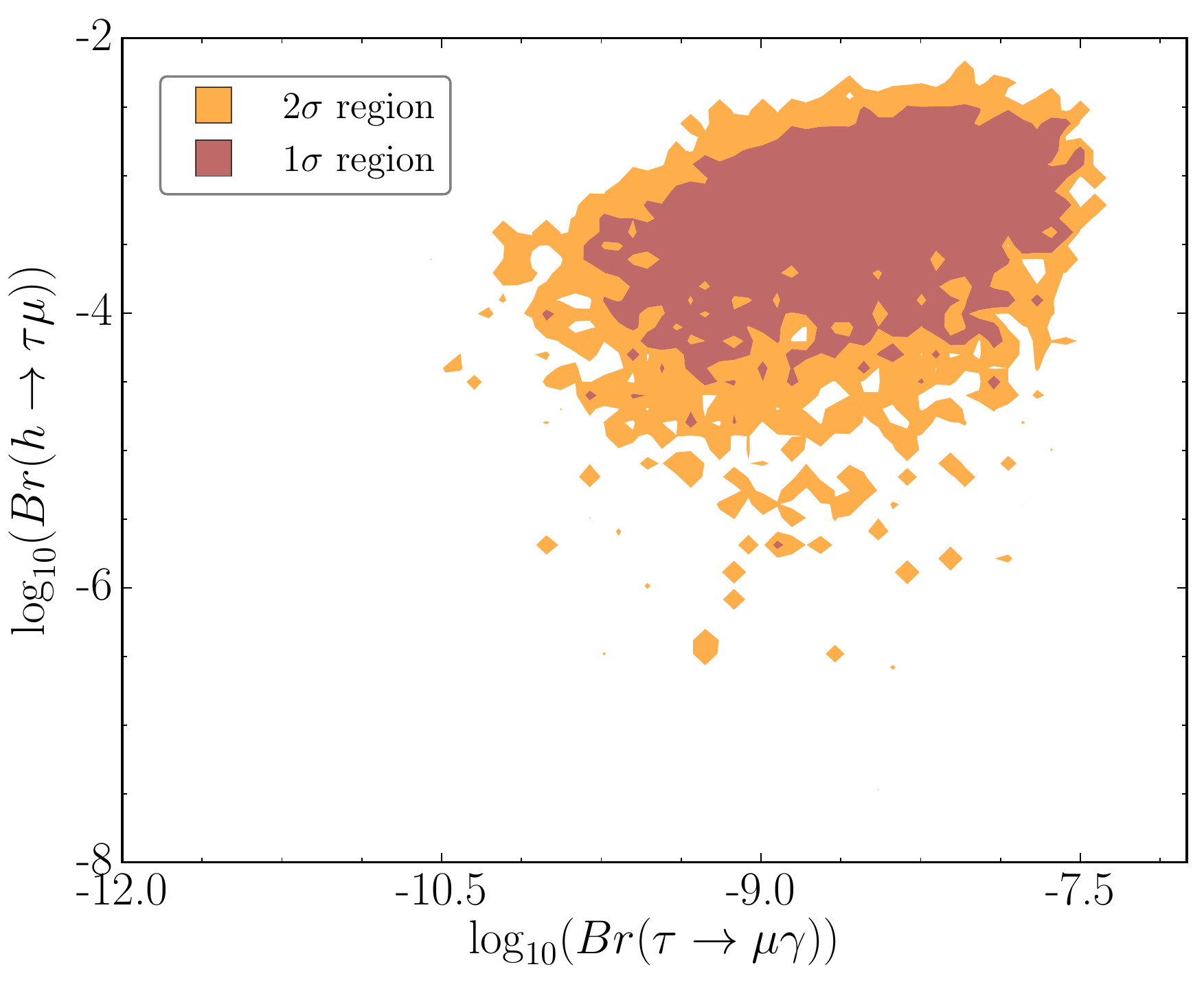}}
\caption{Allowed regions of the branching ratios ${\rm Br}(\tau \to \mu \gamma)$ and ${\rm Br}(h\to \tau \mu)$ for (a) $\mu = 0$, (b) NO, and (c) IO.}\label{fig:tautomugamma}
\end{figure}

In figure~\ref{fig:tautomugamma}, we plot the allowed regions of the branching ratios ${\rm Br}(h\to \tau \mu)$ and ${\rm Br}(\tau \to \mu \gamma)$. For $\mu=0$ (left panel), only \emph{upper} bounds on the CLFV and HLFV processes exist and these are compatible with the results of refs.~\cite{Sierra:2014nqa, Dorsner:2015mja}. For $\mu \neq 0$, one can observe that there are \emph{lower} bounds on the CLFV and HLFV processes in both NO (middle panel) and IO (right panel). That is, reproducing neutrino masses implies that CLFV and HLFV cannot be arbitrarily small. In particular, ${\rm Br}(h\to \tau \mu) \gtrsim 10^{-6}\,(10^{-7})$ in NO (IO). For NO, the upper bound on ${\rm Br}(h\to \tau \mu)$ saturates the experimental one, while for IO, we obtain ${\rm Br}(h\to \tau \mu) \lesssim 5\cdot 10^{-3}$. 
In the future, Belle~II is expected to reach a sensitivity on ${\rm Br}(\tau \to \mu \gamma)$ of $\mathcal{O}(10^{-9})$~\cite{HerediadelaCruz:2016yqi}, which would significantly probe a substantial part of the allowed parameter space of IO and almost the complete allowed region of NO. This is one of the most interesting results of our work. Similarly, if sensitivities of ${\rm Br}(h\to \tau \mu)$ at future colliders reach $10^{-4}\,(10^{-5})$, NO (IO) will be tested at 68~\% C.L.

\begin{figure}[ht!]
\centering
\subfloat[NO]{\includegraphics[width=0.45\textwidth]{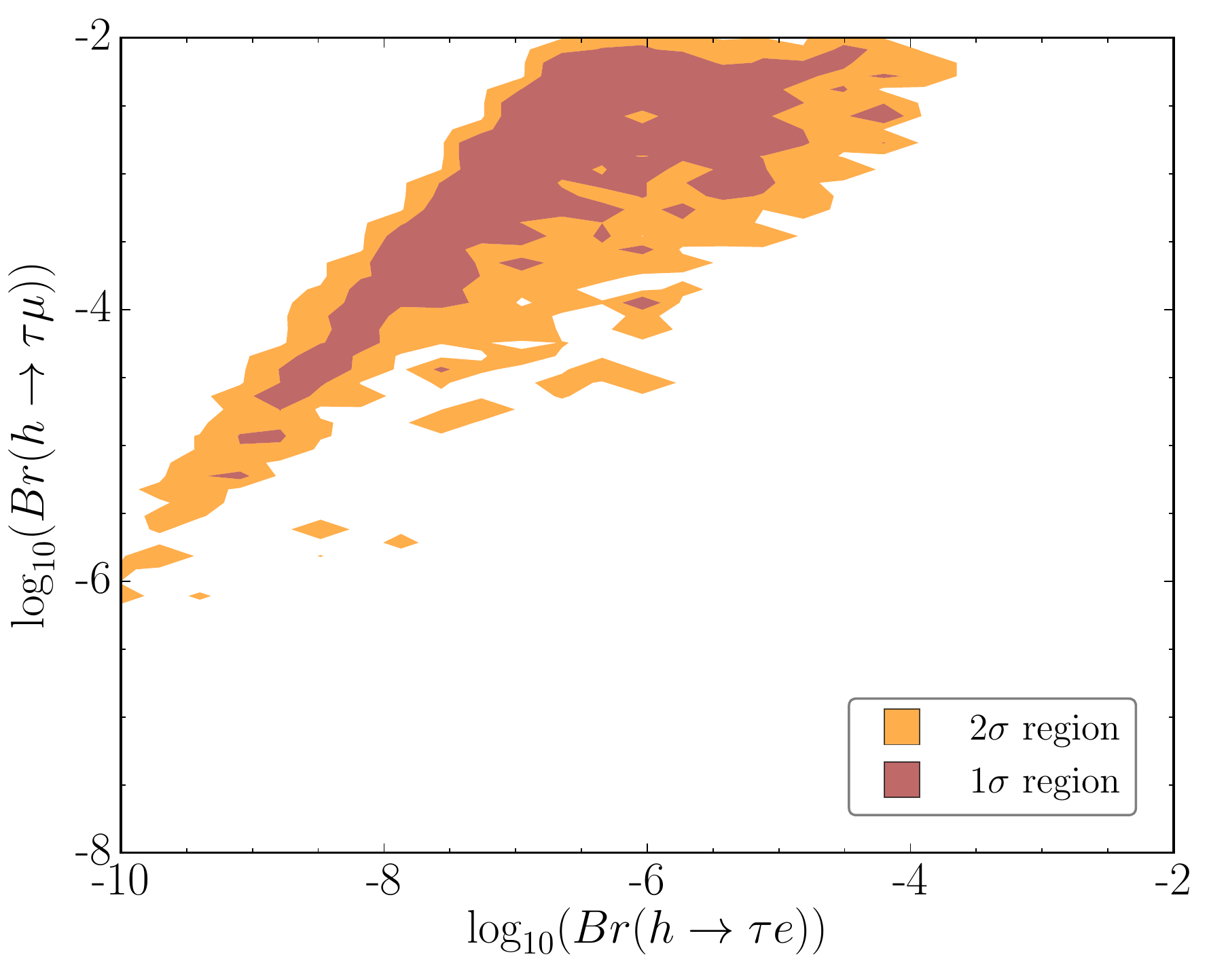}}
\subfloat[IO]{\includegraphics[width=0.45\textwidth]{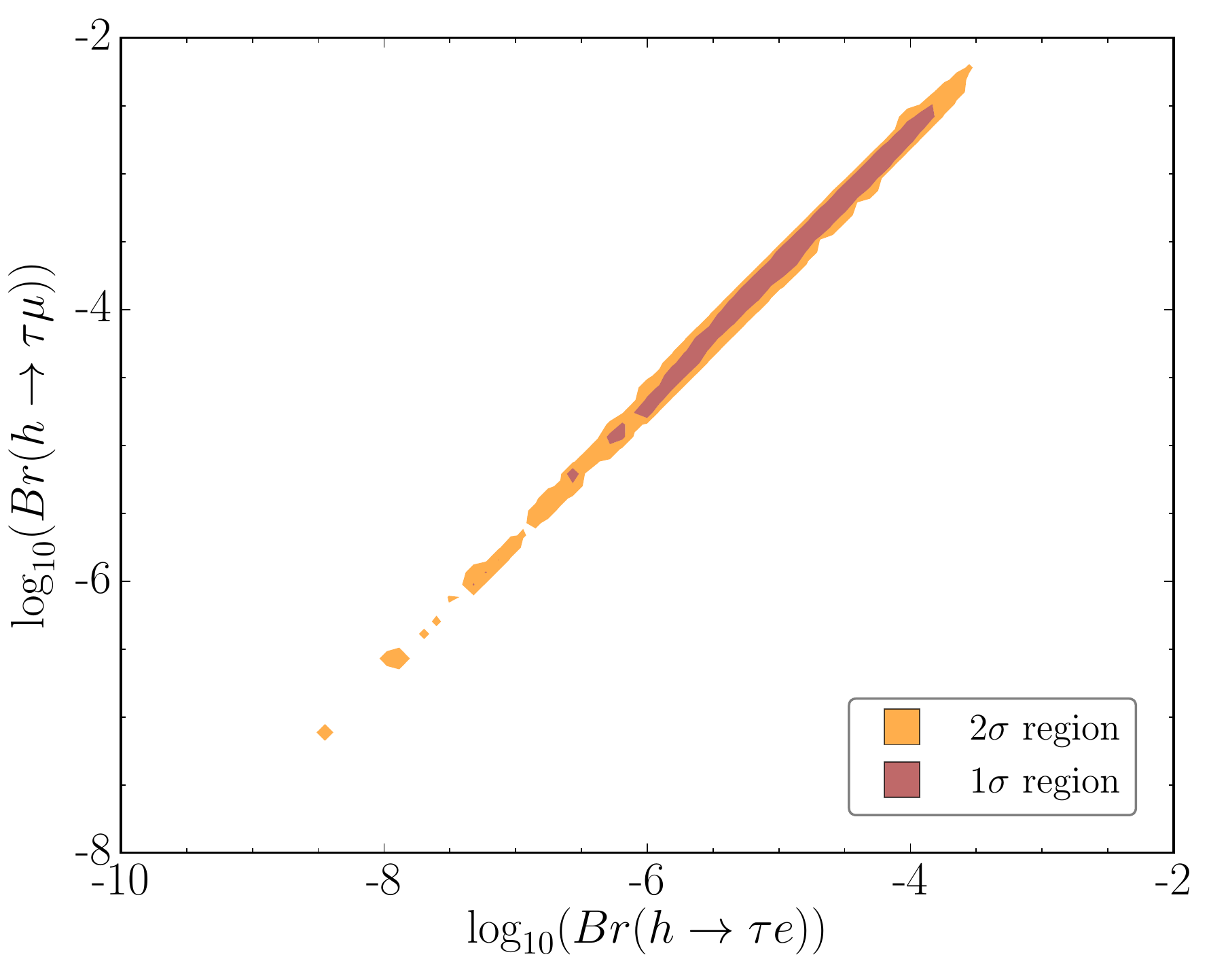}}
\caption{Allowed regions of the branching ratios ${\rm Br}(h\to \tau e)$ and ${\rm Br}(h\to \tau \mu)$ for (a) NO and (b) IO. }\label{fig:taumu_taue}
\end{figure}

In figure~\ref{fig:taumu_taue}, we show the allowed regions of the branching ratios ${\rm Br}(h\to \tau e)$ and  ${\rm Br}(h\to \tau \mu)$ for NO (left panel) and IO (right panel). No correlation exists for $\mu = 0$, while there is a strong correlation for $\mu \neq 0$, stronger for IO than for NO. In general, we find that ${\rm Br}(h\to \tau e) \lesssim 10^{-2}\, {\rm Br}(h\to \tau \mu)$ or even lower for both orderings. Therefore, observations of $h\to \tau e$ will be considerably more challenging than for $h\to \tau \mu$.

\begin{figure}[ht!]
\centering
\subfloat[NO]{\includegraphics[width= 0.45\textwidth]{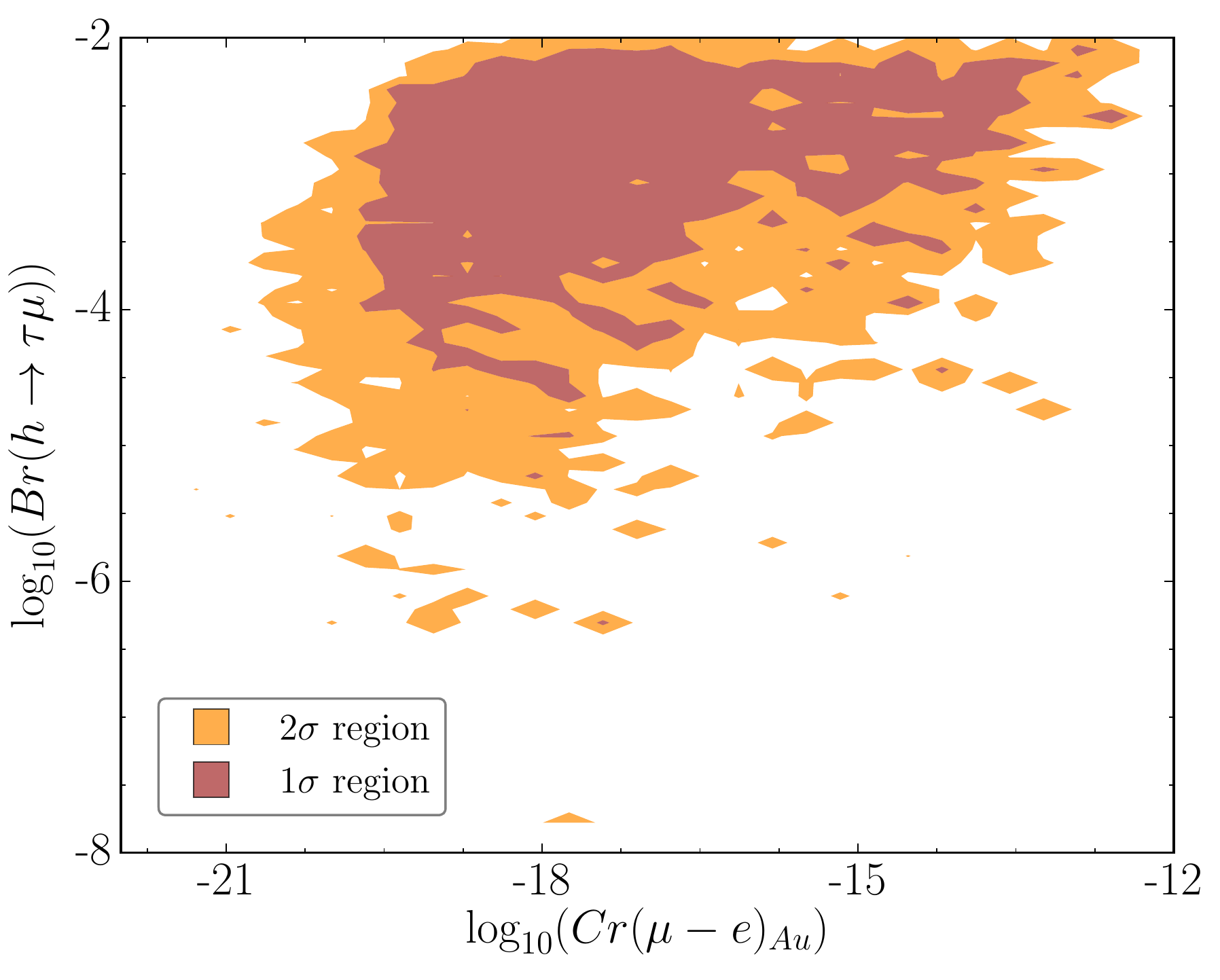}}
\subfloat[IO]{\includegraphics[width= 0.45\textwidth]{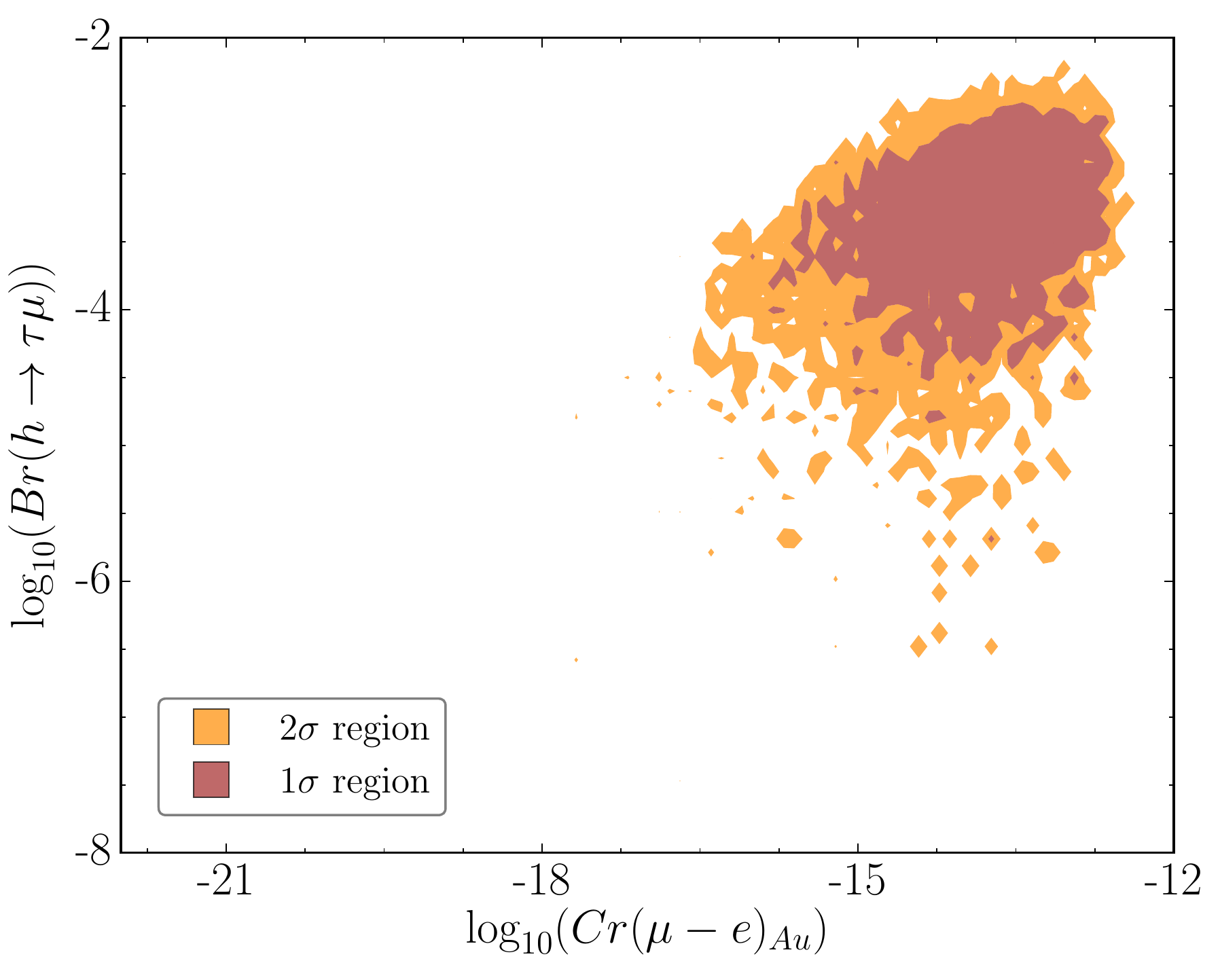}}
\caption{Allowed regions of $\mu e$ conversion and the branching ratio ${\rm Br}(h\to \tau\mu)$ for (a) NO and (b) IO.}\label{fig:mueconv}
\end{figure}

In figure~\ref{fig:mueconv}, we display the allowed regions of the $\mu e$ conversion rate in gold and the branching ratio ${\rm Br}(h\to \tau\mu)$ for NO (left panel) and IO (right panel). As discussed in section~\ref{sec:mueconv}, next generation experiments, using aluminium and titanium, are expected to achieve an improved sensitivity of up to about four orders of magnitude, maybe reaching $\mathcal{O}(10^{-19})$~\cite{Donghia:2016lzt}. The $\mu e$ conversion rates of these materials are of the same order of magnitude as that of gold,\footnote{Quantitatively, the $\mu e$ conversion rates for Al and Ti scale as $\rm{Cr}(\mu \to e)_{\rm Al}\simeq 0.5\cdot \rm{Cr}(\mu \to e)_{\rm Au}$ and $\rm{Cr}(\mu \to e)_{\rm Ti}\simeq 0.8\cdot \rm{Cr}(\mu \to e)_{\rm Au}$, respectively.} and therefore, if a negative result is obtained, IO would be excluded, while there would still be a considerable allowed region for NO. Thus, there is a complementarity between ${\rm Br}(\tau \to \mu\gamma)$  and $\mu e$ conversion. On the other hand, the sensitivity of ${\rm Br}(\tau \to e\gamma)$ is expected to reach about $3\cdot10^{-9}$~\cite{HerediadelaCruz:2016yqi} and ${\rm Br}(\mu \to e\gamma)$ is expected to be improved by one order of magnitude, but these are not able to test the model as thoroughly as ${\rm Br}(\tau \to \mu\gamma)$ and $\mu e$ conversion.

\begin{figure}[ht!]
\centering
\subfloat[NO]{\includegraphics[width= 0.45\textwidth]{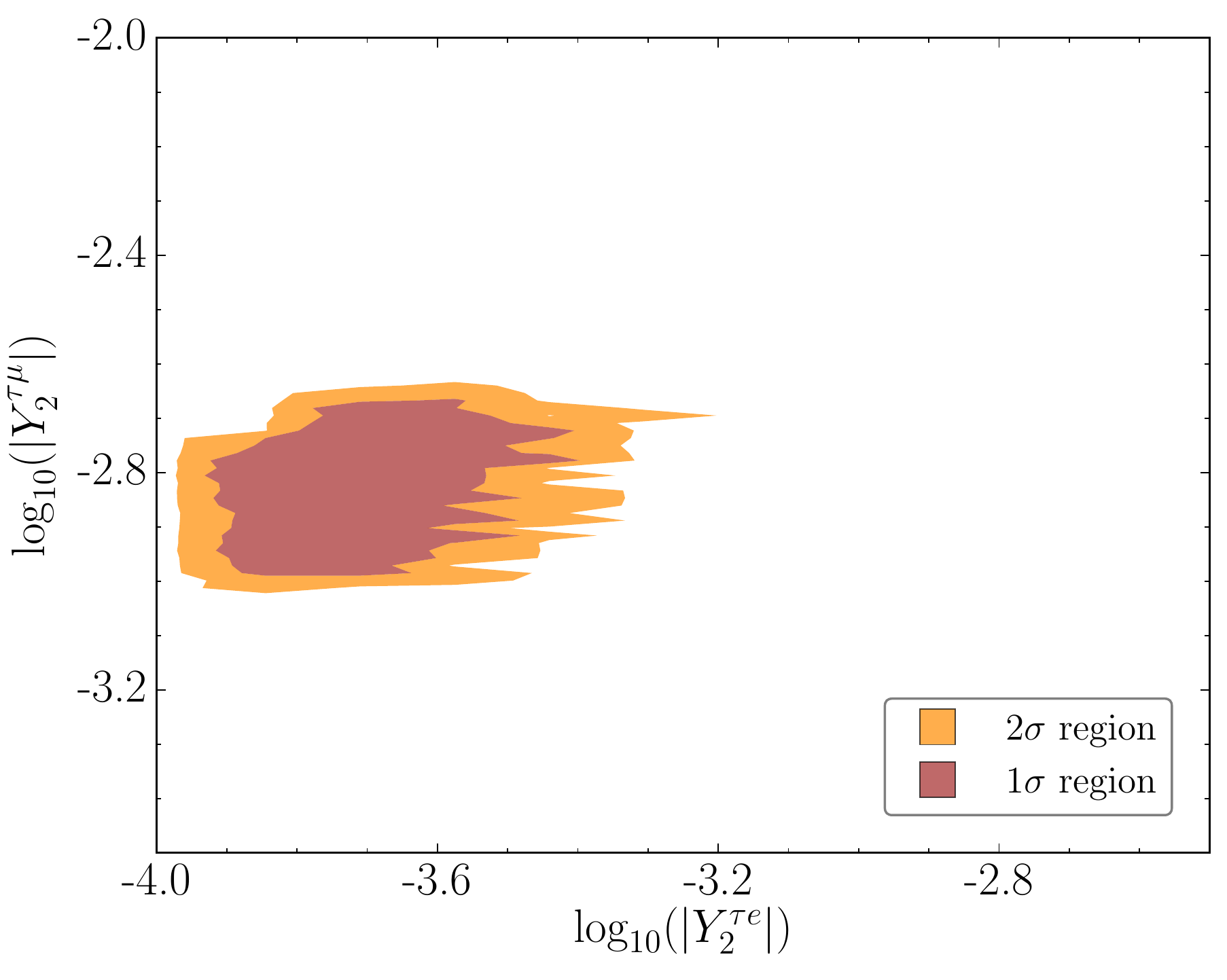}}
\subfloat[IO]{\includegraphics[width= 0.45\textwidth]{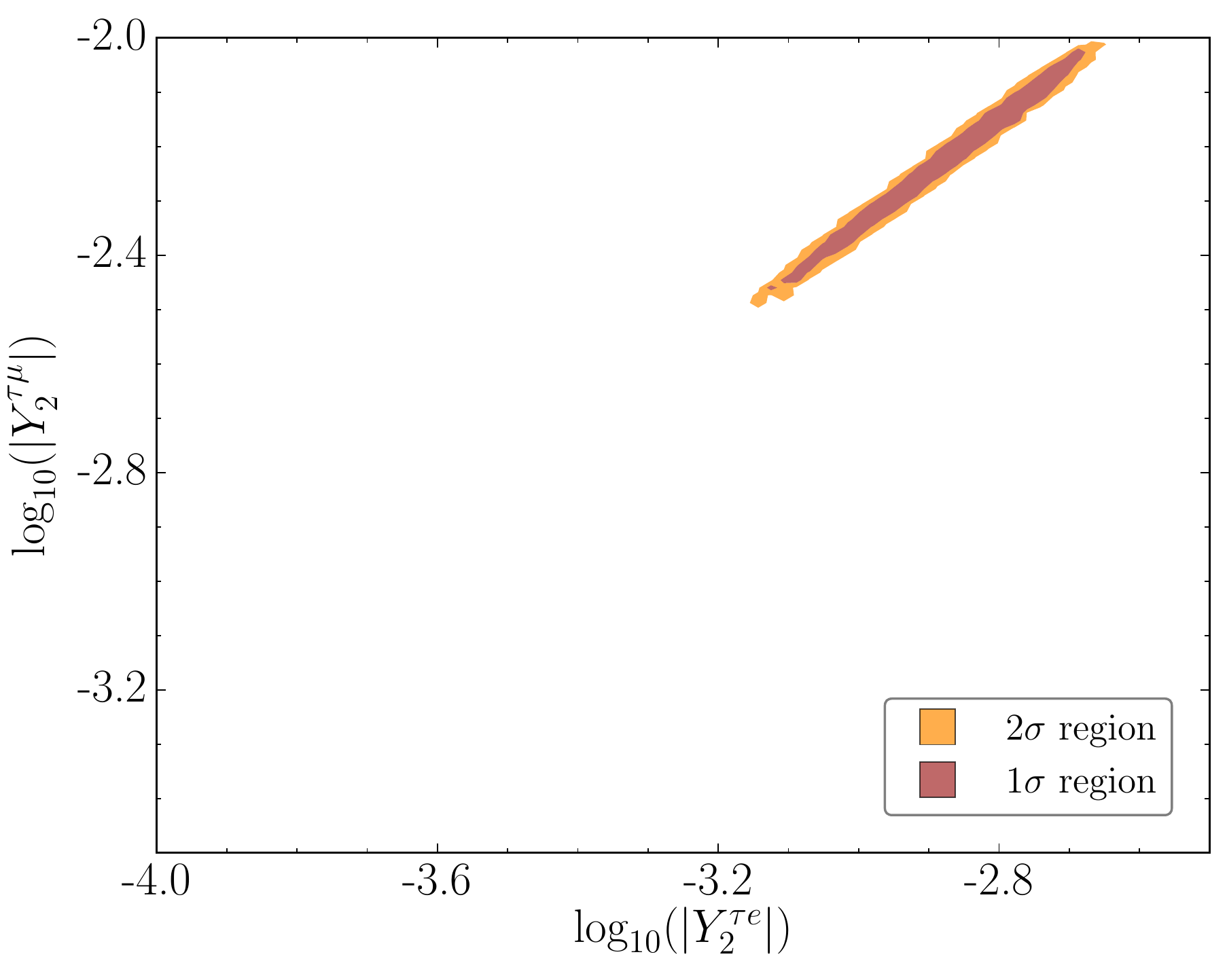}}
\caption{Allowed regions of the Yukawa couplings $Y_2^{\tau e}$ and $Y_2^{\tau \mu}$ for (a) NO and (b) IO.}\label{fig:Yukawa}
\end{figure}

In figure~\ref{fig:Yukawa}, we present the allowed regions for the absolute values of the Yukawa couplings $Y_2^{\tau e}$ and $Y_2^{\tau \mu}$ in both NO (left panel) and IO (right panel). We find that the regions are quite well defined, although they are larger in IO than in NO, and the allowed values are larger for $Y_2^{\tau \mu}$ than for $Y_2^{\tau e}$. Note that the value of $Y_2^{\tau \mu}$, which is a very relevant parameter since it controls the decays of the scalars into $\tau \mu$ (together with $Y_2^{\mu \tau}$), is always larger than around $10^{-3}\,(2.5\cdot 10^{-3})$ in NO (IO).

\begin{figure}[ht!]
\centering
\subfloat[NO]{\includegraphics[width= 0.45\textwidth]{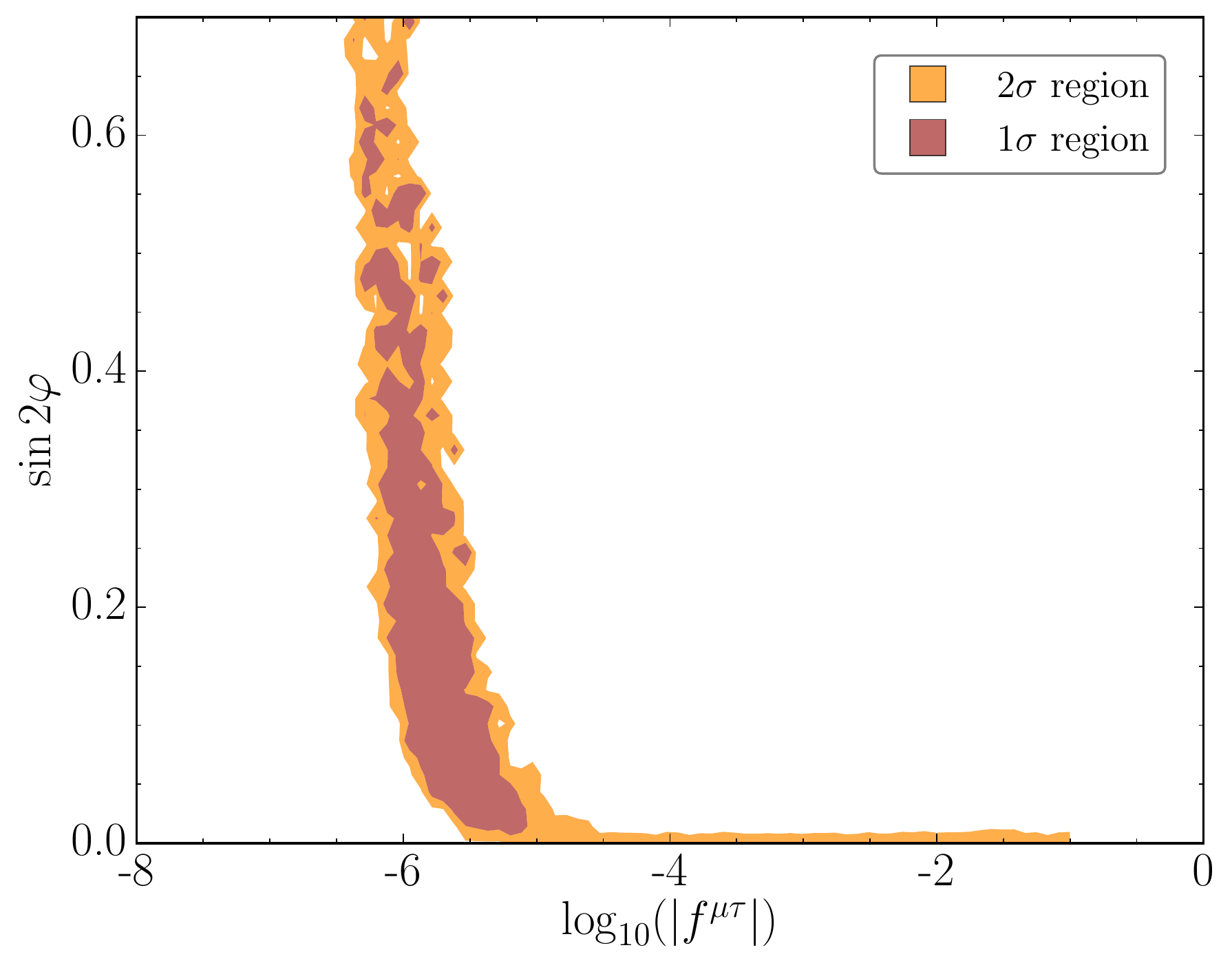}}
\subfloat[IO]{\includegraphics[width= 0.45\textwidth]{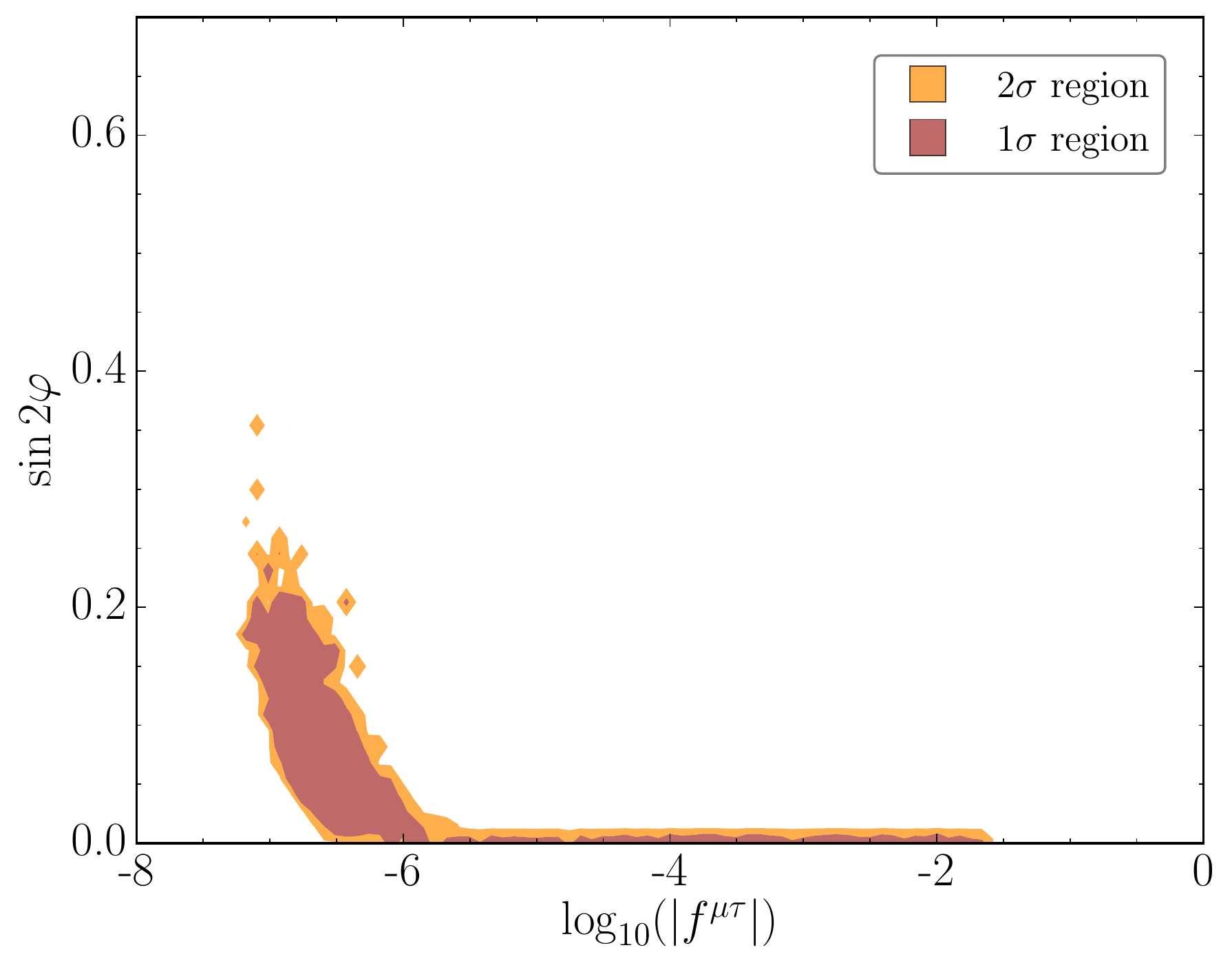}}
\caption{Allowed regions of the Yukawa coupling $f^{\mu\tau}$ and the mixing angle $\sin{2\varphi}$ for (a) NO and (b) IO.}\label{fig:fmutau_s2fi}
\end{figure}

The scale of neutrino masses is controlled by the charged scalar mixing angle $\varphi$, which is proportional to the trilinear coupling $\mu$, see eq.~\eqref{eq:charged_mixing}, and the Yukawa couplings $f$ and $Y_2$. Therefore, in figure~\ref{fig:fmutau_s2fi}, we plot the allowed regions of $|f^{\mu\tau}|$ and $s_{2\varphi}$ for both NO (left panel) and IO (right panel). The range of $|f^{\mu\tau}|$ is similar in both orderings. One can clearly see that $s_{2\varphi}$ is close to zero for $|f^{\mu\tau}| \gtrsim 10^{-5}$, while it grows very fast for $|f^{\mu\tau}| \lesssim 10^{-5}$, reaching values of $0.7\,(0.4)$ for NO (IO). As expected, the naturality condition of eq.~\eqref{eq:mu}, which is added to the $\chi^2$ function, restricts $\mu$ to be smaller than about $3\,(30)$~TeV for $\kappa=1\,(10)$ at $2\sigma$.

The Yukawa couplings $f^{\mu\tau}$ and $f^{e\tau}$ are always in the range $[10^{-7}, 0.1]$ and have similar allowed regions in both NO and IO. In fact, they are highly correlated with $f^{\mu\tau} \simeq f^{e\tau}\,(0.1 f^{e\tau})$ in NO (IO). Their allowed $1\sigma$ C.L.~regions lie roughly below $10^{-5}$, suppressing all interactions mediated by the antisymmetric Yukawa coupling $f$ of the singly-charged scalar singlet. Therefore, in order to describe neutrino masses and leptonic mixing, $Y_2$ should be much larger than $f$.

\begin{figure}[ht!]
\centering
\subfloat[$\mu=0$]{\includegraphics[width= 0.33\textwidth]{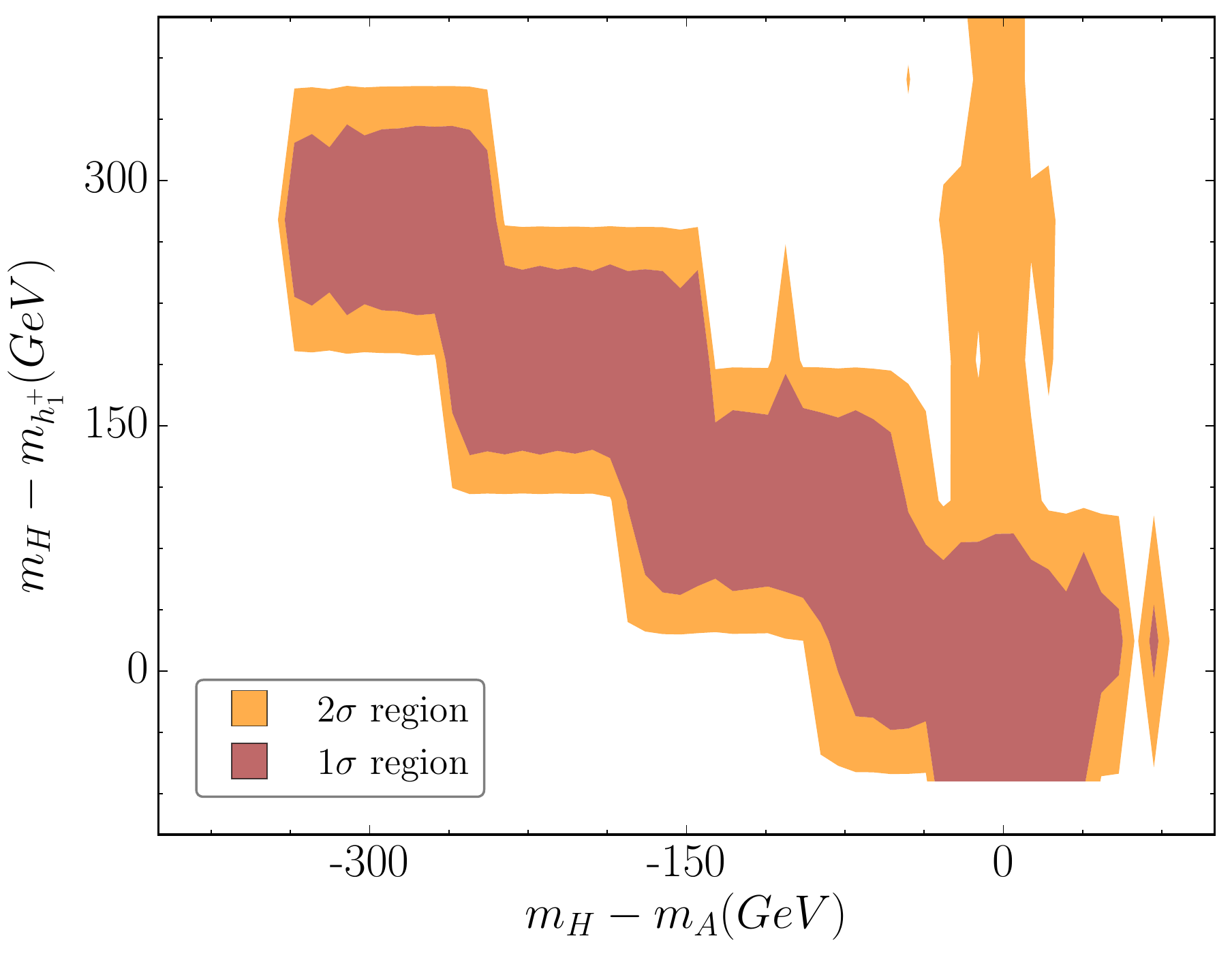}}
\subfloat[NO]{\includegraphics[width= 0.33\textwidth]{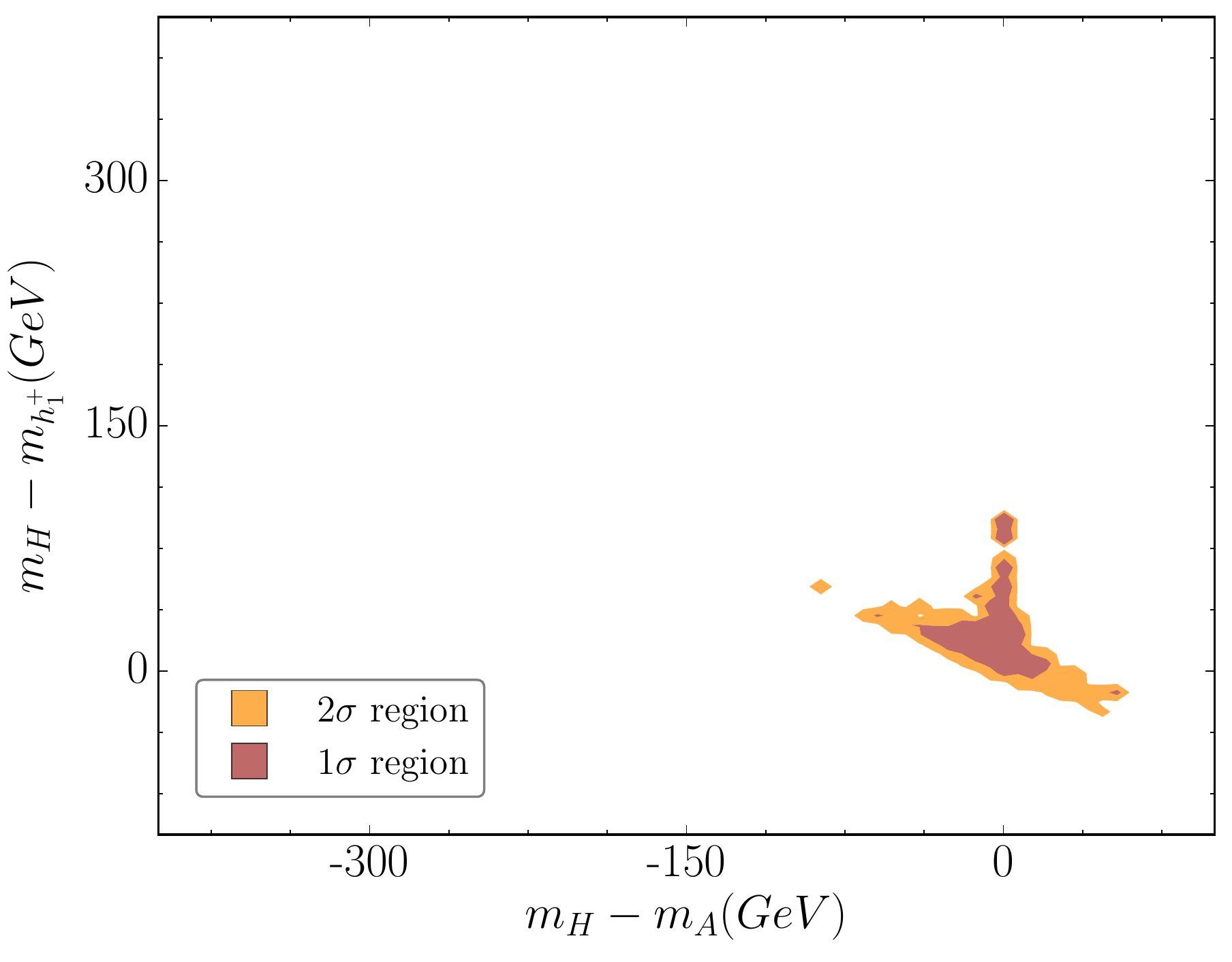}}
\subfloat[IO]{\includegraphics[width= 0.33\textwidth]{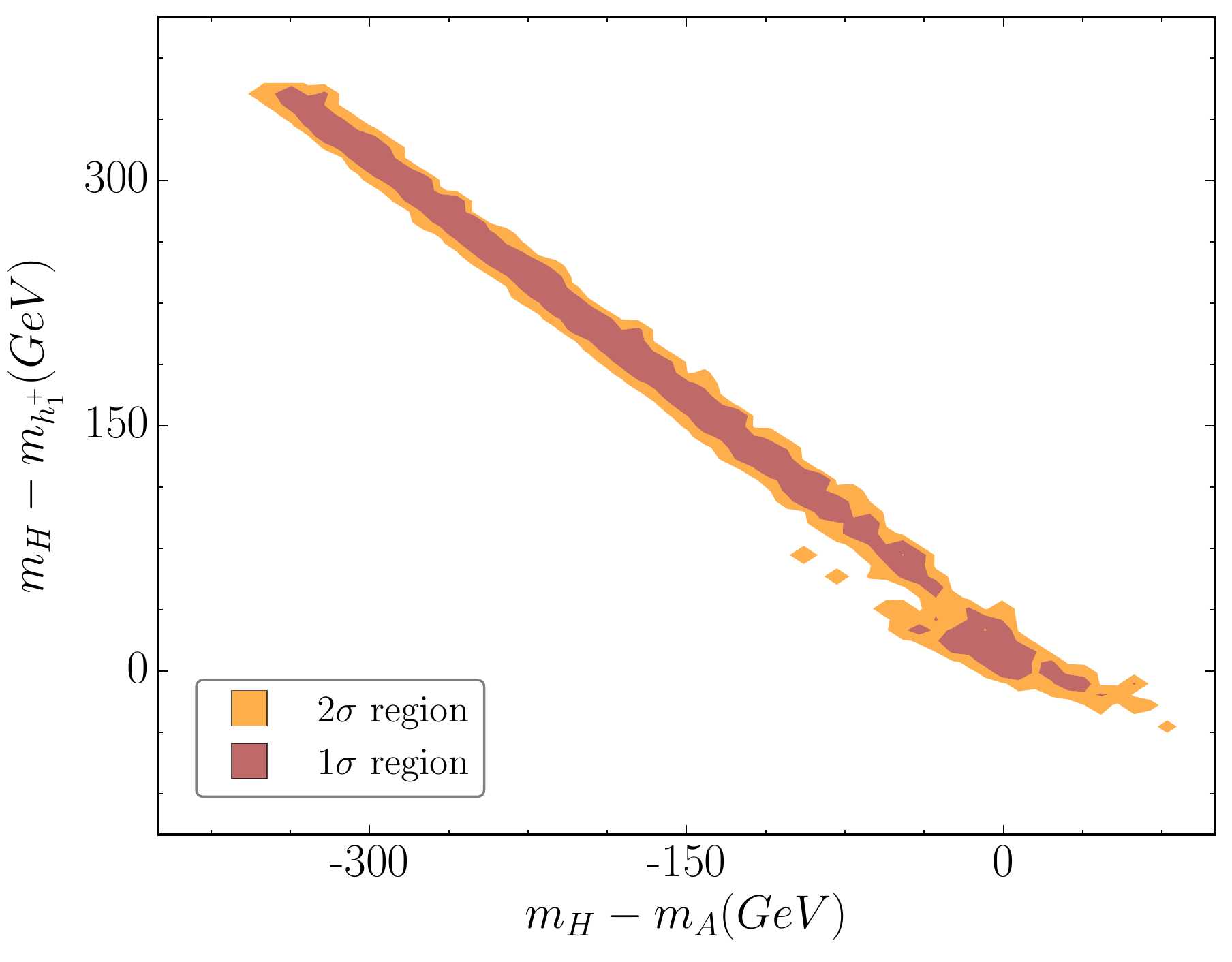}}
\caption{Splittings of the scalar masses: Allowed regions of the mass differences $m_H - m_A$ and $m_H - m_{h_1^+}$ for (a) $\mu = 0$, (b) NO, and (c) IO.}\label{fig:scalarmass}
\end{figure}

In figure~\ref{fig:scalarmass}, we show the allowed regions of the Higgs scalar mass differences $m_H-m_A$ and $m_H-m_{h_1^+}$. These affect the size of the parameter $T$, see appendix~\ref{sec:ewpt}. We find similar results to those in ref.~\cite{Benbrik:2015evd}. At $1\sigma$ C.L.~and at low scalar masses, $m_H-m_{h_1^+}$ is roughly equal to $m_H-m_A$, thus canceling the contributions to $T$. At $2\,\sigma$ C.L., for the two cases $\mu=0$ and IO, $m_H$ and $m_A$ can be close to each other and still fulfill $m_H  \geq m_{h_1^+}$. The allowed ranges of $m_A$ and $m_H$ are the same, but the two are not completely correlated, especially for IO. In fact, it is possible to have $m_A\simeq 100$~GeV, while $100~{\rm GeV}\lesssim m_H\lesssim 500~{\rm GeV}$.
In addition, we find that the mass of the lightest charged scalar is $m_{h^+_1} \lesssim 0.9\,(1.7)$~TeV at $1\,(2)\,\sigma$ C.L.~in NO, while $m_{h^+_1} \lesssim 0.7\,(1.1)$~TeV at $1\,(2)\,\sigma$ C.L.~in IO. The other charged scalar of the model, i.e.~$h_2^+$, is always heavier than $h_1^+$, see eq.~\eqref{eq:charged}, and can reach values of $\mathcal{O}(100)$~TeV in both orderings. The larger the $m_{h_2^+}$, the smaller the $s_{2\varphi}$, see eq.~\eqref{eq:charged_mixing}, and therefore, the smaller the neutrino masses.\footnote{The heavy scalars contribute to $m_h$ at one-loop level and could, in principle, pose a problem for naturality, especially $h^+_2$ as it is the heaviest one. However, their contributions to the Higgs boson self-energy are suppressed by both the loop factor and the quartic couplings of the scalar potential and $m_h$ is therefore \emph{natural} for the values used.} 

\begin{figure}[ht!]
\centering
\subfloat[NO]{\includegraphics[width= 0.45\textwidth]{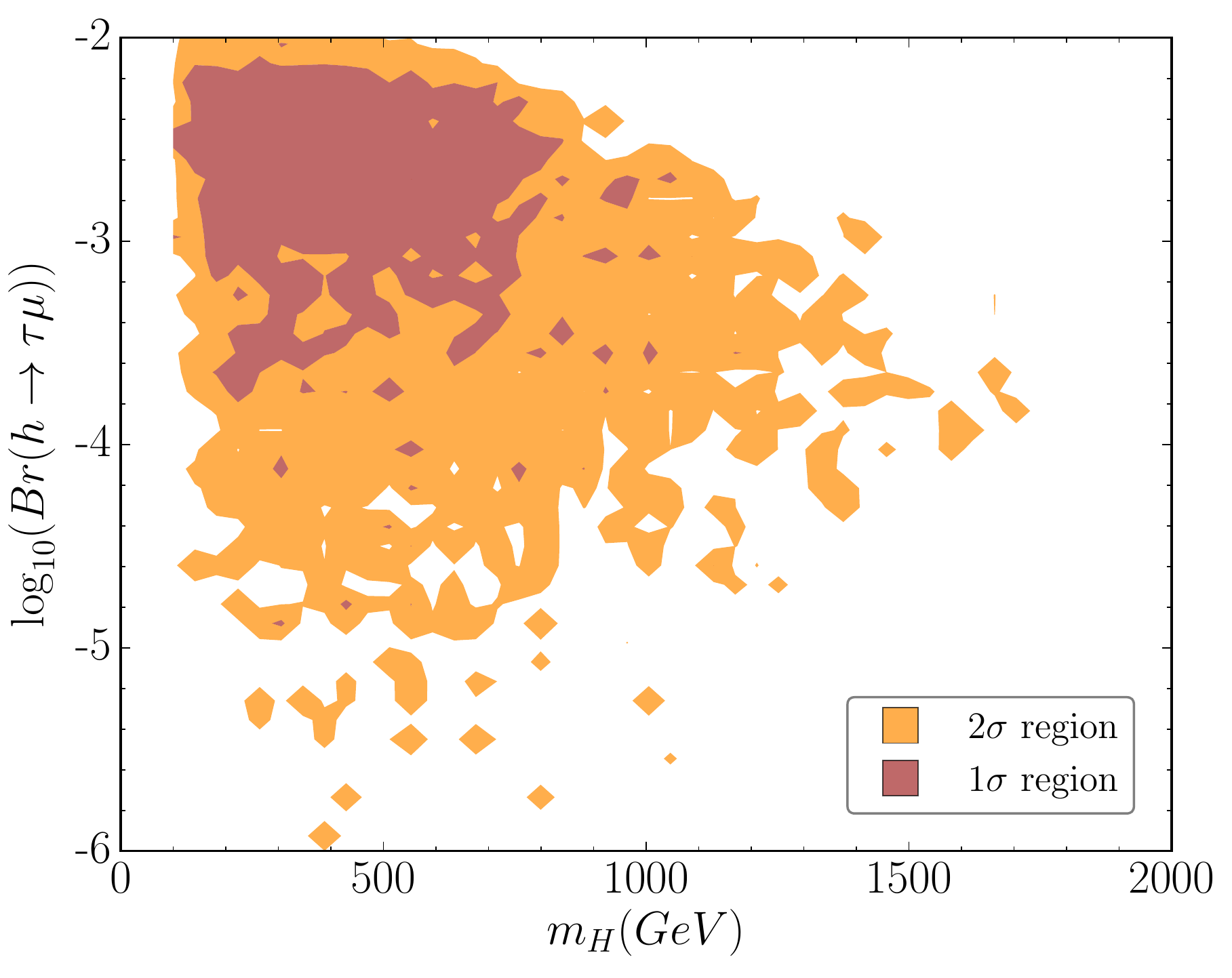}}
\subfloat[IO]{\includegraphics[width= 0.45\textwidth]{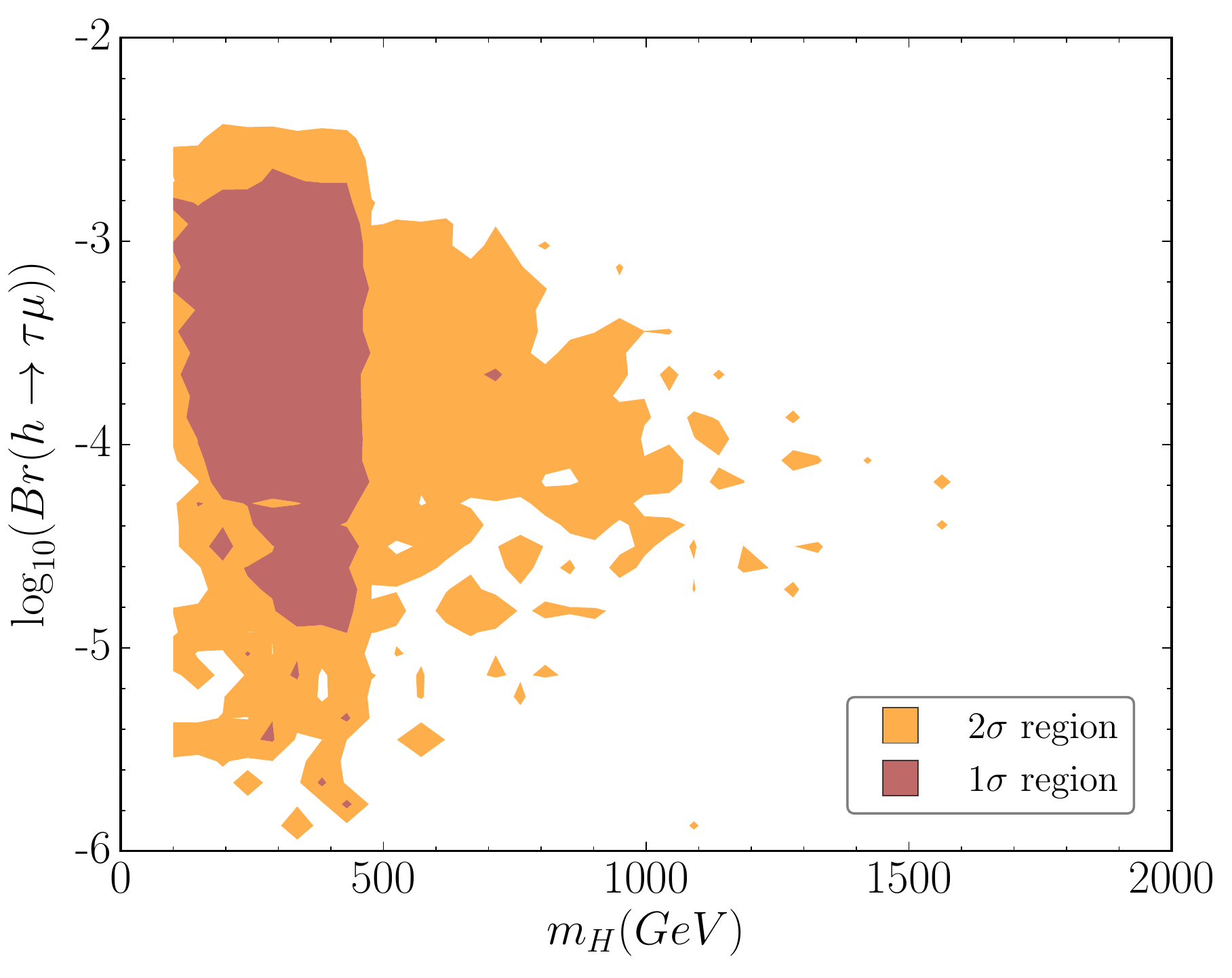}}
\caption{Allowed regions of the mass $m_H$ and the branching ratio ${\rm Br}(h \to \tau \mu)$ for (a) $\mu = 0$, (b) NO, and (c) IO.}\label{fig:mHhtaumu}
\end{figure}

From eq.~\eqref{BRH2}, we know that ${\rm Br}(h \to \tau \mu)$ is proportional to $1/m^4_H$, i.e., it decouples with the CP-even scalar mass. In figure~\ref{fig:mHhtaumu}, we display the allowed regions of the mass $m_H$ and the branching ratio ${\rm Br}(h \to \tau \mu)$.  For NO (left panel), $m_H \lesssim 0.9\,(1.7)$~TeV at $1\,(2)\,\sigma$ C.L., while for IO (right panel), $m_H \lesssim 0.7\,(1.1)$~TeV at $1\,(2)\,\sigma$ C.L. Therefore, if an extra CP-even scalar (and close by CP-odd and charged scalars, see figure~\ref{fig:scalarmass}) is observed below $0.9\,(0.7)$~TeV, then one expects ${\rm Br}(h \to\tau \mu) \gtrsim 10^{-4}\,(10^{-5})$ for NO (IO) at $1\sigma$ C.L. The heavy CP-even scalar $H$ could also have sizable decays into $\tau \mu$, depending on the scalar spectrum. 

\begin{figure}[ht!]
\centering
\subfloat[$\mu=0$]{\includegraphics[width= 0.33\textwidth]{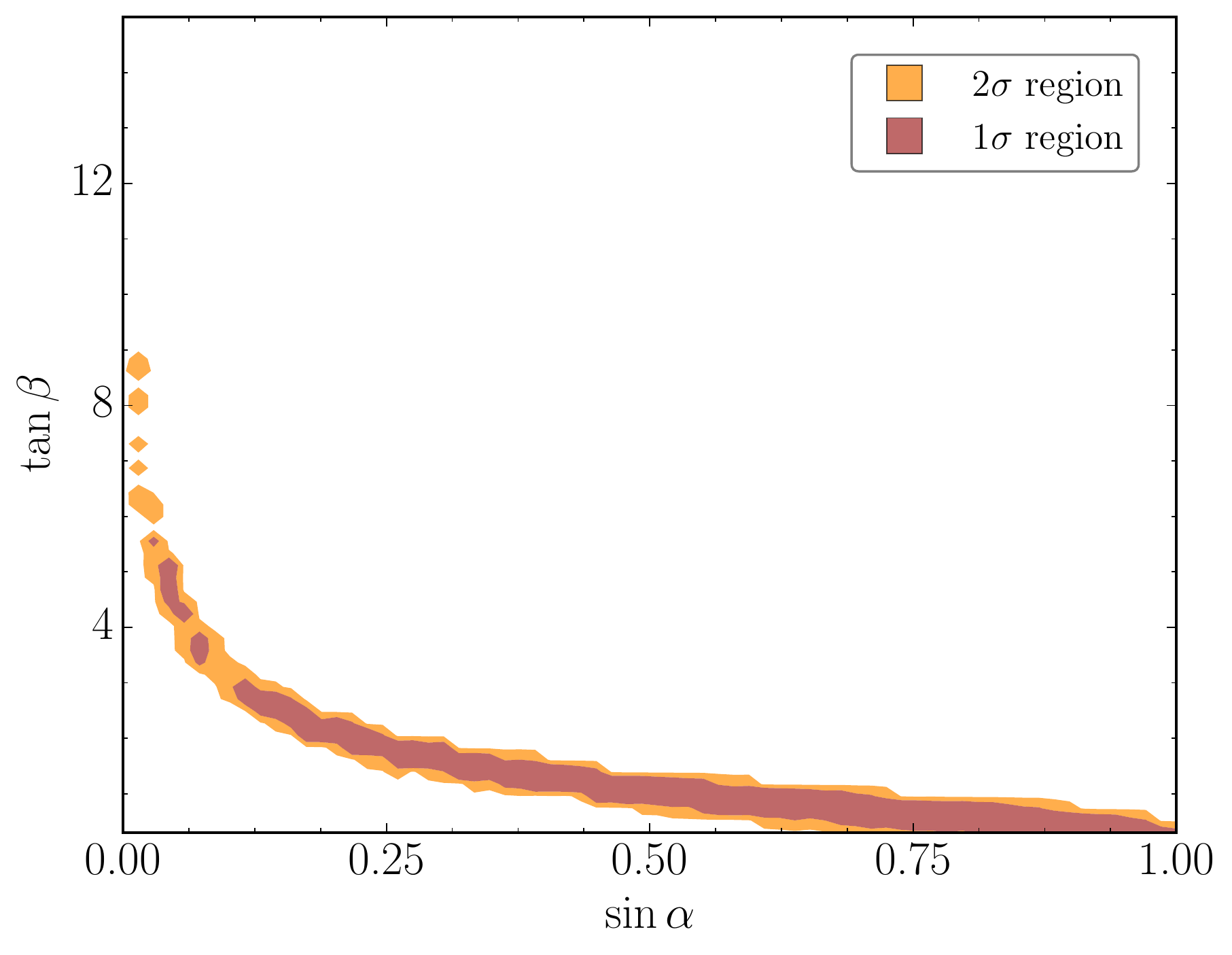}}
\subfloat[NO]{\includegraphics[width= 0.33\textwidth]{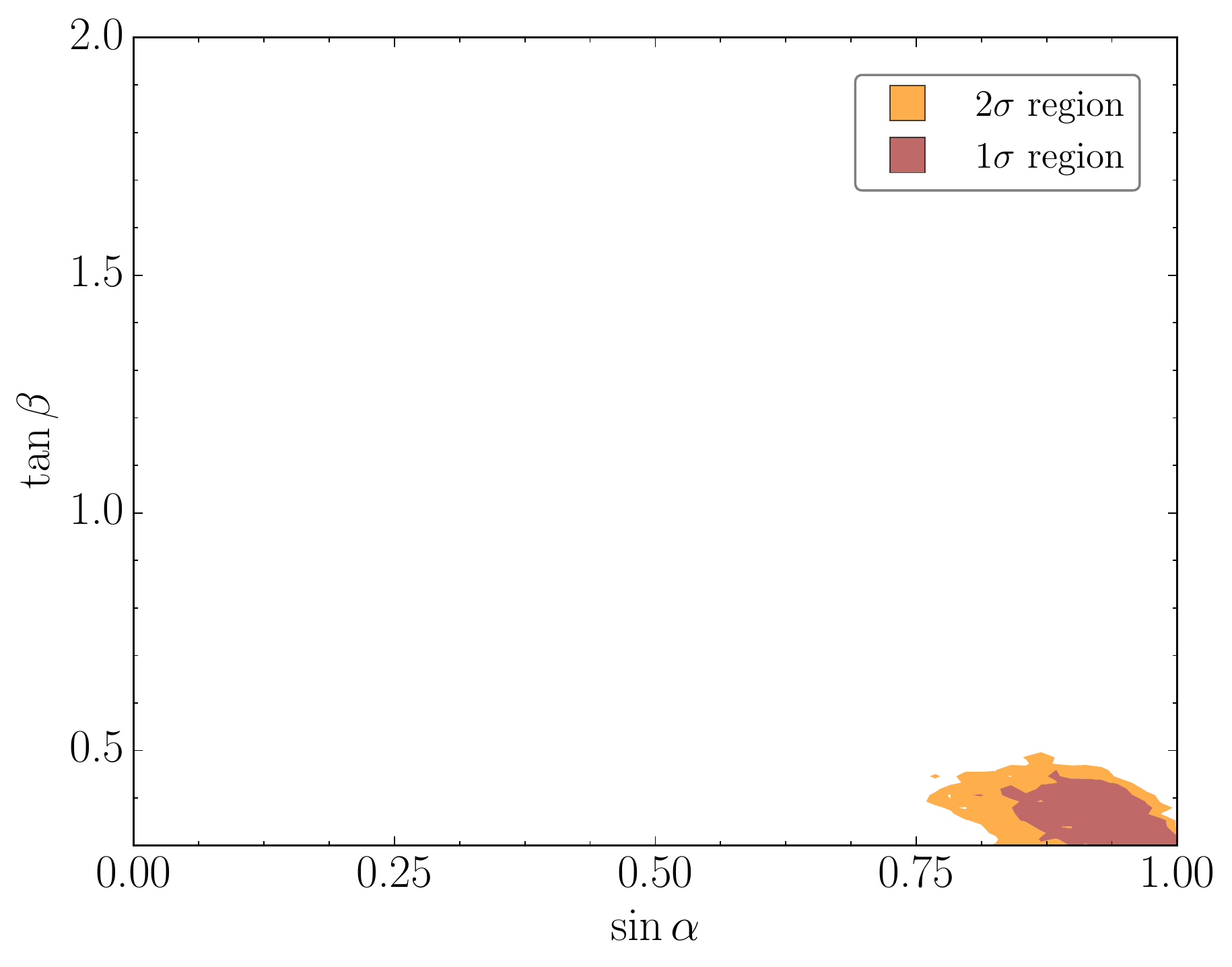}}
\subfloat[IO]{\includegraphics[width= 0.33\textwidth]{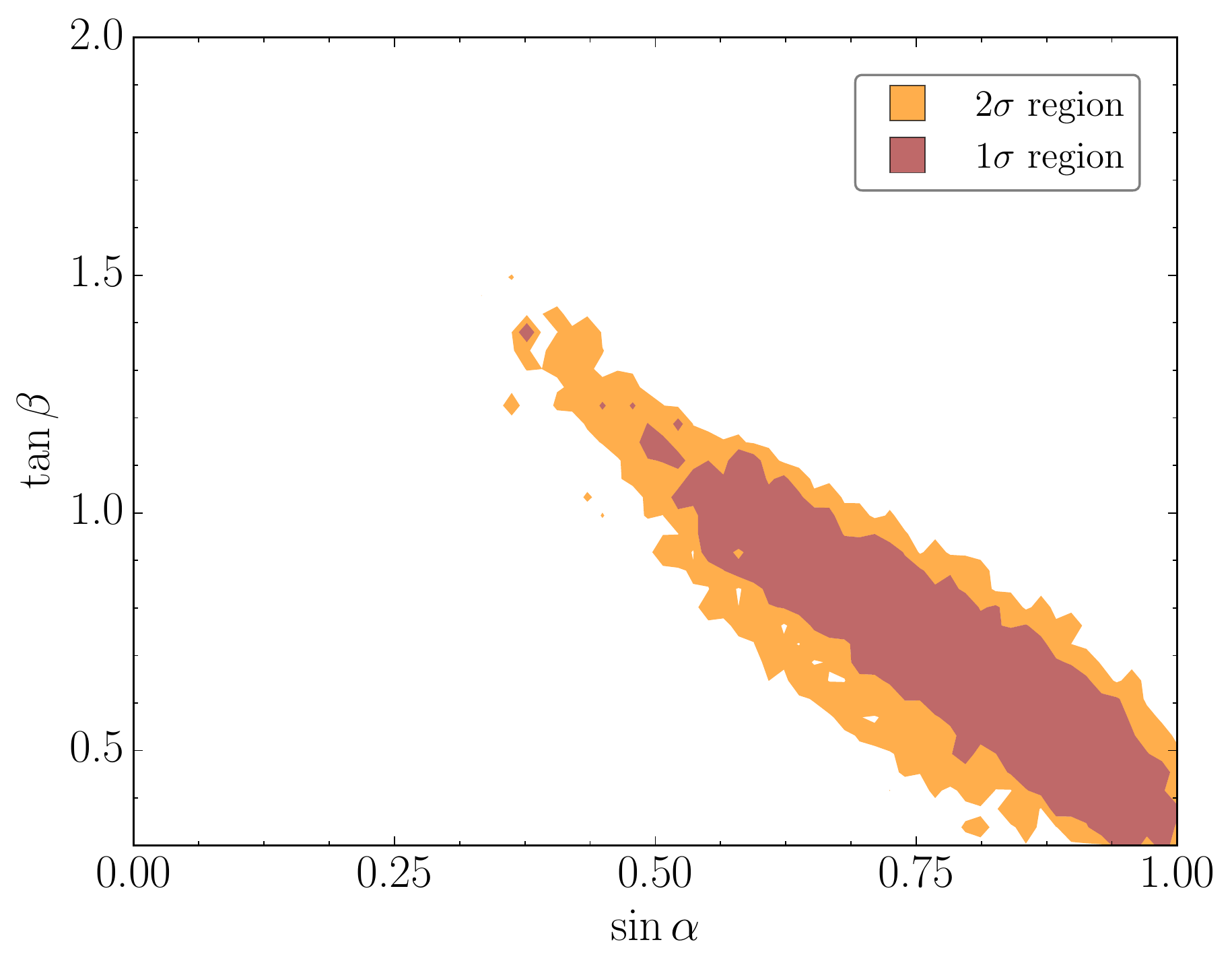}}
\caption{Allowed regions of the mixing angles $\sin{\alpha}$ and $\tan{\beta}$ for (a) $\mu = 0$, (b) NO, and (c) IO.}\label{fig:tbsa}
\end{figure}

In figure~\ref{fig:tbsa}, we show $\tan{\beta}$ as a function of $s_\alpha$ for $\mu = 0$ (left panel), NO (middle panel), and IO (right panel). Having Higgs boson decays close to the observed ones implies being close to the decoupling limit, i.e.~$s_{\beta-\alpha}\to 1$, and therefore, $\tan{\beta}$ and $s_\alpha$ are strongly correlated. For massless neutrinos, $\tan{\beta}$ can reach values up to 15 and $s_\alpha$ can approach zero. However, if neutrino masses are introduced, the value of $\tan{\beta}$ is severely constrained to smaller values in both orderings and the allowed range of $s_\alpha$ is also reduced. The upper bound on $\tan{\beta}$ is clearly more severe for NO, where $\tan{\beta}$ is driven to the smallest possible values (i.e.~below 0.5), while for IO, it can reach values up to 1.4. We have also performed a run forcing the value of the unphysical parameter $\tan{\beta}$ to be large, i.e.~$40\lesssim\tan{\beta}\lesssim 50$, and we find that the fit becomes significantly worse, leading to a value for the minimum of the $\chi^2$ function of $\mathcal{O}(1000)$, and thus excluding this scenario.

\begin{figure}[ht!]
\centering
\subfloat[NO]{\includegraphics[width= 0.45\textwidth]{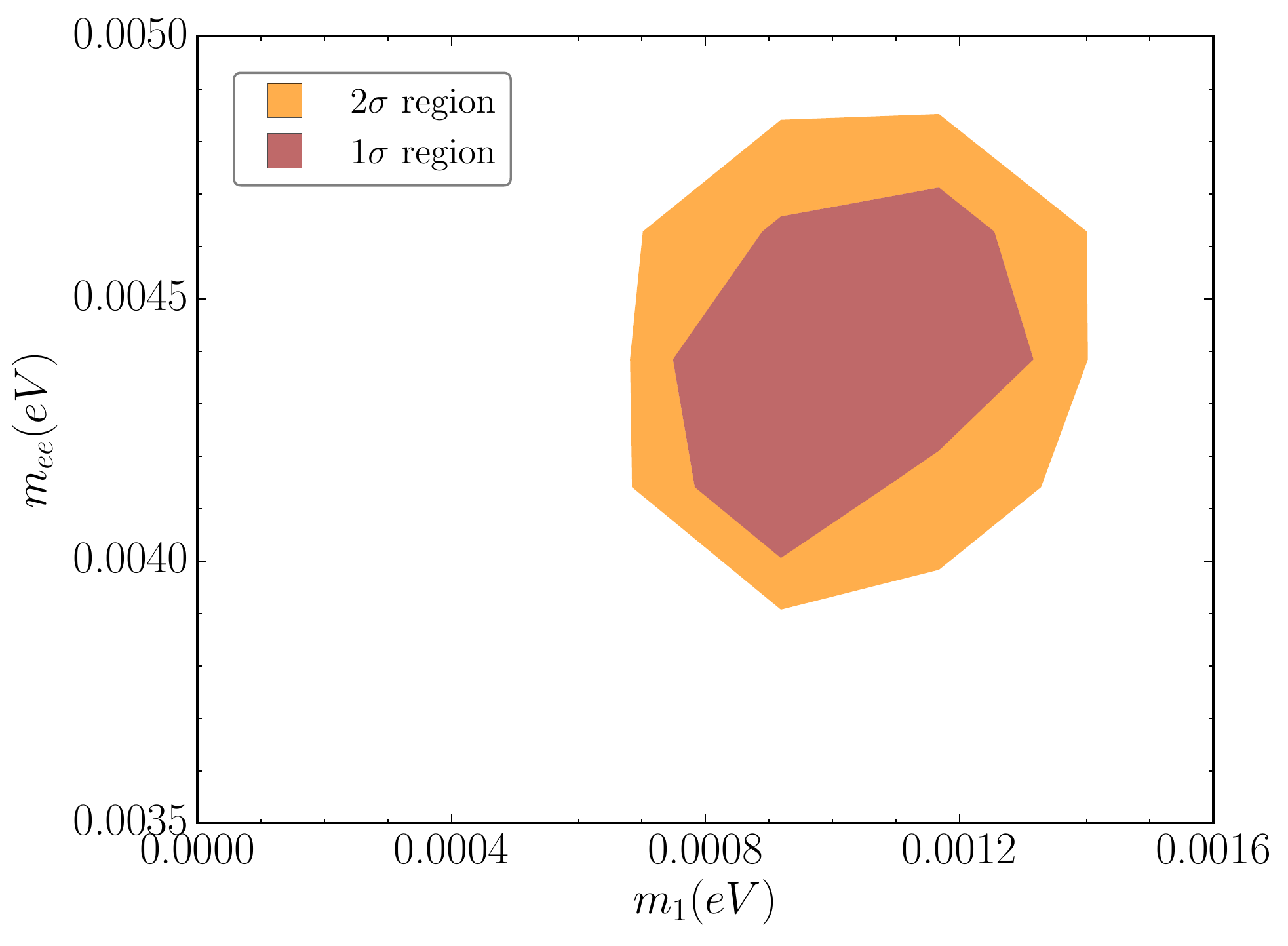}}
\subfloat[IO]{\includegraphics[width= 0.431\textwidth]{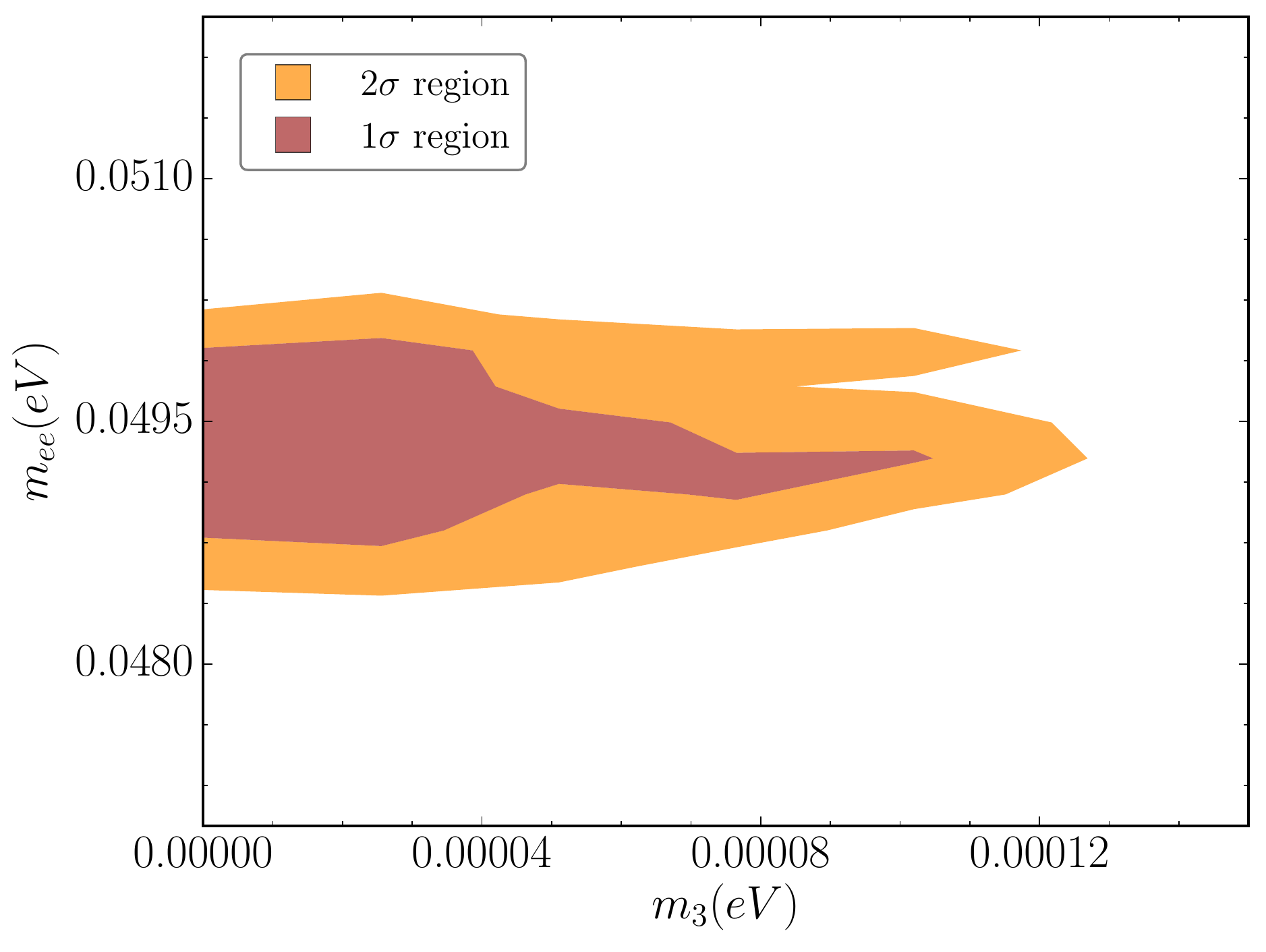}}
\caption{Allowed regions of the mass of the lightest neutrino mass eigenstate ($m_1$ for (a) NO and (b) $m_3$ for IO) and the effective neutrino mass parameter $m_{ee}$ appearing in neutrinoless double beta decay.}\label{fig:smallmass}
\end{figure}

In figure~\ref{fig:smallmass}, we present the allowed regions of the effective mass parameter $m_{ee}$ that appears in neutrinoless double beta decay and the smallest neutrino mass ($m_1$ for NO and $m_3$ for IO) for NO (left panel) and IO (right panel). We can see that the smallest neutrino mass is several orders of magnitude smaller in IO than in NO. In fact, in IO, it can be massless, whereas this is not the case in NO. This is consistent with the fact that the fit in NO is bad when $Y_2^{\mu \tau}=0$, rendering one neutrino massless. In NO, we obtain $|m_{ee}| \simeq (4-5)\,\mathrm{meV}$, while in IO, $|m_{ee}|$ is one order of magnitude larger, i.e.~about 50~meV, and thus, it will be possibly probed in planned neutrinoless double beta decay experiments. Furthermore, the mass of the lightest neutrino mass eigenstate is less than $1.4\cdot 10^{-3}$~eV ($1.3\cdot 10^{-4}$~eV) in NO (IO). Note that in NO there is a lower bound on this mass, $6\cdot 10^{-4}$ eV, while IO is compatible with a massless neutrino.

In addition, we mention the phases of the leptonic mixing matrix $U$. We find that the preferred value for the Dirac CP-violating phase $\delta$ is close to $2\pi$ in NO and close to $\pi$ in IO, thus implying no leptonic CP violation in our scenario of the Zee model. The hint of $\delta \sim 3\pi/2$ from global fits to neutrino oscillation data, if confirmed, can therefore not be accommodated in any of the two orderings. We also find that the values of both Majorana CP-violating phases $\phi_1$ and $\phi_2$ are around $2\pi$ in both orderings.

As expected, the strong limits from other LFV processes imply that the NSI parameters $\chi^{\rm m}_{\tau\tau}$ and  $\varepsilon^{\rm m}_{\tau\tau}$ defined in eq.~\eqref{eq:NSI} are very suppressed in both orderings, i.e.~$\chi^{\rm m}_{\tau\tau},\varepsilon^{\rm m}_{\tau\tau}\lesssim 10^{-8}$, and therefore well below future experimental sensitivity. The NSI parameters $\varepsilon^{\rm \rho \sigma}_{\alpha \beta}$, given in eq.~\eqref{eq:NSIdef}, are also generated in the Zee--Babu model, where they reach values of about $10^{-4}$~\cite{Ohlsson:2009vk, Herrero-Garcia:2014hfa}. However, in the Zee model, NSIs turn out to be smaller, since neutrino masses are generated at one-loop level, while in the Zee--Babu model, the latter arise at two loops. 

Furthermore, we check how our results change when imposing the fine-tuning parameter $\kappa$ to be 10 instead of 1, see eq.~\eqref{eq:mu}. For both values of $\kappa$, the $1\sigma$ C.L.~region is close to the upper limit on $\mu$, even though the entire range down to $\mu=1$~GeV is allowed at $2\sigma$ C.L. The value of $\mu$ is not significantly correlated to the value of $\tan{\beta}$. We find that the allowed ranges for the scalar masses (where the upper bounds determine the neutrino masses) depend critically on $\kappa$, having larger allowed mass ranges the larger the value of $\kappa$. Quantitatively, for $\kappa=1$ at $1\,(2)\,\sigma$ C.L., the upper bounds on the masses are $m_ A, m_H, m_{h_1^+} = 0.9\,(1.7)$~TeV for NO and $m_A, m_H, m_{h_1^+} = 0.7\,(1.1)$~TeV for IO, whereas for $\kappa=10$ at $1\,(2)\,\sigma$ C.L., the upper bounds are $m_A, m_H, m_{h_1^+} = 1.6\,(2.5)$~TeV for NO and $m_A, m_H, m_{h_1^+} = 0.9\,(1.4)$~TeV for IO.

Finally, we summarize some of our main results in table~\ref{tab:results}, where we display the $2\sigma$ C.L.~regions for some of the most interesting observables and parameters for NO (for $\kappa=1,10$), and IO (for $\kappa=1,\,10$). We emphasize once more that for $\mu=0$ only upper bounds on the CLFV processes (and lower bounds on the scalar masses) exist, whereas for $\mu \neq 0$, there are \emph{lower} bounds for the CLFV processes (and \emph{upper} bounds on the scalar masses). This means that the scalar sector cannot be arbitrarily heavy if neutrino masses are to be reproduced. The precise upper bound depends crucially on $\mu$, see eq.~\eqref{eq:mu}. For $\mu=0$, the masses could be arbitrarily large, unlike the case of having $\mu \neq 0$ and reproducing neutrino masses, which imposes that they are below about 2~TeV. We do not display the ranges for all observables, such as neutrino masses and leptonic mixing parameters, as their contributions to $\chi_{\rm min}^2$ are shown in figure~\ref{fig:chi2}. In the Zee model (including $\mu=0$), the value of the muon AMM, which has not been included in the fit, is several orders of magnitude smaller than the experimental one. This implies a deviation of about $3.5\sigma$. The allowed ranges for the scalar masses $m_A$ and $m_H$ are the same for both values of $\kappa$, but the two are not completely correlated (especially in IO). It is possible to simultaneously have $m_A\simeq 100$~GeV and $m_H$ varying in the range $100~{\rm GeV}\lesssim m_H\lesssim 500~{\rm GeV}$. We conclude by stating that NO will be tested in the next generation CLFV searches, specially with $\tau \to \mu \gamma$ and $\mu e$ conversion, as well as searches for $h \to \tau \mu$ and the other new scalars at colliders.
\begin{table}
\footnotesize 
\centering
    \begin{tabular}{ c |c  c | c  c |}
        \cline{2-5}
       &   \multicolumn{2}{ c |}{NO} & \multicolumn{2}{ c |}{IO} \\ \cline{2-5}
         	\cline{1-5}  \multicolumn{1}{|c|}{Quantity} & \multicolumn{1}{c}{$\kappa=1$ } & \multicolumn{1}{c|}{$\kappa=10$ } & \multicolumn{1}{c}{$\kappa=1$} & \multicolumn{1}{c|}{$\kappa=10$} \\
		\cline{1-5}  \multicolumn{1}{|c|}{$ \chi^2_{\rm min}$}  & 10.7 & 11.0 & 21.7 & 21.5 \\
		\cline{1-5} \multicolumn{1}{|c|}{${\rm Br}(h\to \tau \mu)$} & $[1 \cdot10^{-6}, 1 \cdot10^{-2}]$& $[1 \cdot10^{-6}, 1 \cdot10^{-2}]$ & $[2\cdot 10^{-7}, 4\cdot 10^{-3}]$ & $[1 \cdot 10^{-7}, 5\cdot 10^{-3}]$ \\
		\cline{1-5} \multicolumn{1}{|c|}{${\rm Br}(h\to \tau e)$} & $[1 \cdot10^{-10}, 2\cdot 10^{-4}]$& $[1 \cdot 10^{-10}, 2\cdot 10^{-4}]$& $[6\cdot 10^{-9}, 3 \cdot10^{-4}]$ &$[3\cdot 10^{-9}, 3 \cdot10^{-4}]$  \\
		\cline{1-5} \multicolumn{1}{|c|}{${\rm Br}(\tau\to \mu \gamma)$} &$[8\cdot 10^{-10}, 3\cdot 10^{-8}]$  & $[1 \cdot 10^{-10}, 3\cdot 10^{-8}]$ & $[3\cdot 10^{-11}, 3\cdot 10^{-8}]$  & $[3\cdot 10^{-11}, 3\cdot 10^{-8}]$ \\
		\cline{1-5} \multicolumn{1}{|c|}{${\rm Br}(\mu\to e \gamma)$} &$[1 \cdot10^{-21}, 6\cdot 10^{-13}]$  & $[3\cdot 10^{-22}, 6\cdot 10^{-13}]$ & $[1 \cdot10^{-31}, 1\cdot 10^{-12}]$  & $[1 \cdot10^{-34}, 1\cdot 10^{-12}]$ \\
		\cline{1-5} \multicolumn{1}{|c|}{${\rm Cr}(\mu \to e)$} & $[1 \cdot10^{-21}, 4\cdot 10^{-13}]$  & $[1 \cdot10^{-21}, 4\cdot 10^{-13}]$& $[3\cdot 10^{-17}, 3\cdot 10^{-13}]$ & $[3\cdot 10^{-17}, 3\cdot 10^{-13}]$ \\
		\cline{1-5} \multicolumn{1}{|c|}{$m_A, m_H$~$[\mathrm{TeV}]$}  & $<1.7$& $<2.5$& $<1.1$ & $<1.4$ \\
		\cline{1-5} \multicolumn{1}{|c|}{$m_{h_1^+}$~$[\mathrm{TeV}]$}  & $<1.7$& $<2.5$& $<1.1$ & $<1.4$\\
		\cline{1-5} \multicolumn{1}{|c|}{$\sin(\beta-\alpha)$} & $[0.98,1.0]$ & $[0.98,1.0]$ & $[0.97,1.0]$ & $[0.97,1.0]$ \\
        \hline
    \end{tabular}
     \caption{\label{tab:results} Some results of our numerical scan. We show the ranges of the 95~\% C.L.~regions of different observables and parameters for NO and IO for two values $\kappa=1,10$ for the naturality upper bound on the trilinear coupling $\mu$, see eq.~\eqref{eq:mu}.}
\end{table}

\section{Summary and conclusions}
\label{sec:conc}

It is well known that there is LFV in the neutrino sector and this is also expected in the charged-lepton sector. In this work, we have studied the Zee model in detail, which is a simple extension of the SM that can accommodate neutrino masses and leptonic mixing if at the same time sizable signals in LFV processes are generated. We have performed a full numerical scan of the parameter space for three different cases (i) $\mu=0$, which implies massless neutrinos, (ii) NO, and (iii) IO. We have found that neutrino masses and leptonic mixing can be easily accommodated in NO, whereas IO is disfavored in comparison to NO due to the difficulty to fit the leptonic mixing angles $\theta_{12}$ and $\theta_{23}$ as well as the Dirac CP-violating phase $\delta$. In fact, if $\theta_{23}$ turns out to be in the first octant, only NO would be allowed in the Zee model. Note also that none of the orderings can reproduce the hint of $\delta\simeq 3\pi/2$ from global fits to neutrino oscillation data \cite{Esteban:2016qun,NuFIT3.0}.

If expected sensitivities are achieved in $\tau \to \mu \gamma$ and no signal is observed, a significant portion of the allowed parameter space for NO would be ruled out. This would put the Zee model under severe pressure, requiring an extension, e.g.~involving the Yukawa couplings that give rise to terms proportional to $m_e$ and possibly large hierarchies among them, and relaxing the naturality demands on the trilinear coupling $\mu$ considerably, as ${\rm Br} (\tau \to \mu \gamma)\sim 10^{-9}$ would still rule out NO for $\kappa\sim 10$. If no signals are observed in future $\mu e$ conversion experiments, which are expected to increase their sensitivities by several orders of magnitude, the allowed regions in the parameter space will be strongly reduced, and IO will be basically excluded. 

We have analyzed if the predicted rates of the Zee model for HLFV decays are observable at the LHC and future colliders. We have found that $\mathrm{Br}(h\to \tau \mu)$ can be at the percent level, whereas $\mathrm{Br}(h\to \tau e)$ is at least two orders of magnitude smaller. If no signals are observed for ${\rm Br}(h\to \tau \mu)$ at future colliders at the $10^{-5}$ level, both orderings will be excluded at $1\sigma$ C.L. In general, we find that the expected sensitivities for CLFV will be more constraining in the near future. However, both CLFV and HLFV will have significant impact on the allowed parameter space of the Zee model.

In the model, neutrinoless double beta is due to only light neutrinos. Therefore, as usual, current experiments will be only sensitive to IO, while NO will only be tested if a further-order-of-magnitude improvement is achieved. In the Zee model, NSIs are always very suppressed, the strong limits from other CLFV processes and the fact that neutrino masses need to be generated at one-loop level.

In general, we have found that the masses of the new scalars should be at most a few TeV, which implies that they, especially the charged scalars that are pair-produced via Drell--Yan processes, can be searched for at the LHC. In particular, the masses of the neutral scalars and the charged scalar $h_1^+$ are below $2.5$~TeV, and typically they are lower than that, for both NO and IO. The phenomenology of the scalar sector is very model-dependent, like in general for 2HDMs, although it is possible to have sizable decays of the heavy neutral scalars into $\tau \mu$, correlated with the light Higgs ones.

We conclude by emphasizing that the general Zee model studied in this work is fully testable in the near future by combining different LFV processes. In particular, both orderings should be completely tested by CLFV and HLFV processes in the forthcoming years, as well as collider searches of the new scalars. Furthermore, if a signal in $h\to \tau \mu$ is observed at the LHC or in a future collider, the Zee model will be one of the best-motivated scenarios to accommodate it and at the same time describe neutrino masses and leptonic mixing.

\section*{Acknowledgements}

We thank Nuria Rius, Arcadi Santamaria, and Thomas Schwetz for useful discussions and comments on the manuscript, Mikael Twengstr\"{o}m for technical advice on numerical computations, and Andrew Fowlie for providing help with {\sc Superplot}. We also acknowledge Thomas Schwetz for suggesting to show the individual contributions to the $\chi^2$ function and for discussions regarding global fits of neutrino oscillation data. Numerical computations were performed on resources provided by the Swedish National Infrastructure for Computing (SNIC) at PDC Center for High Performance Computing (PDC-HPC) at KTH Royal Institute of Technology in Stockholm, Sweden under project numbers PDC-2016-8, PDC-2016-27, PDC-2016-36, PDC-2016-60, and PDC-2016-82. This work was supported by the University of Adelaide and the Australian Research Council through the ARC Centre of Excellence for Particle Physics at the Terascale (CoEPP) (CE110001104).

\appendix

\section{Electroweak precision tests}
\label{sec:ewpt}

The Peskin--Takeuchi parameters $S$, $T$, and $U$ give a parametrization of the new physics contributions to electroweak radiative quantities, in particular to gauge boson self-energies~\cite{Peskin:1990zt,Peskin:1991sw}. We follow closely ref.~\cite{Haber:2010bw}, extending their results for the 2HDM to the Zee model by adding the extra contributions stemming from the singly-charged scalar singlet $h^+$. We refer the reader to ref.~\cite{Haber:2010bw} for additional details on the evaluations of the one-loop self-energies.

In the Zee model, the parameter $T$ is given by\footnote{Using ref.~\cite{Grimus:2007if}, one would obtain a different factor in front of the first term on the third line: $- 2 s^2_\varphi c^2_\varphi \mathcal{F}(m_{h_1^+}^2,m_{h_2^+}^2)$.}
\begin{align} 
T &= \dfrac{1}{16\pi^2 \alpha_{\rm em} v^2} \left\lbrace  c^2_\varphi \left[ c^2_{\beta-\alpha} \mathcal{F}(m_{h_1^+}^2,m_{h}^2) + s^2_{\beta-\alpha}\mathcal{F}(m_{h_1^+}^2,m_{H}^2) + \mathcal{F}(m_{h_1^+}^2,m_{A}^2) \right] \right. \nonumber\\
&+ s^2_\varphi \left[ c^2_{\beta-\alpha} \mathcal{F}(m_{h_2^+}^2,m_{h}^2) + s^2_{\beta-\alpha}\mathcal{F}(m_{h_2^+}^2,m_{H}^2) + \mathcal{F}(m_{h_2^+}^2,m_{A}^2) \right] \nonumber\\ 
&- \frac{1}{2}\,s^2_\varphi c^2_\varphi\, \mathcal{F}(m_{h_1^+}^2,m_{h_2^+}^2) - c^2_{\beta-\alpha} \mathcal{F}(m_{h}^2,m_{A}^2) - s^2_{\beta-\alpha} \mathcal{F}(m_{H}^2,m_{A}^2) \nonumber\\
&+ \left. 3 c^2_{\beta-\alpha} \left[ \mathcal{F}(m_{Z}^2,m_{H}^2) - \mathcal{F}(m_{W}^2,m_{H}^2) - \mathcal{F}(m_{Z}^2,m_{h}^2) + \mathcal{F}(m_{W}^2,m_{h}^2) \right] \right\rbrace \,, \label{eq:T}
\end{align}
where $\alpha_{\rm em} \equiv e^2/(4\pi)$ is Sommerfeld's fine-structure constant\footnote{Note that $\pi \alpha_{\rm em} v^2 = m_W^2 s_W^2$, where $s_W = \sin \theta_W$ and $\theta_W$ being the Weinberg angle.} and the symmetric auxiliary function $\mathcal{F}$ is defined as
\begin{equation} \label{Fdef}
\mathcal{F}(m_1^2,m_2^2) =\mathcal{F}(m_2^2,m_1^2) \equiv \frac{m_1^2+m_2^2}{2} -\frac{m_1^2m_2^2}{m_1^2-m_2^2}\ln\frac{m_1^2}{m_2^2}\,.
\end{equation}
Similarly, for the parameter $S$, following ref.~\cite{Haber:2010bw} and adding the singly-charged contributions to the different gauge boson self-energies, we obtain
\begin{align}  \nonumber
S&= \frac{1}{\pi m_Z^2} \biggl\{s^2_{\beta-\alpha}
\mathcal{B}_{22}(m_Z^2,m_{H}^2,m_{A}^2)
\nonumber +c^2_{\beta-\alpha}
\bigl[\mathcal{B}_{22}(m_Z^2,m_{h}^2,m_{A}^2)
 + \mathcal{B}_{22}(m_Z^2,m_Z^2,m_{H}^2)\\ \nonumber
&-\mathcal{B}_{22}(m_Z^2,m_Z^2,m_{h}^2)- m_Z^2\mathcal{B}_{0}(m_Z^2,m_Z^2,m_{H}^2)
+m_Z^2\mathcal{B}_{0}(m_Z^2,m_Z^2,m_{h}^2) \bigl]\\
&
+\frac{c^2_\varphi}{2}(c_{2\varphi}-3)\,\mathcal{B}_{22}(m_Z^2,{m^2_{h_1^+}},{m^2_{h_1^+}}) 
-\frac{s^2_\varphi}{2}(c_{2\varphi}+3)\,\mathcal{B}_{22}(m_Z^2,{m^2_{h_2^+}},{m^2_{h_2^+}}) \nonumber\\
&
+2 s^2_\varphi c^2_\varphi\, \mathcal{B}_{22}(m_Z^2,{m^2_{h_1^+}},{m^2_{h_2^+}})
\biggr\}\,, \label{eq:S}
\end{align}
and for the combination $S+U$, we find\footnote{We believe there are typos in the last two terms of ref.~\cite{Haber:2010bw}, which should have opposite signs.}
\begin{align}  \nonumber
S+ U &= \frac{1}{\pi m_W^2} \nonumber\\
&\times \biggl\{ c^2_\varphi \left[\mathcal{B}_{22}(m_W^2,{m^2_{h_1^+}},m_{A}^2) +s^2_{\beta-\alpha}\mathcal{B}_{22}(m_W^2, m^2_{h_1^+},m_{H}^2) +c^2_{\beta-\alpha}\mathcal{B}_{22}(m_W^2,m_{h}^2,m^2_{h_1^+})\right]\nonumber\\
&+s^2_\varphi \left[\mathcal{B}_{22}(m_W^2,{m^2_{h_2^+}},m_{A}^2) +s^2_{\beta-\alpha} \mathcal{B}_{22}(m_W^2, m^2_{h_2^+},m_{H}^2) +c^2_{\beta-\alpha}\mathcal{B}_{22}(m_W^2,m_{h}^2,m^2_{h_2^+})\right]\nonumber \\
&+c^2_{\beta-\alpha}\bigl[\mathcal{B}_{22}(m_W^2,m_W^2,m_{H}^2)-\mathcal{B}_{22}(m_W^2,m_W^2,m_{h}^2)\nonumber\\
&-m_W^2\mathcal{B}_{0}(m_W^2,m_W^2,m_{H}^2)+m_W^2\mathcal{B}_{0}(m_W^2,m_W^2,m_{h}^2)\bigr] \nonumber\\
&-2 c^2_\varphi \,\mathcal{B}_{22}(m_W^2,{m^2_{h_1^+}},{m^2_{h_1^+}})-2 s^2_\varphi \,\mathcal{B}_{22}(m_W^2,{m^2_{h_2^+}},{m^2_{h_2^+}})
\biggr\}\,, \label{eq:S+U}
\end{align}
where the renormalized auxiliary functions $\mathcal{B}_{22}$ and $\mathcal{B}_0$ are defined as
\begin{align} 
\mathcal{B}_{22}(q^2,m_1^2,m_2^2) &\equiv
B_{22}(q^2,m_1^2,m_2^2)-B_{22}(0,m_1^2,m_2^2)\,,\label{B22} \\
\mathcal{B}_{0}(q^2,m_1^2,m_2^2) &\equiv B_{0}(q^2,m_1^2,m_2^2)-B_{0}(0,m_1^2,m_2^2) \label{B0}
\end{align}
with the Passarino--Veltman functions $B_{22}$ and $B_{0}$~\cite{Passarino:1978jh}, arising from
two-point self-energies. 
Using eqs.~\eqref{eq:S} and \eqref{eq:S+U}, one can readily obtain an expression for the parameter $U$. 
In the limit $s_\varphi=0$, it can be easily checked that one recovers the 2HDM results of the electroweak precision tests of ref.~\cite{Haber:2010bw}.
Finally, in dimensional regularization, the two functions $B_{22}$ and $B_{0}$ read~\cite{Haber:2010bw}
\begin{align} 
B_{22}(q^2,m_1^2,m_2^2) &= \tfrac{1}{4}(\Delta+1)(m_1^2+m_2^2-\tfrac{1}{3}q^2)-\frac{1}{2}\int^1_0 X \ln(X-i\epsilon) \, {\rm d}x \,, \label{BB22}\\
B_{0}(q^2,m_1^2,m_2^2) &= \Delta-\int^1_0 \ln(X-i\epsilon) \, {\rm d}x \, \label{BB0}
\end{align}
with
\begin{equation}
X \equiv m_1^2 x + m_2^2(1-x) -q^2x(1-x)\,, \qquad
\Delta \equiv \frac{2}{4-d} + \ln (4\pi) - \gamma
\end{equation}
in $d$ space-time dimensions, where $\gamma \simeq 0.577$ is the Euler--Mascheroni constant. Note that $B_{22}$ and $B_{0}$ are symmetric in their last two arguments. In appendix~\ref{sec:PV}, we derive explicit analytical expressions for $B_{22}$ and $B_0$ as well as $\mathcal{B}_{22}$ and $\mathcal{B}_0$.

\section{Explicit analytical expressions for the Passarino--Veltman functions $B_0$ and $B_{22}$ and the renomalized auxiliary functions $\mathcal{B}_{0}$ and $\mathcal{B}_{22}$}
\label{sec:PV}

Following the seminal work by Passarino and Veltman~\cite{Passarino:1978jh} closely, it holds that\footnote{Note that we use a different sign convention for $B_{22}$ than ref.~\cite{Passarino:1978jh}.}
\begin{align}
B_{22}(q^2,m_1^2,m_2^2) &= \tfrac{1}{6} \big[ A_0(m_1^2) + \left( m_1^2 + m_2^2 - \tfrac{1}{3} q^2 \right) + 2 m_2^2 B_0(q^2,m_1^2,m_2^2) \nonumber\\
&+ (m_2^2 - m_1^2 + q^2) B_1(q^2,m_1^2,m_2^2) \big]\,, \label{eq:B22rel}
\end{align}
where the additional Passarino--Veltman function $A_0$ is given by
\begin{equation}
A_0(m^2) = m^2 \left( \Delta + 1 - \ln m^2 \right)\,. \label{eq:A0}
\end{equation}
Furthermore, we obtain
\begin{align}
B_0(q^2,m_1^2,m_2^2) &= \Delta + 2 - \ln q^2 - \ln(1-x_1) - \ln(1-x_2) + x_1 \ln \frac{x_1-1}{x_1} + x_2 \ln \frac{x_2-1}{x_2}\,, \label{eq:B0PV}\\
B_1(q^2,m_1^2,m_2^2) &= -\frac{1}{2} \Delta - \frac{1}{2} (x_1 + x_2) - \frac{1}{2} + \frac{1}{2} \ln q^2 \nonumber\\
&+ \frac{1}{2} \left[ \ln(1-x_1) + \ln(1-x_2) - x_1^2 \ln \frac{x_1-1}{x_1} - x_2^2 \ln \frac{x_2-1}{x_2} \right]\,, \label{eq:B1PV}
\end{align}
where $x_1$ and $x_2$ are the roots of the equation $q^2 x^2 + (m_1^2-m_2^2-q^2) x + m_2^2 = 0$. Inserting eqs.~\eqref{eq:A0}, \eqref{eq:B0PV}, and \eqref{eq:B1PV} into eq.~\eqref{eq:B22rel}, it follows after some tedious calculations that the functions $B_{22}$ and $B_0$ can be written as
\begin{align}
B_{22}(q^2,m_1^2,m_2^2) &= \frac{1}{12} \biggl\{ -\frac{(m_1^2-m_2^2)^2}{q^2} + 7 (m_1^2+m_2^2) - \frac{Y^3}{2q^4} \ln \frac{m_1^2+m_2^2-q^2+Y}{m_1^2+m_2^2-q^2-Y} \nonumber\\
&+ \left[ \frac{(m_1^2-m_2^2)^3}{q^4} - \frac{3(m_1^4-m_2^4)}{q^2} \right] \ln \frac{m_1}{m_2} \nonumber\\
&+ \left[\Delta - \ln(m_1 m_2)\right] \left[3(m_1^2+m_2^2) - q^2\right] - \frac{8}{3} q^2 \biggr\}\,, \label{eq:B22PV}\\
B_0(q^2,m_1^2,m_2^2) &= \Delta + 2 + \frac{Y}{2q^2} \ln \frac{m_1^2+m_2^2-q^2+Y}{m_1^2+m_2^2-q^2-Y} - \frac{1}{q^2} (m_1^2 - m_2^2) \ln \frac{m_1}{m_2} - \ln(m_1 m_2)\,, \label{eq:B0PV2}
\end{align}
where
\begin{equation}
Y \equiv \sqrt{[(m_1+m_2)^2-q^2][(m_1-m_2)^2-q^2]}\,.
\end{equation}

In fact, it is possible to find closed-form expressions even for the auxiliary functions $\mathcal{B}_{22}$ and $\mathcal{B}_{0}$. Inserting eqs.~\eqref{eq:B22PV} and \eqref{eq:B0PV2} into eqs.~\eqref{B22} and \eqref{B0}, respectively, and using the functions $B_{22}$ and $B_0$ evaluated at $q^2 = 0$, i.e.
\begin{align}
B_{22}(0,m_1^2,m_2^2) &= \frac{1}{8} (2 \Delta + 3) (m_1^2+m_2^2) - \frac{1}{2} \frac{m_1^4 \ln m_1 - m_2^4 \ln m_2}{m_1^2-m_2^2} \,, \\
B_0(0,m_1^2,m_2^2) &= \Delta + 1 - \frac{m_1^2 \ln m_1^2-m_2^2 \ln m_2^2}{m_1^2-m_2^2} \,,
\end{align}
we obtain for $m_1,m_2 > q > 0$ and $(m_1-m_2)^2 > q^2$
\begin{align}
\mathcal{B}_{22}(q^2,m_1^2,m_2^2) &= \frac{1}{12} \biggl\{ -\frac{(m_1^2-m_2^2)^2}{q^2} + \frac{5}{2} (m_1^2+m_2^2) - \frac{Y^3}{2q^4} \ln \frac{m_1^2 + m_2^2 -q^2 + Y}{m_1^2 + m_2^2 - q^2 - Y} \nonumber\\
&+ \left[ \frac{(m_1^2-m_2^2)^3}{q^4} - \frac{3(m_1^4-m_2^4)}{q^2} + \frac{3(m_1^4+m_2^4)}{m_1^2-m_2^2} \right] \ln \frac{m_1}{m_2} \nonumber\\
&- \left[ \frac{8}{3} + \Delta - \ln (m_1 m_2) \right] q^2 \biggr\} \,, \label{BBB22}\\
\mathcal{B}_0(q^2,m_1^2,m_2^2) &= 1 + \frac{Y}{2q^2} \ln \frac{m_1^2 + m_2^2 - q^2 + Y}{m_1^2 + m_2^2 - q^2 -Y} + \frac{(m_1^2+m_2^2)q^2 - (m_1^2-m_2^2)^2}{(m_1^2-m_2^2)q^2} \ln \frac{m_1}{m_2} \,, \label{BBB0}
\end{align}
whereas for $m_1,m_2 > q > 0$ and $(m_1-m_2)^2 < q^2$, we have to make the replacements $Y \to i Y'$ and
$$
\ln \frac{m_1^2 + m_2^2 -q^2 + Y}{m_1^2 + m_2^2 - q^2 - Y} \to i \arctan \frac{(m_1^2+m_2^2-q^2) Y'}{(m_1^2+m_2^2-q^2)^2 - 2 m_1^2 m_2^2}
$$
in eqs.~\eqref{BBB22} and \eqref{BBB0}, respectively, where $Y' \equiv \sqrt{[(m_1+m_2)^2-q^2][q^2-(m_1-m_2)^2]}$.
In the case when $m_1 = q$ and $m_2 = m$ (which is useful for computing eqs.~\eqref{eq:S} and \eqref{eq:S+U}), we have for $0 < m/2 < q < m$
\begin{align}
\mathcal{B}_{22}(q^2,q^2,m^2) &= \frac{1}{12} \biggl\{ \frac{3}{2} q^2 + \frac{9}{2} m^2 - \frac{m^4}{q^2} - \frac{m^3Z^3}{q^4} \arctan \frac{Z}{m} \nonumber\\
&+ \left[ \frac{(m^2-q^2)^3}{q^4} - \frac{3(m^4-q^4)}{q^2} + \frac{3(m^4+q^4)}{m^2-q^2} \right] \ln \frac{m}{q} \nonumber\\
&- \left[ \frac{8}{3} + \Delta - \ln(mq) \right] q^2 \biggr\} \,, \label{BBB22qqm}\\
\mathcal{B}_0(q^2,q^2,m^2) &= 1 - \frac{mZ}{q^2} \arctan \frac{Z}{m} - \frac{m^2}{q^2} \frac{m^2-3q^2}{m^2-q^2} \ln \frac{m}{q} \,, \label{BBB0qqm}
\end{align}
where
\begin{equation}
Z \equiv \sqrt{4q^2-m^2}\,,
\end{equation}
whereas for $0 < q < m/2$, we have to make the replacements $Z \to i Z'$ and
$$
\arctan \frac{Z}{m} \to i {\rm \, artanh\,} \frac{Z'}{m} = \frac{i}{2} \ln \frac{m+Z'}{m-Z'}
$$
in eqs.~\eqref{BBB22qqm} and \eqref{BBB0qqm}, respectively, where $Z' \equiv \sqrt{m^2-4q^2}$. Similarly, in the case when $m_1 = m_2 = m > q > 0$, we find
\begin{align}
\mathcal{B}_{22}(q^2,m^2,m^2) &= \frac{1}{12} \left[ 8 m^2 - \frac{2W^3}{q} \arctan \frac{q}{W} - \left( \frac{8}{3} + \Delta - 2 \ln m \right) q^2 \right] \,, \label{BBB22qmm}\\
\mathcal{B}_0(q^2,m^2,m^2) &= 2 - \frac{2W}{q} \arctan \frac{q}{W} \,,
\end{align}
where
\begin{equation}
W \equiv \sqrt{4m^2-q^2}\,.
\end{equation}

\subsection{Comments on cancellation of dimensionful logarithms and divergent terms in $S$ and $S+U$}

Note that the terms proportional to the dimensionful logarithm $\ln(m_1 m_2)$ in eq.~\eqref{BBB22} as well as in eqs.~\eqref{BBB22qqm} and \eqref{BBB22qmm} sum up to contributions for both $S$ in eq.~\eqref{eq:S} (contribution of seven $\mathcal{B}_{22}$ functions) and $S+U$ in eq.~\eqref{eq:S+U} (contribution of ten $\mathcal{B}_{22}$ functions) that are proportional to $c_\varphi^2 [ \ln(m_A/m_{h_1^+}) + \ln(m_H/m_{h_1^+}) ] + s_\varphi^2 [ \ln(m_A/m_{h_2^+}) + \ln(m_H/m_{h_2^+}) ]$, which is indeed a result of dimensionless logarithms. Furthermore, it should be noted that all terms on the form $-(\Delta +1) q^2/12$ (which contains a divergency) in eq.~\eqref{B22} with eq.~\eqref{BB22} cancel exactly for both $S$ and $S+U$ in eqs.~\eqref{eq:S} and \eqref{eq:S+U}, respectively, as they must. This also holds true for the divergent terms in $T$ in eq.~\eqref{eq:T}.

\providecommand{\href}[2]{#2}\begingroup\raggedright\endgroup

\end{document}